\documentclass[useAMS,usenatbib]{mn2e}
\usepackage{graphicx}
\usepackage{epsfig}
\usepackage{amssymb}
\usepackage{lscape}
\usepackage{ulem}
\usepackage{txfonts}
\usepackage{lipsum}

\title[ANN Quenching]{An Artificial Neural Network Approach For Ranking Quenching Parameters In Central Galaxies}

\author[Teimoorinia, Bluck \& Ellison] {Hossen Teimoorinia$^{1}$, Asa F. L. Bluck$^{1,2}$ \&
Sara L. Ellison$^{1}$\\
$^{1}$Department of Physics \& Astronomy, University of Victoria, 3800 Finnerty Road, Victoria, British Columbia, V8P 1A1, Canada. \\
$^{2}$ Institute of Astronomy, Department of Physics, ETH Zurich, Wolfgang-Pauli-Strasse 27, Zurich, CH-8093, Switzerland. \\}

\def\LaTeX{L\kern-.36em\raise.3ex\hbox{a}\kern-.15em
    T\kern-.1667em\lower.7ex\hbox{E}\kern-.125emX}

\begin{document}

\label{firstpage}

\maketitle

\begin{abstract}

We present a novel technique for ranking the relative importance of galaxy properties in the process of quenching star formation. Specifically, we develop an artificial neural network (ANN) approach for pattern recognition and apply it to a population of over 400,000 central galaxies taken from the Sloan Digital Sky Survey Data Release 7. We utilise a variety of physical galaxy properties for training the pattern recognition algorithm to recognise star forming and passive systems, for a `training set' of $\sim$100,000 galaxies. We then apply the ANN model to a `verification set' of $\sim$100,000 different galaxies, randomly chosen from the remaining sample. The success rate of each parameter singly, and in conjunction with other parameters, is taken as an indication of how important the parameters are to the process(es) of central galaxy quenching. 
 We find that central velocity dispersion, bulge mass and B/T are excellent predictors of the passive state of the system, indicating that properties related to the central mass of the galaxy are most closely linked to the cessation of star formation.  Larger scale galaxy properties (total or disk stellar masses), or those linked to environment (halo masses or $\delta_5$) perform significantly less well.  Our results are plausibly explained by AGN feedback driving the quenching of central galaxies, although we discuss other possibilities as well.

\end{abstract}


\section{Introduction}

Explaining why galaxies stop forming stars is a challenging problem in modern astrophysics. The fact that galaxies are observed to come in two broad `types' in the local Universe is evidenced by the bimodality of several fundamental galaxy properties, including star formation rate (SFR), integrated galaxy colour, and morphology (e.g. Strateva et al. 2001, Brinchmann et al. 2004, Baldry et al. 2004, Driver et al. 2006, Baldry et al. 2006, Cameron \& Driver 2009, Cameron et al. 2009,  Wuyts et al. 2011, Peng et al. 2010, 2012,  Wake et al. 2012). A compelling picture of how and why galaxies form into these distinct classes is emerging from the theoretical perspective of hierarchical assembly of dark matter haloes, and galaxy formation and feedback within these structures 
(e.g. Cole et al. 2000, Springel et al. 2005, Bower et al. 2006, Croton et al. 2006, De Lucia et al. 2006, De Lucia \& Blaizot 2007, Somerville et al. 2008, Bower et al. 2008, Guo et al. 2011, Henriques et al. 2014, Vogelsberger et al. 2014a,b, Schaye et al. 2015). However, many of the details, including exactly what set of processes cause the quenching of galaxies, is still debated (e.g. Bell et al. 2012, Carollo et al. 2013, Woo et al. 2013, Bluck et al. 2014, Dekel et al. 2014, Knobel et al. 2014, Bluck et al. 2015, Woo et al. 2015, Tacchella et al. 2015, Peng et al. 2015).

The fraction of passive (non-star forming galaxies) in a given population has been found to depend strongly on both the stellar mass of the galaxy and the local density in which it resides (Baldry et al. 2006, Peng et al. 2010). The natural division of galaxies by whether or not they are the most massive `central' galaxy or less massive `satellite' galaxies in a given dark matter halo, has yielded further insight on this issue, with Peng et al. (2012) finding that central galaxies have a passive fraction mostly correlated with their stellar mass and satellites being more affected by local density. In addition to mass and local density, the structure or morphology of a galaxy also has a strong impact on the passive fraction (e.g. Driver et al. 2006, Cameron et al. 2009, Cameron \& Driver 2009, Mendel et al. 2013, Bluck et al. 2014). 
More recent work has found that the central density or mass of the galactic bulge can provide a particularly tight constraint on the passive fraction (e.g. Cheung et al. 2012, Fang et al. 2013, Bluck et al. 2014, Lang et al. 2014, Omand et al. 2014, Woo et al. 2015). However, there is also evidence that the mass of the group or cluster dark matter halo, calculated via indirect abundance matching techniques, is a tighter constraint on the passive fraction of centrals than stellar mass (Woo et al. 2013), but not bulge mass or centralised velocity dispersion (Bluck et al. 2014, 2015, Woo et al. 2015). 

There are several viable quenching mechanisms suggested theoretically. Galaxy merging offers an initially tempting explanation because it can, in principle, explain the bimodality in SFR (or colour) and morphology (or structure) simultaneously. Galaxies with recent (major) mergers will have their disk components disrupted and diminished and their bulges enhanced (e.g. Toomre \& Toomre 1972, Barnes \& Hernquist 1992, Cole et al. 2000), although if the merger is gas rich disks may reform (e.g. Burkert \& Naab 2004, Springel \& Hernquist 2005, Hopkins et al. 2013). 
Additionally, the merging galaxy will initially also have elevated star formation (as seen observationally in, e.g., Ellison et al. 2008, Scudder et al. 2012, Hung et al. 2013, Patton et al. 2013, Ellison et al. 2013) and hence presumably gas consumption (although the observational evidence for this link is mixed, e.g. Ellison et al. 2015 and references therein), potentially leading to a significant lowering of SFR due to a lack of further fuel for star formation (as seen in recent simulations, e.g. Moreno et al. 2015). 
However, if the galaxy remains connected to the Universe, gas replenishment will inevitably occur from cooling of the hot gas halo, cold gas streams, and minor gas rich mergers. Therefore, merging by itself cannot account for truly (or permanently) passive systems, additional processes will be needed. This is true generally for any quenching mechanism which `strips' gas from a galaxy but does not prevent further gas inflow, i.e. `strangling' the galaxy (see Peng et al. 2015 for a discussion).

For centrals, it is clear that a source of heat and/or mechanical disruption will be necessary to prevent cooling or accretion of gas onto a galaxy in order for it to cease forming stars. This can be achieved in numerous ways, e.g. through energetic feedback from active galactic nuclei (AGN) (e.g. McNamara et al. 2000, Nulsen et al. 2005, Hopkins et al. 2006a,b, Croton et al. 2006, Bower et al. 2008, Hopkins et al. 2008, Dunn et al. 2010, Hopkins et al. 2010, Fabian 2012), supernovae and stellar winds (e.g. Dalla \& Schaye 2008, Guo et al. 2012, Vogelsberger et al. 2014a, Schaye et al. 2015), or by stabilizing virial shocks in haloes above some critical dark matter mass (e.g. Dekel \& Birnboim 2006, Dekel et al. 2009, Woo et al. 2013, Dekel et al. 2014). 
One other alternative is that the gas is in fact present and continues to be replenished, but somehow cannot be generated into new stellar populations, possibly due to stabilizing torques applied across giant molecular clouds from centrally concentrated mass sources (e.g. Martig et al. 2009). This latter option, however, does not appear to have strong observational support since passive galaxies are most frequently found to lack cold gas reservoirs, which must be explained by other feedback mechanisms that can by themselves account for the lack of ongoing star formation in massive galaxies (e.g. Catinella et al. 2010, Saintonge et al. 2011, Genzel et al. 2015).

Due to the relative motion of satellite galaxies through the dark matter potential of the group or cluster, and across the hot gas halo, there are several additional routes available for the quenching of satellite galaxies compared to centrals. Processes such as galaxy - galaxy and host halo tidal interactions, ram pressure stripping, removal of the hot gas halo and subsequent stifling of gas supply from cooling, and pre-processing in groups prior to cluster infall can all result in the quenching of satellite galaxies (e.g. Balogh et al. 2004, Cortese et al. 2006, Moran et al. 2007, van den Bosch et al. 2007,2008, Tasca et al. 2009, Peng et al. 2012, Hirschmann et al. 2013, Wetzel et al. 2013). 
These environmental processes work in concert with the mass-correlating central galaxy quenching mechanisms outlined above. Thus, the quenching of satellite galaxies is likely to be a much more complex process than that of centrals. In this first work on applying ANN techniques to galaxy quenching we focus on the simpler central galaxy population, with a publication on satellite galaxy quenching to follow (Bluck et al., in prep.).

We can potentially identify the dominant central galaxy quenching mechanism, from the contenders outlined above, by investigating which galaxy properties are most closely correlated with the passive fraction. For example, the total energy available for feedback on a galaxy released via an AGN will be roughly proportional to the mass of the central supermassive black hole (Soltan 1982, Silk \& Rees 1998, Fabian 1999, Bluck et al. 2011, 2014) and hence to the central velocity dispersion and bulge mass (Magorrian et al. 1998, Ferrarese \& Merritt 2000, Gebhardt et al. 2000, Haring \& Rix 2004, Hopkins et al. 2007, McConnell \& Ma 2013). Alternatively, the total energy released from supernovae over the lifetime of a galaxy will be roughly proportional to the total stellar mass of the galaxy, as integrated star formation rate (e.g. Croton et al. 2006, Guo et al. 2011). 
Further, the energy available from virial shocks in dark matter haloes will be proportional to the the gravitational potential, i.e. the dark matter halo mass, and hence also to the total stellar mass of the group or cluster (Dekel \& Birnboim 2006, Dekel et al. 2009, Woo et al. 2013, 2015).

Several attempts have been made to identify which galaxy properties are most closely linked to quenching for central and satellite galaxies (e.g. Peng et al. 2010, 2012, Woo et al. 2014, Bluck et al. 2014, 2015, Woo et al. 2015). However, these studies are typically only able to consider one or two variables at a time, motivating the need for a more inclusive and sophisticated analysis methodology. Artificial neural networks (ANNs) are a powerful tool for analysing large and complex datasets and exposing patterns in non-linear physical systems in industry, engineering and the biological sciences (see Wichchukit \& O'Mahony 2010 and references therein). Their application to astrophysics has so far been somewhat limited, although there are some noticeable exceptions and successes (e.g. Andreon et al. 2001, Ball et al. 2004, Teimoorinia et al. 2012, Teimoorinia \& Ellison 2014). 
In this work we apply ANN pattern recognition techniques to the multi-variate ranking of parameters that distinguish star forming from passive galaxies. Our aim is to use these rankings to provide observational evidence for or against the dominant quenching mechanisms of central galaxies.

The paper is structured as follows: Section 2 describes our data and sample selection. Section 3 outlines the details of our ANN method and analysis methodology as applied to the SDSS. Section 4 presents our results for centrals, including single and multi-variables. We discuss what drives central galaxy quenching in light of our results in Section 5, and conclude by giving a summary of our contribution in Section 6. We present an investigation of the potential for sample biases and systematics to affect the results in the Appendix. Throughout we assume a $\Lambda$CDM cosmology with \{$\Omega_{M}$, $\Omega_{\Lambda}$, $H_{0}$\} = \{0.3, 0.7, 70 ${\rm km~s^{-1}~Mpc^{-1}}$\}, and adopt AB magnitude units.

\section{Data}

\subsection{Overview}

Our data source is the Sloan Digital Sky Survey Data Release 7 (SDSS DR7, Abazajian et al. 2009) spectroscopic sample. We form a sub-sample of 414915 central galaxies with stellar masses in the range 9 $< {\rm log}(M_{*}/M_{\odot}) <$ 12 at $z_{\rm spec} <$ 0.2. Full details of this sample, and on the stellar masses, morphologies and structures, star formation rates, and environments of these galaxies are given in Bluck et al. (2014) Section 2, and references therein. What follows in this sub-section is a brief overview of the most important details.

The star formation rates for our sample are derived in Brinchmann et al. (2004), with adaptions made in Salim et al. (2007). These are based on spectroscopic emission lines for star forming galaxies with strong emission lines which are not identified as AGN, and via an empirical relationship between the strength of the 4000 \AA \hspace{0.1cm} break (D$_{n}$4000) and the specific star formation rate of a galaxy (sSFR = SFR/$M_{*}$) for non-star forming (weak or non-emission line) galaxies and AGN. AGN are determined by the Kauffmann et al. (2003) line cut applied to the Baldwin, Phillips \& Terlevich (BPT, 1981) emission line diagram, at a S/N $>$ 1. For the strong emission line galaxies which are not AGN, the SFRs are based on H$\alpha$, H$\beta$, [OIII] and [NII] line strengths. 
For both methods for deducing SFRs a fibre correction is applied, based on the colour and magnitude of light not contained within the spectroscopic fibre. 

Rosario et al. (2015) have demonstrated that the D$_{n}$4000 SFRs can be quite inaccurate; however, in this work we only aim to separate star forming from passive systems. Thus, the high error associated with SFRs in passive systems does not significantly impact our ability to identify them as passive. This is a more complex issue for AGN, where many galaxies could be star forming and still have their SFRs determined from the D$_{n}$4000 method. To combat this, we test the effect of removing AGN from our sample in \S A2. We find that this does not alter any of our results or conclusions, and hence that our rankings are stable to possible inaccuracies in the SFRs.

Stellar masses for our sample, and for the component disks and spheroids, are computed in Mendel et al. (2014) via fitting the observed $ugriz$ magnitudes to model spectral energy distributions (SEDs). For the components, a dual S\'{e}rsic ($n_{s}$ = 4 bulge, $n_{s}$ = 1 disk) model is applied in each of the Sloan wave-bands, and combined to form a stellar mass for the bulge and disk components via SED fitting. Details on the bulge-to-total light fitting can be found in Simard et al. (2011), which is based on a GIM2D decomposition (Simard et al. 2002), and details on the mass determination is provided in Mendel et al. (2014). From this, we define the galaxy structure to be:

\begin{equation}
B/T = \frac{M_{\rm bulge}}{M_{*}} =  \frac{M_{\rm bulge}}{M_{\rm bulge} + M_{\rm disk}}
\end{equation}

\noindent where $M_{*}$ is the total stellar mass of the galaxy, taken here as the sum of the component bulge and disk masses. Note that since this is a mass ratio, it is not affected by ongoing star formation and hence provides an independent measure of galaxy structure. This would not be the case with a $B/T$ parameter {\it by light} based on a single optical wave-band or a classic S\'{e}rsic index (also based on a given wave-band).  Bulge effective radius is also taken from the public catalogues released in Simard et al. (2011).

Halo masses are estimated from an abundance matching technique applied to the total stellar mass of the group or cluster in which each central galaxy resides. These are taken from the SDSS group catalogues of Yang et al. (2007, 2008, 2009). At $M_{\rm halo} > 10^{12} M_{\odot}$ over 90 \% of galaxies are correctly assigned to groups in model data from the Millennium Simulation (Springel et al. 2005). Within these groups, the most massive galaxy is defined as the central and all other group members are defined as satellites of that central. This is the same sample of estimated halo masses used in other recent quenching papers (e.g. Woo et al. 2013, Bluck et al. 2014, 2015, Woo et al. 2015).

Velocity dispersions in our sample are derived from the widths of absorption lines, made public in Bernardi et al. (2003), with updates to the method added in Bernardi et al. (2007). We discard all velocity dispersions which are derived from line widths with a S/N $<$ 3.5. We also remove all cases where $\sigma_{\rm err} >$ 50 km s$^{-1}$ (only a few percent of the sample). Further, for some analyses, we restrict the sample to $\sigma >$ 70 km s$^{-1}$, due to the instrumental resolution of the SDSS, although this has very little impact on our final results (see Section A4). This leaves us with $\sim$ 80 \% of our original sample which pass these data quality cuts. For our main analyses we include the low velocity dispersions in our sample to avoid biasing our input data such that only bulge dominated galaxies are included at low stellar masses. 
In principle this can lead to a lower predictive power of velocity dispersion, since measurements with higher uncertainty are used, but we test for this explicitly in the Appendices and find that our results and conclusions are unaffected. 

We then apply an aperture correction, so that all velocity dispersions are computed at the same effective aperture. Specifically, we use the formula in Jorgensen et al. (1995), defining the central velocity dispersion as:

\begin{equation}
{\rm CVD} \equiv  \sigma_{e/8} = \Big(\frac{R_{e}/8}{R_{ap}}\Big)^{-0.04} \sigma_{ap}
\end{equation}

\noindent where $\sigma_{ap}$ is the measured velocity dispersion in the aperture. $R_{e}$ is the bulge (or elliptical) effective radius and $R_{ap}$ is the radius of the aperture in the same units. This aperture correction typically only affects the velocity dispersion by $\sim$ 10\%.

We use three qualitatively different metrics of environment in this work: 1) group halo masses, 2) central - satellite divisions (both of which are described above) and 3) local densities. For the local densities, we utilise the normalised surface galaxy density evaluated at the $n^{th}$ nearest neighbour, based on values computed in Baldry et al. (2006). The local densities are calculated as:

\begin{equation}
\delta_{n} = \frac{\Sigma_{n}}{\langle \Sigma_{n}(z \pm \delta_{z}) \rangle} \hspace{0.2cm} 
\end{equation}

\noindent where 

\begin{equation}
\Sigma_{n} = \frac{n}{\pi r^{2}_{p,n}}
\end{equation}

\noindent $r_{p,n}$ is the projected distance (in physical units) to the $n^{th}$ nearest galaxy neighbour. $\langle \Sigma_{n}(z \pm \delta_{z}) \rangle$ is the mean value of the local density parameter at the redshift range in question. This effectively normalises the density parameter accounting for the flux limit of the SDSS. Thus, a galaxy residing in a perfectly average density of space (at a given redshift) would have log($\delta_{n}$) = 0, with galaxies residing in under-densities having negative values and galaxies residing in over-densities having positive values of this parameter. In this work we set $n$ = 5; however, none of our results are strongly affected by this choice, with identical rankings achieved for $n$ = 3 \& 10.

\subsection{Defining `Passive'}

In order to train the ANN codes to identify passive and star forming systems, we must first have a clear definition of what constitutes a passive (or star forming) galaxy. In this work we follow the prescription for defining passive in Bluck et al. (2014) Section 3. We start by selecting only star forming emission line galaxies, which are not identified as AGN in the BPT emission line diagnostic diagram. Specifically, we select out only those galaxies which are designated as star forming by the Kauffmann et al. (2003) line cut on the BPT diagram, and additionally have a S/N $>$ 3 in all of the relevant emission lines (H$\alpha$, H$\beta$, [OIII] and [NII]). The SFR - $M_{*}$ relationship is uni-modal for this sub-sample (see Fig. 5 in Bluck et al. 2014). We then calculate the distance any given galaxy resides at from this `star forming main sequence'. Quantitatively, we calculate:

\begin{equation}
\Delta{\rm SFR} = {\rm log} \Big( \frac{{\rm SFR}(M_{*},z)} {\rm{median}(\rm{SFR_{SF}}(M_{*} \pm \delta M_{*},z \pm \delta z))} \Big) 
\label{eq-DSFR}
\end{equation}

\noindent where SFR$_{SF}$ is the the star formation rate of the star forming sub-sample matched at the redshift and stellar mass of each galaxy. The matching thresholds are set to 0.005 for redshift and 0.1 dex for stellar mass and are then increased (in increments of 0.005 and 0.1 dex, respectively) if necessary until a minimum of five star forming `control' galaxies are found for each galaxy, or else the hard limits of 0.02 and 0.3 dex are reached. In most cases there are $>$ 200 controls available per galaxy and less that one percent of galaxies are excluded from the sample due to lack of controls.

The distribution of $\Delta$SFR is highly bimodal (as with the more familiar colour bimodality, e.g. Strateva et al. 2001), and it has a clear minimum at $\Delta$SFR = -1, i.e. at a SFR a factor of ten below the star forming main sequence (see Fig. 1). This provides a natural constraint to separate passive from star forming galaxies. The minimum of this distribution does not vary as a function of mass, morphology, or local density, hence it is a very stable and universally applicable definition for passive (see Fig. 7 in Bluck et al. 2014). Thus, we define passive and star forming galaxies to be:

\begin{tabular}{l|l}
PA: & $\Delta$SFR $\leq$ -1 \\
SF: & $\Delta$SFR $>$ -1 \\
\end{tabular}

\noindent In some of the analyses that follow in this paper we consider the possibility of a third classification, that of the `green valley'. The $\Delta$SFR limits for this configuration are given by:

\begin{tabular}{l|l}
PA: & $\Delta$SFR $\leq$ -1.2 \\ 
GV: & -1.2 $<$ $\Delta$SFR $<$ -0.6 \\ 
SF: & $\Delta$SFR $\geq$ -0.6 \\ 
\end{tabular}

\noindent None of our conclusions depend critically on whether we adopt two or three star forming classifications for our sample (see Section A3). It is important to stress at the outset that our approach implicitly assumes that there are only two (or three including the green valley) star formation states a galaxy can be in. This is a reasonable simplification given the extent of the bimodality of $\Delta$SFR; however, our approach in this paper will not be sensitive to subtle trends in sSFR or green valley migration as is evidenced in some other works (e.g. Schawinski et al. 2014, Woo et al. 2015).

\subsection{ANN Input Parameters}

There is a wide variety of possible galaxy properties we could include in our ANN analysis of star forming and passive systems, however there are a few constraints that must be met. First, it is crucial to avoid using galaxy parameters which are trivially related to the SFR or colour of a galaxy. This rules out using magnitudes, colours, luminosities, as well as SFR variants (e.g. sSFR, $\Delta$SFR) as input parameters. Also structural parameters based on single magnitudes will be highly biased by ongoing star formation in the optical, hence, we must avoid using B/T or $n_{s}$ parameters, if they are based on luminosities as opposed to stellar masses.

We choose eight different galaxy parameters, all of which are not trivially linked to star formation, but are connected to various proposed theoretical mechanisms for quenching central galaxies. They represent a wide range in scale and hence may help to resolve which of the leading theories for galaxy quenching are most likely to be correct, and to what degree they can be impacting the evolution of central galaxies. The physical parameters of the central galaxies used in this work are shown in Table \ref{tab-data}. 
Note that there are parameters connected to the galaxy environments ($M_{\rm halo}$, $\delta_{5}$), the outer regions of galaxies ($M_{\rm disk}$), the whole galaxy ($M_{*}$, B/T) and the inner regions of galaxies (CVD, $M_{\rm bulge}$, $Re$). This should provide a valuable test as to the scale and range of the quenching process for centrals.

\label{sec-data}

\begin{table}
\begin{center}
\caption{The physical parameters of the central galaxies used in this work.}
\begin{tabular}{c|l|l|l}
\hline\hline
\# & Symbol & Description & Scale$^{*}$\\
\hline

1  & CVD  & Central Velocity Dispersion & $\sim$ 1 kpc \\
2  & M$_{\rm Bulge}$& Bulge Stellar Mass & 0.5 -- 4 kpc \\
3  & Re & Bulge Effective Radius & 0.5 - 4 Kpc\\
4  & B/T & Bulge-to-Total Stellar Mass Ratio & 0.5 -- 8 kpc\\
5  & M$_{*}$& Total Stellar Mass & 2 -- 8 kpc \\
6  & M$_{\rm Disk}$& Disk Stellar Mass & 4 -- 10 kpc \\
7  & M$_{\rm Halo}$& Group Halo Mass & 0.1 - 1 Mpc\\
8  & $\delta_{5}$& Local Density Parameter & 0.5 - 3 Mpc\\
\hline
\end{tabular}
\label{tab-data}
\end{center}
$^{*}$ Approximate 1 $\sigma$ range from centre of galaxy. For photometric quantities half-light radii are used.
\end{table}

\section{The Method}
\label{sec-method}

\subsection{ANN}

In many situations linear models are not sufficient to capture complex phenomena, and thus non-linear models such as artificial neural networks (ANNs) are necessary. ANNs are among the most powerful tools in pattern recognition problems. They consist of simple mathematical units which are connected to each other in different layers and in different, often highly complicated, ways. In a multi-layer network, each layer adds its own level of non-linearity. So, naturally, a single layer network cannot produce the non-linearity that can be seen through multiple layers. A two-layer network is strong enough to handle a multi-parameter problem such as our classification problem in this paper and is frequently applied in similar works (e.g. Ellison et al. 2016). The specific configurations are chosen based on the nature of the problem under study, and in this way ANNs can learn to detect regularities, correlations and patterns in certain sets of data. Current applications of ANNs in astronomy include star-galaxy discrimination and galaxy classification (e.g. Cortiglioni et al. 2001; Andreon et al. 2001; Ball et al. 2004; Teimoorinia 2012; Teimoorinia \& Ellison 2014), but their power in data analysis has been largely untapped. 

Generally, input parameters (e.g. parameters in  Table \ref{tab-data}) are connected to the first layer of a network with some mathematical units (nodes) which are called neurons. The first layer can be connected to a second layer (with some new neurons, in different and complicated ways) and, at the end, the second layer are connected to the output layer. In a binary classification, the output layer contains only two nodes. Through iteration between inputs and outputs, the parameters of the mathematical nodes (weights and biases) can be fixed to optimise solving the classification problem. In this way we will have a trained network. In fact, the aim in training steps is to minimise the difference between the predicted and observed values by a performance function such as, e.g., a mean square error function. A trained network should then be validated (during the training steps or after training) by an independent data set to test performance of the trained network and also to avoid over-fitting problems. Overfitting is then evident by the result for a training set being good but for a validation set being unacceptable.

An ANN model is generally `learned' from a set of training data where, in a supervised learning mode, the training data is labelled with the `correct' answers. Since the aim of finding a model is to provide useful predictions in future situations, questions about choosing a model are important, especially when we do not know much about the underlying nature of the process being studied. ANNs offer a powerful solution to this problem by allowing the analysis algorithm to form its own model `organically' from iterations with the training set. In many cases, we may wish to learn a mapping from D-dimensional inputs to scalar, or G-dimensional, outputs. In other words, both the inputs and outputs may be multidimensional. 
In these complicated situations few techniques in the machine learning area are as effective as ANN minimisation analyses. These approaches are highly effective for many complex problems, such as finding a patterns in large datasets and in classifying non-linear multi-dimensional data between predetermined sets or classes (as in this work).

In a classification problem the general goal of the ANN is for the algorithm to `learn' a decision boundary or a threshold. Here, we have two (or three including the green valley, see Section A3) predetermined classes for the star forming states of SDSS galaxies: passive (PA) and star forming (SF). Generally, once a model is found by the network we perform a classification by comparing the posterior class probabilities, i.e. $\rm{P(SF|\textbf{x}})$ and $\rm{ P(PA|\textbf{x})}$, in which \textbf{x} is our multidimensional input data. Thus, from Bayes theorem, we have:

\begin{equation}
\rm{P(PA|\textbf{x})=  \frac{P(\textbf{x}|PA)P(PA)}{P({\textbf{x}})} = \frac{ P(\textbf{x}|PA)P(PA)}{P(\textbf{x}|PA)P(PA)+P(\textbf{x}|SF)P(SF)}}
\label{eq-prob}
\end{equation}

\noindent The above equation can be written as:

\begin{equation}
\rm{P(PA|\textbf{x})= f(g(\textbf{x}))=\frac{1}{1+e^{-g(\textbf{x})}}}
\label{eq-fx}
\end{equation}

\noindent In which f is a sigmoid (or activation) function and g is given by:

\begin{equation}
\rm{g(\textbf{x})=ln\Big(\frac{P(\textbf{x}|PA)P(PA)}{P(\textbf{x}|SF)P(SF)}\Big)}
\label{eq-gx}
\end{equation}

\noindent In an our ANN approach we use a two-layered network, specifically modelling the data as: 

\begin{equation}
\rm{P(PA|\textbf{x})=f\big(\sum\limits_{j=1} w^{(2)}_j f\big(\sum\limits_{i=1} w^{(1)}_{i,j} x_i+b^{(1)}_i\big)+b^{(2)}_j\big)}
\label{eq-ann}
\end{equation}
 
\noindent in which w and b are weights and biases of the network in different layers that are fixed by the training steps. The suffix (1) indicates the first layer and (2) indicates the second layer. In this way we construct a model of the class probability given the measurement (as in Bishop 1995). 

Our results are stable to issues of over-fitting because we use a neural network model with typically ten neurons applied to a training set of 100,000 galaxies as input data. Moreover, we have many unused galaxies from training with which we can verify the fit on an independent `validation set' of $\sim$ 100,000 different galaxies. The results from this study are always identical for both of these sets. We also apply an early stopping technique in which the training set is itself split into two sub-sets (70\% training and 30\% validation) to test the performance of the two sets at an early stage of development. If they show different behaviours we can identify issues and retrain accordingly. 
Finally, we repeat the training several times and exclude the worst cases (where a global minimum solution is not found) from our final analysis. In this manner we always concentrate on the `best' possible results from our network model, taking the average over these as our performance indicator. Multiple application of our ANN procedure on the same problem also ensures that our results are converged, and hence have settled in a global minimum solution. In the next sub-section we give an example of our ANN approach applied to a simplified dataset to illustrate our analysis techniques.

\begin{figure}
\centering
\includegraphics[width=8.5cm,height=4.5cm,angle=0]{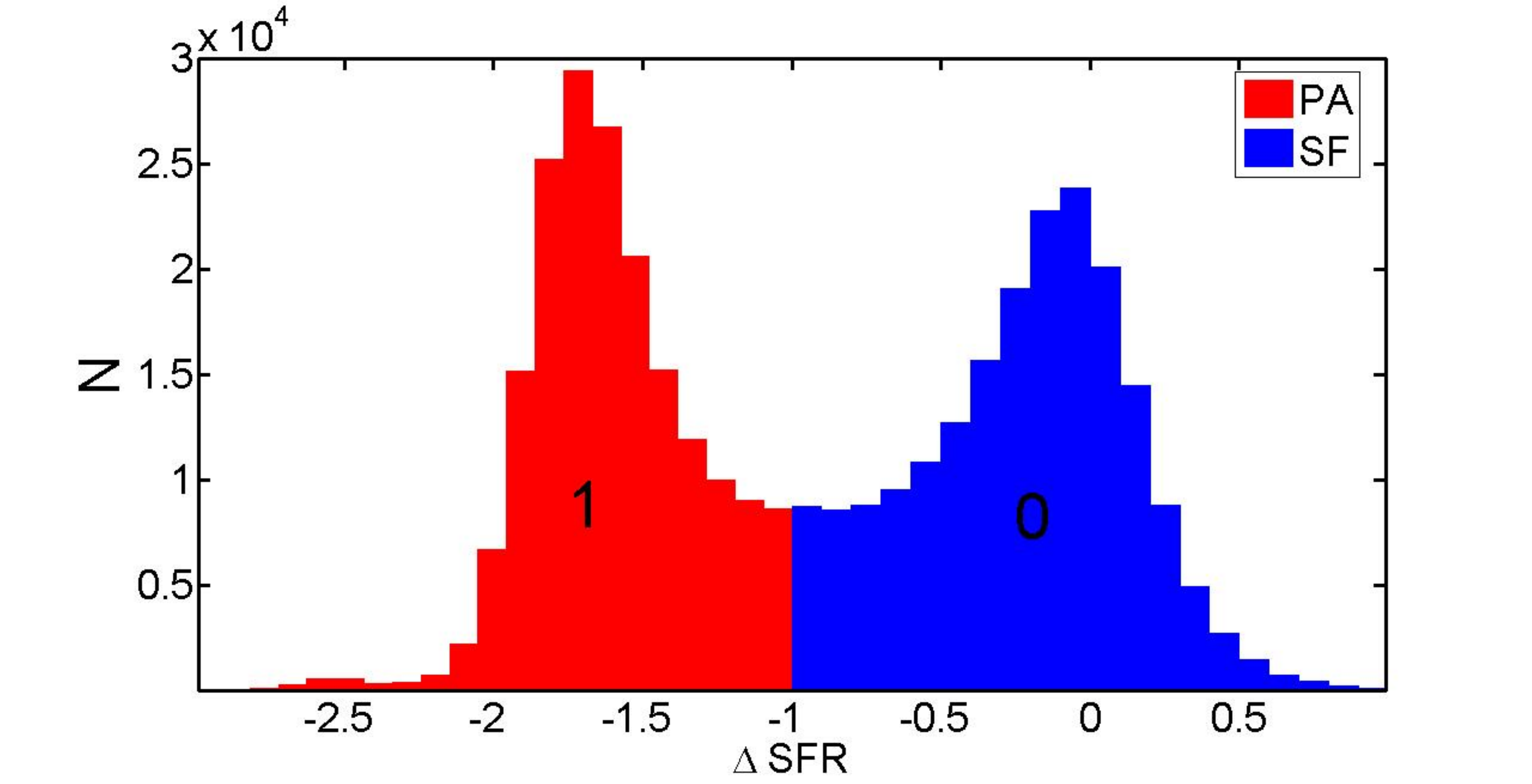}
\includegraphics[width=8.5cm,height=4.5cm,angle=0]{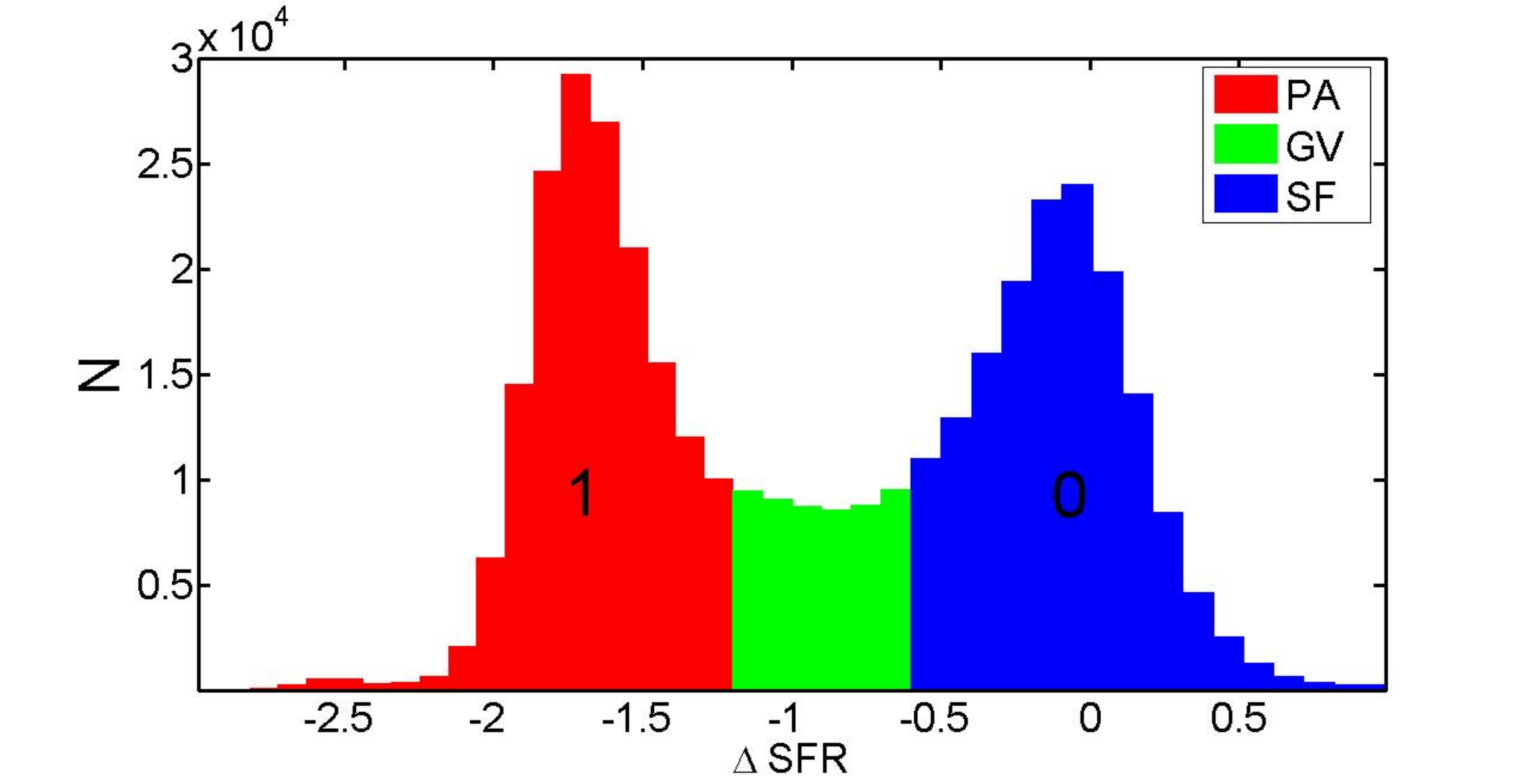}
\caption{Distribution of $\Delta$SFR values, defined in equation \ref{eq-DSFR}. Top panel: All the passive galaxies ($\Delta$SFR $<$ -1) are labeled by a value of 1 for the purposes of our ANN minimisation. We assign to all star forming galaxies ($\Delta$SFR $\geq$ -1) a value 0. The output of the ANN procedure will thus be a probability (between 0 and 1) for how likely each galaxy is to be passive or star forming, given the input data. Bottom panel: the same as the top panel but showing the green valley galaxies as a separate class, which are excluded from some analyses.}
\label{fig-hist-01}
\end{figure}

\begin{figure}
\centering
\includegraphics[width=8cm,height=4.4cm,angle=0]{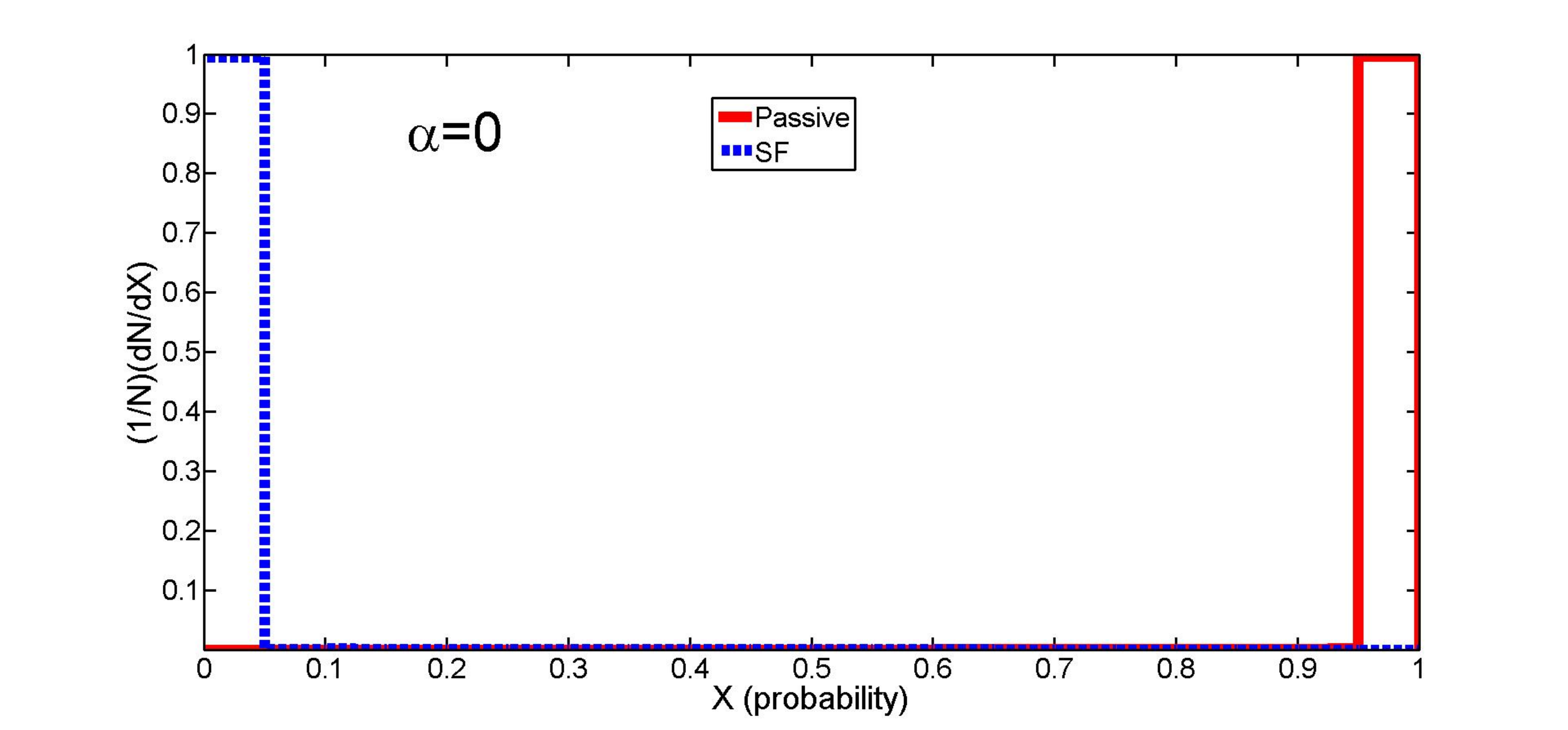}
\caption{Output ANN probabilities for galaxies being star forming (X = 0) or passive (X = 1) for two categories: originally determined passive galaxies (the red line) and star forming galaxies (the blue dashed line). This shows the perfect case of equation \ref{eq-dsfr} (for $\alpha=0$), i.e. where we give the ANN codes all the relevant information to assign the passive state of each system. Unsurprisingly, this yields a 100\% accurate classification. The original classification of the data is shown in Figure \ref{fig-hist-01}.}
\label{fig-dist-area-1}
\end{figure}

\subsection{ANN Performance Test and Example}

In Figure \ref{fig-hist-01} we show the distribution of $\Delta$SFR for our sample of central galaxies. A cut at the minimum of this distribution ($\Delta$SFR = -1) cleanly separates the galaxies into two different groups (see Section 2.2, and Bluck et al. 2014 for full details). This is an example of a binary classification. In this kind of problem, a classifier can classify input data into two desired classes. The input values can be different physical properties (e.g. the physical galaxy parameters in Table \ref{tab-data}) with different combinations, i.e. single or multiple variables. However, the target data are always just two different labels: SF $|$ PA. 
For statistical purposes, we designate these possibilities by two real numbers, 0 and 1. In this case, one can associate an output value of zero to star forming galaxies and an output of value of 1 to passive galaxies. In practice, the output of the network will be the estimated probability that the input pattern (from the data) belongs to one of the two categories. 

To test our method (and illustrate our analysis technique) we use the definition of `passive', from $\Delta$SFR, as an input data to the ANN. But we distort it in a coherent manner to ascertain the effect of noise (or randomness) on the pattern recognition. Specifically, we define the transform:

\begin{equation}
\Delta\rm{SFR} \longmapsto \Delta\rm{SFR} +\alpha R
\label{eq-dsfr}
\end{equation}

\noindent In which R is a random number between -1 and 1. In other words, with increasing $\alpha$ we add more random `noise' to our input data `signal' ($\Delta$SFR).  

In a binary classification, the output of a classifier will be two different probability distributions (i.e. how likely each galaxy is to belong in each category). A trivial example is when $\alpha=0$, i.e. when we give the ANN training code all of the information it needs to decide unambiguously whether or not each galaxy belongs in the passive or star forming sample. In this case a classifier should be able to classify the data perfectly. Thus, no overlap (or misclassifications) of the two distributions is expected. We show the result of this test in Figure \ref{fig-dist-area-1}. As expected 100\% of the data is correctly classified into the two categories: star forming (blue line) or passive (red line).

\begin{figure}
\centering
\includegraphics[width=8cm,height=4.4cm,angle=0]{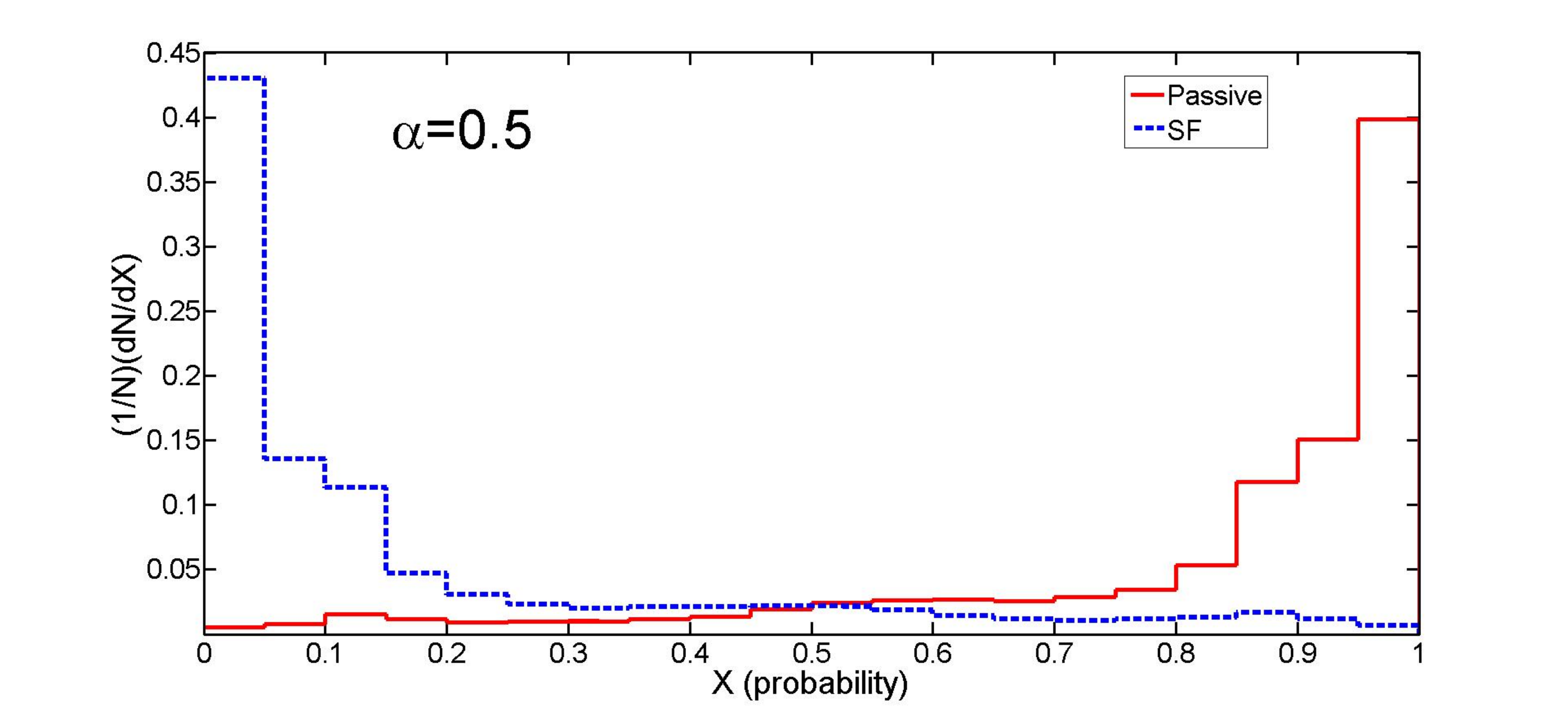}
\includegraphics[width=8cm,height=4.4cm,angle=0]{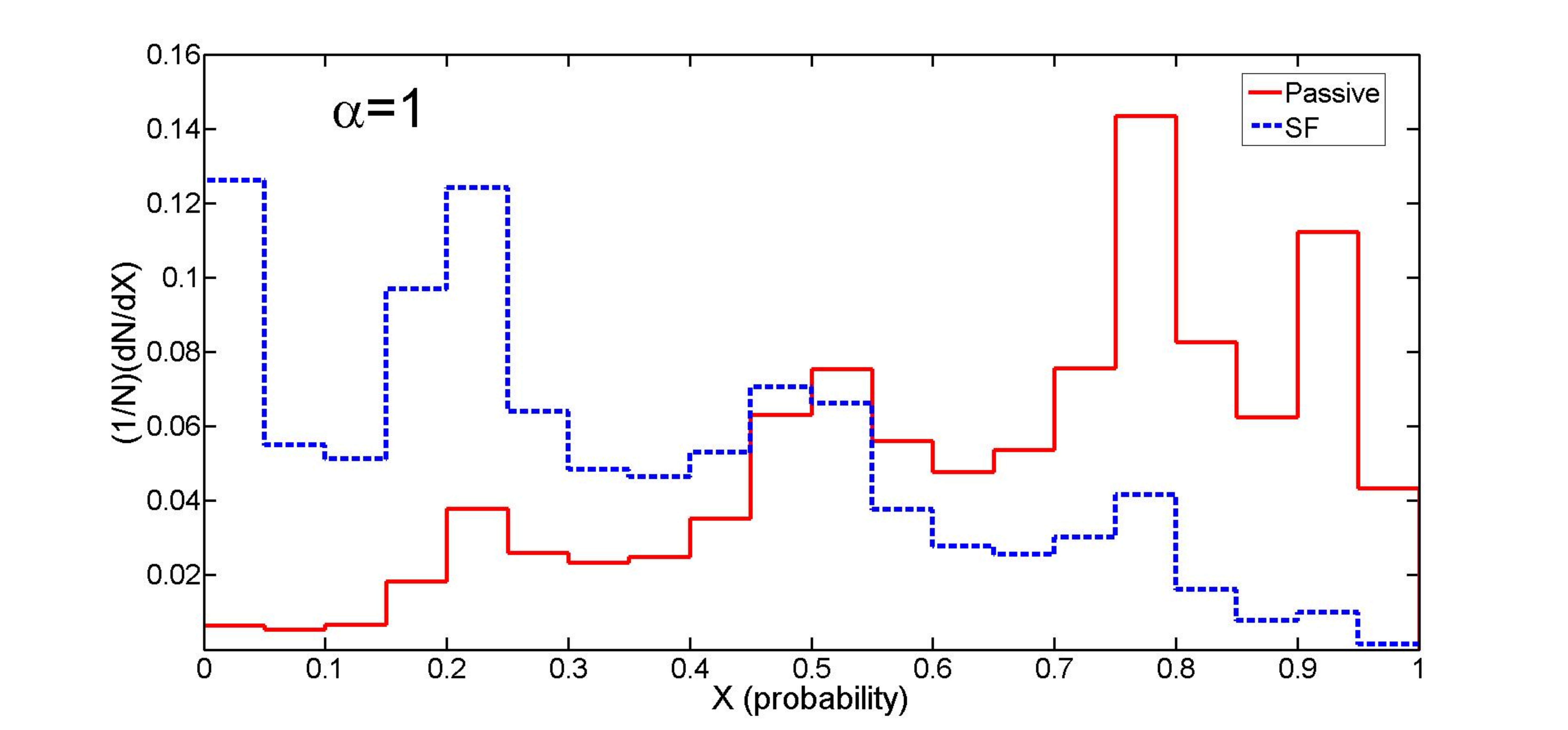}
\includegraphics[width=8cm,height=4.4cm,angle=0]{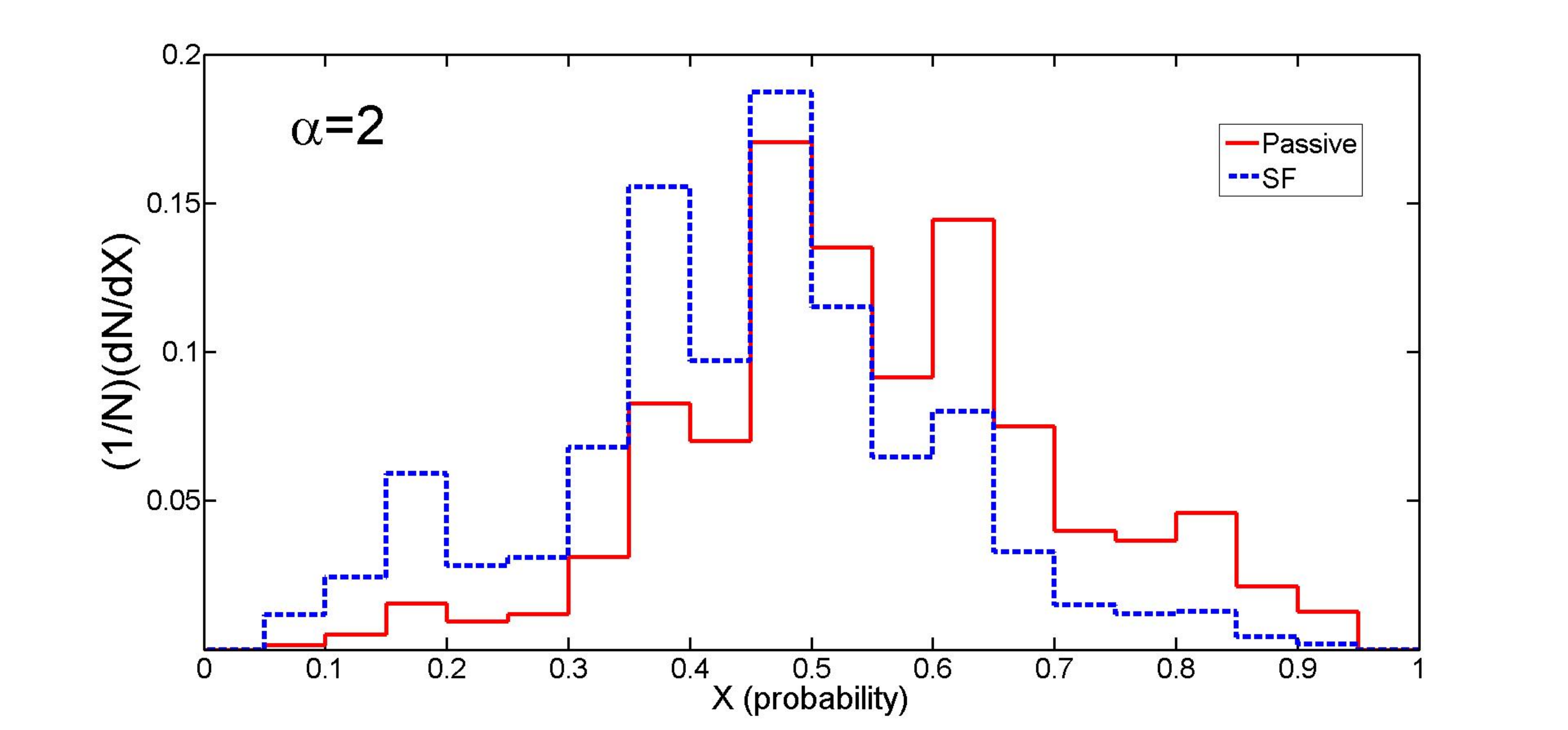}
\includegraphics[width=8cm,height=4.4cm,angle=0]{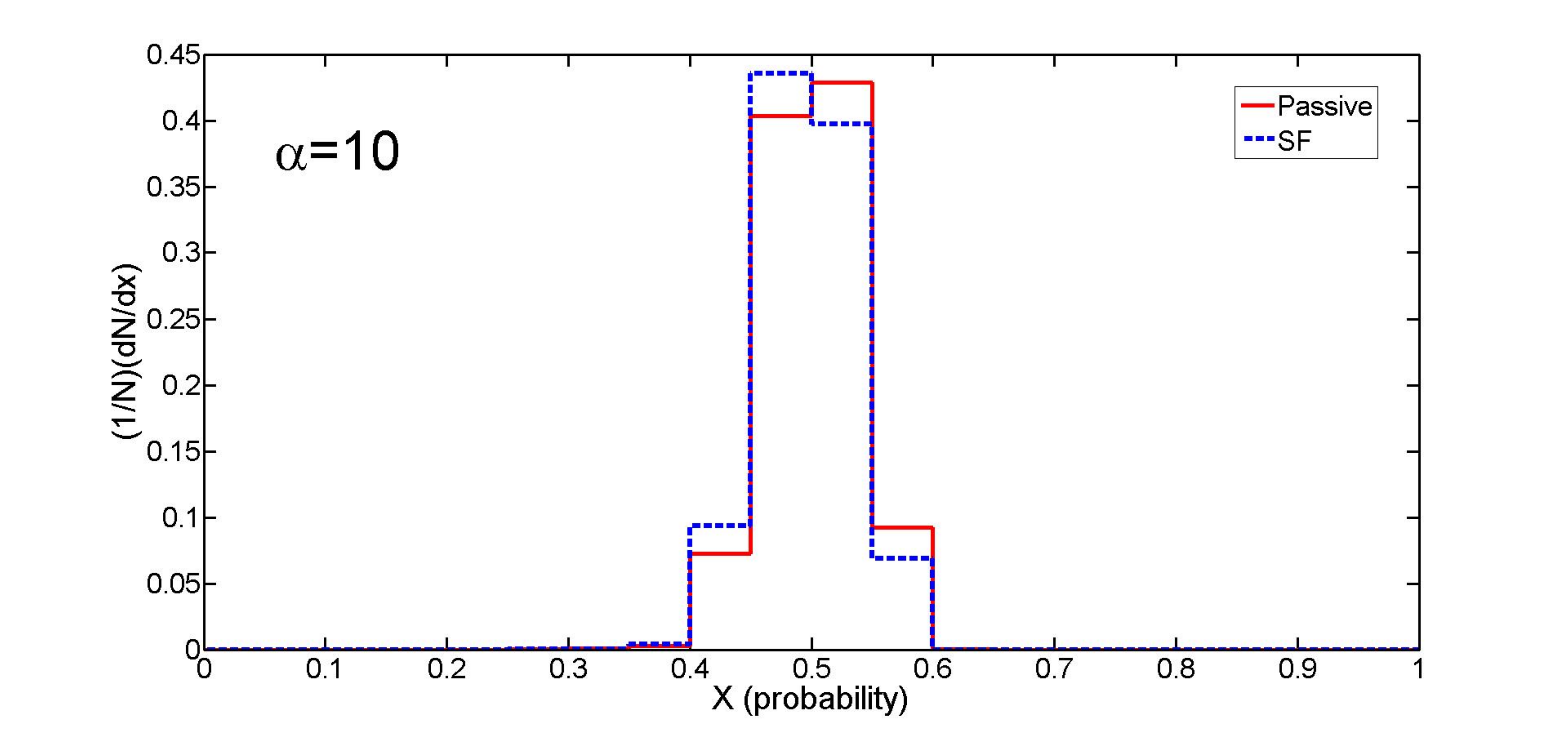}
\caption{Output ANN probability distributions for four example cases of our randomness parameter, $\alpha$. In each plot the X-axis shows the probability that each galaxy is passive based on the best fit minimisation procedure from the ANN (where 0 = SF, 1 = PA). The red lines are for originally classified passive galaxies, with the blue lines being for originally classified star forming galaxies. The ideal case (for $\alpha$ = 0) is shown in Figure \ref{fig-dist-area-1}. The Y-axis shows the normalised number of galaxies in each probability binning, summing to one. As can be seen, the distributions become less separated as we increase $\alpha$ from top to bottom, indicating less success in predicting the passive fraction by the ANN methods as we increase the noise or randomness of the input data. These distributions can be used as a comparison to the equivalent plots for the science parameters in Figure \ref{fig-area}.}
\label{fig-dist-area-2}
\end{figure}

We increase the value of $\alpha$ to see how the output of the network depends on increased noise, or randomness in the input data. We train the networks on a `training set' of 50,000 passive and 50,000 star forming galaxies, randomly chosen from our parent sample. We then apply the newly formed model to an independent `validation set' of 50,000 different passive and 50,000 different star forming galaxies. We find that our network rankings are converged, i.e. training or verifying on larger samples or running the codes for longer leads to no significant changes in the results or rankings. The output of our trained network on the independent validation set for different values of $\alpha$ is shown in Figure \ref{fig-dist-area-2}. As can be seen, with increasing $\alpha$ the ability of the ANN to distinguish between the two categories becomes diminished. In fact when we increase $\alpha$ by a factor of ten the two distributions become almost indistinguishable. These histograms can be used as a useful comparison 
to the real data analysis later on, see Section 4. In each case we can assign a performance to our classification, which we describe in detail in the next sub-section.

\subsection{Receiver Operating Characteristic (ROC) }
\label{sec-roc}

\begin{figure}
\centering
\includegraphics[width=9cm,height=4.43cm,angle=0]{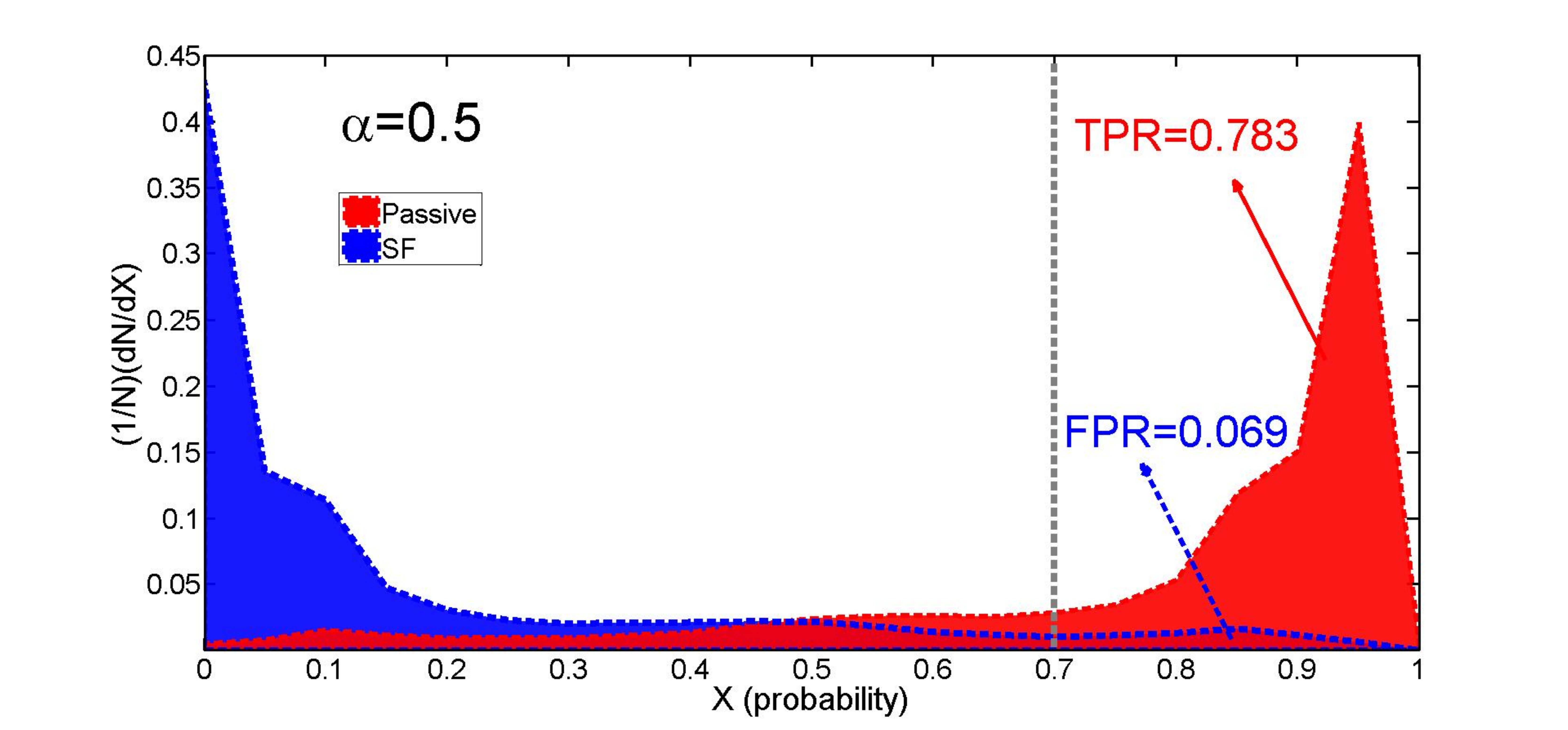}
\caption{Output ANN probability distribution for $\alpha$ = 0.5 case (where 0 = SF, 1 = PA) for originally classified star forming (blue) and passive (red) galaxies. The vertical grey dashed line at X = 0.7 shows a randomly selected threshold. For this threshold, the red shaded area to the right of the line gives the True Passive Rate (TPR = 0.783), and the blue shaded area to the right of the line gives the False Passive Rate (FPR = 0.069). Note that in general FPR + TPR $\neq$ 1, since the sum of the red area and the sum of the blue area (from X = 0 - 1) is unity, not the sum of the blue and the red areas across any given threshold.}
\label{fig-dist-area-2-2}
\end{figure}

A Receiver Operating Characteristic or ROC plot is a statistical tool used to measure the performance of a binary classifier (e.g., Fawcett 2006). To demonstrate how we determine the performance of our ANN classifier we re-plot the distribution related to the value of $\alpha=0.5$ in an area format in Figure \ref{fig-dist-area-2-2}. The red and blue areas show the probability distributions for passive (PA) and star forming (SF) galaxies, respectively. Here, we focus on the passive galaxies which are originally labeled with value 1, although an equivalent formulation of this statistic based on the star forming sample (originally labelled as 0) is also possible. These will give equivalent results because of the binary nature of our experimental setup, i.e. P(PA) = 1 - P(SF). 

\begin{figure}
\centering
\includegraphics[width=8cm,height=6.2cm,angle=0]{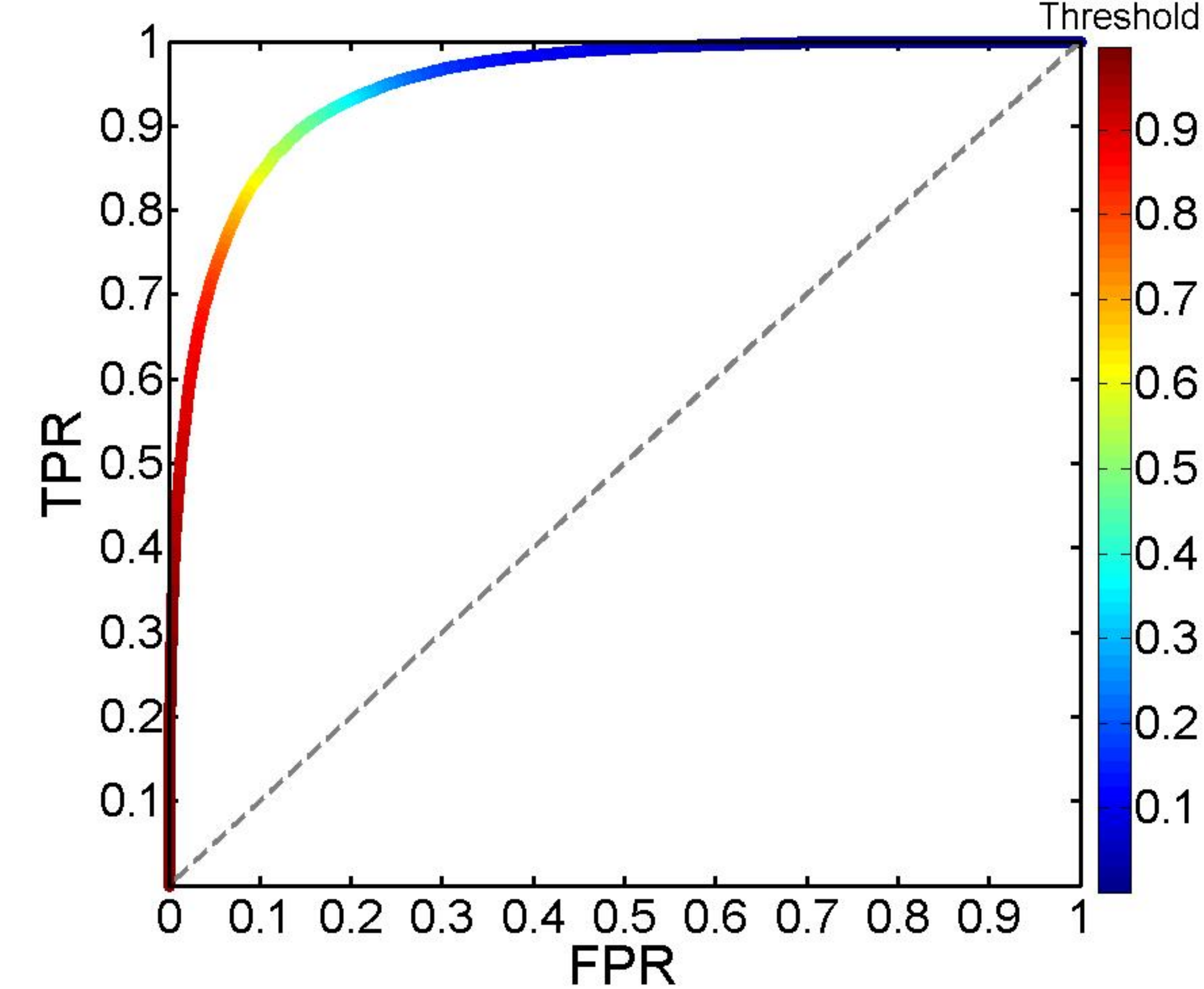}
\caption{A Receiver Operating Characteristic (ROC) plot obtained from the ANN output probability distribution of Figure \ref{fig-dist-area-2-2}, for $\alpha$ = 0.5. Specifically, we plot the True Passive Rate (TPR) vs. the False Passive Rate (FPR), see Section 3.3. We change the threshold from 1 to 0, systematically obtaining different values for TPR and FPR. As an example, the point [0.20 0.85] is obtained from threshold X = 0.5. The thresholds are indicated by the colour of the ROC curve line, labelled by the colour bar. The dashed grey line indicates the result for a random variable, with area under the ROC curve, AUC = 0.5.}
\label{fig-Thresh-2-2}
\end{figure}

On the right hand side of any selected threshold (decision boundary) we will have two relevant percentage values. For example, on the right hand side of the vertical dashed line in Figure \ref{fig-dist-area-2-2} (at a threshold at X = 0.7), the fraction of galaxies that are correctly classified as passive is 0.783. 
We call this the True Passive Rate (hereafter TPR), thus, we have TPR = 0.783. However, there are also some star forming galaxies in this region of the probability distribution which are misclassified as passive galaxies. We call this fraction the False Passive Rate (FPR). For our example threshold at X = 0.7, FPR = 0.069. Thus, for any selected threshold we will have two values: TPR and FPR. The ROC graph is obtained by plotting TPR vs. FPR for all possible thresholds (0 -- 1). Figure \ref{fig-Thresh-2-2} shows this for $\alpha=0.5$. This curve can be used to quantify the performance of our classification (as in Bradley 1997). Higher areas under the ROC curve (hereafter AUC) indicate a better performance of the network in determining the correct star forming or passive state of galaxies.  

\begin{figure}
\centering
\includegraphics[width=8cm,height=6.2cm,angle=0]{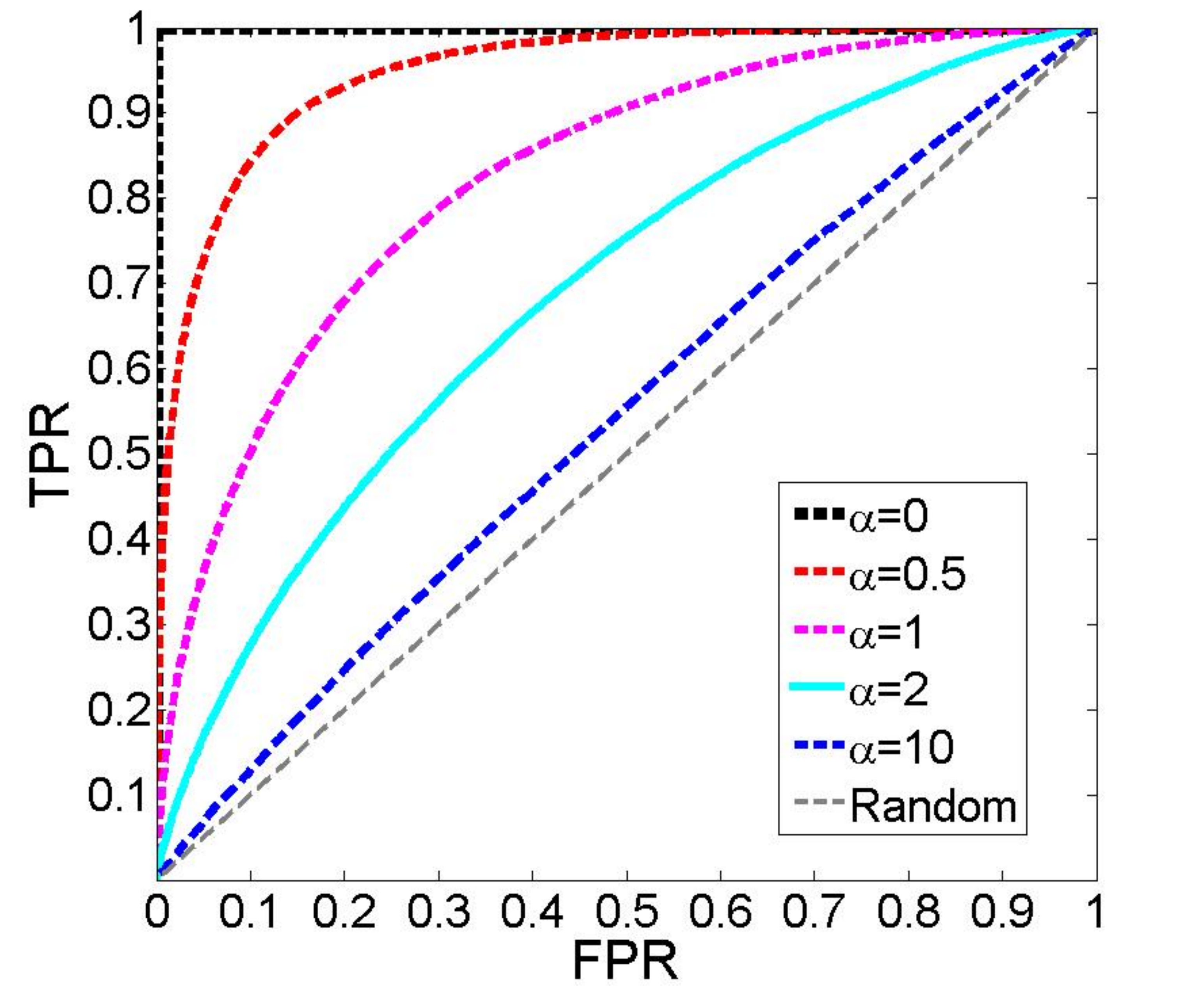}
\caption{Receiver Operating Characteristic (ROC) curves obtained from the ANN output probability distributions for varying values of the randomness parameter, $\alpha$, shown in Figure \ref{fig-dist-area-2}. Specifically, we plot the True Passive Rate (TPR) vs. the False Passive Rate (FPR), see Section 3.3. For $\alpha$ = 0 the performance is perfect with AUC = 1, the AUC then decreases systematically with increasing $\alpha$, up to a theoretical limit of AUC = 0.5 as $\alpha$ $\rightarrow$ $\infty$.}
\label{fig-roc-alpha}
\end{figure}

We plot ROC curves related to different values of $\alpha$ in Figure \ref{fig-roc-alpha}. The black dashed line is for the perfect classification (where $\alpha=0$), which yields an AUC = 1. A sample of completely random numbers ($\alpha \rightarrow \infty)$ will generate the (diagonal) grey dashed line, with AUC = 0.5. All other values of $\alpha$ will yield an AUC performance between these extremes. So, from random to perfect classification the value of AUC changes from 0.5 to 1, respectively. In the engineering literature (e.g. Hosmer \& Lemeshow 2000) the AUC values correspond to success `labels', see Table \ref{tab-auc}.

\begin{table}
\begin{center}
\caption{An interpretation of the AUC parameter in engineering (by Hosmer \& Lameshow 2000)}
\begin{tabular}{c|l}
\hline\hline
AUC Range & Description \\
\hline
1.0 & Perfect Discrimination \\
0.9 -- 1.0 & Outstanding Discrimination \\
0.8 -- 0.9 & Excellent Discrimination \\
0.7 -- 0.8 & Acceptable Discrimination \\
0.5 -- 0.7 & Unacceptable Discrimination \\
0.5 & No Discrimination (Random) \\ 
\hline
\end{tabular}
\label{tab-auc}
\end{center}
\end{table}

\begin{figure}
\centering
\includegraphics[width=9cm,height=5cm,angle=0]{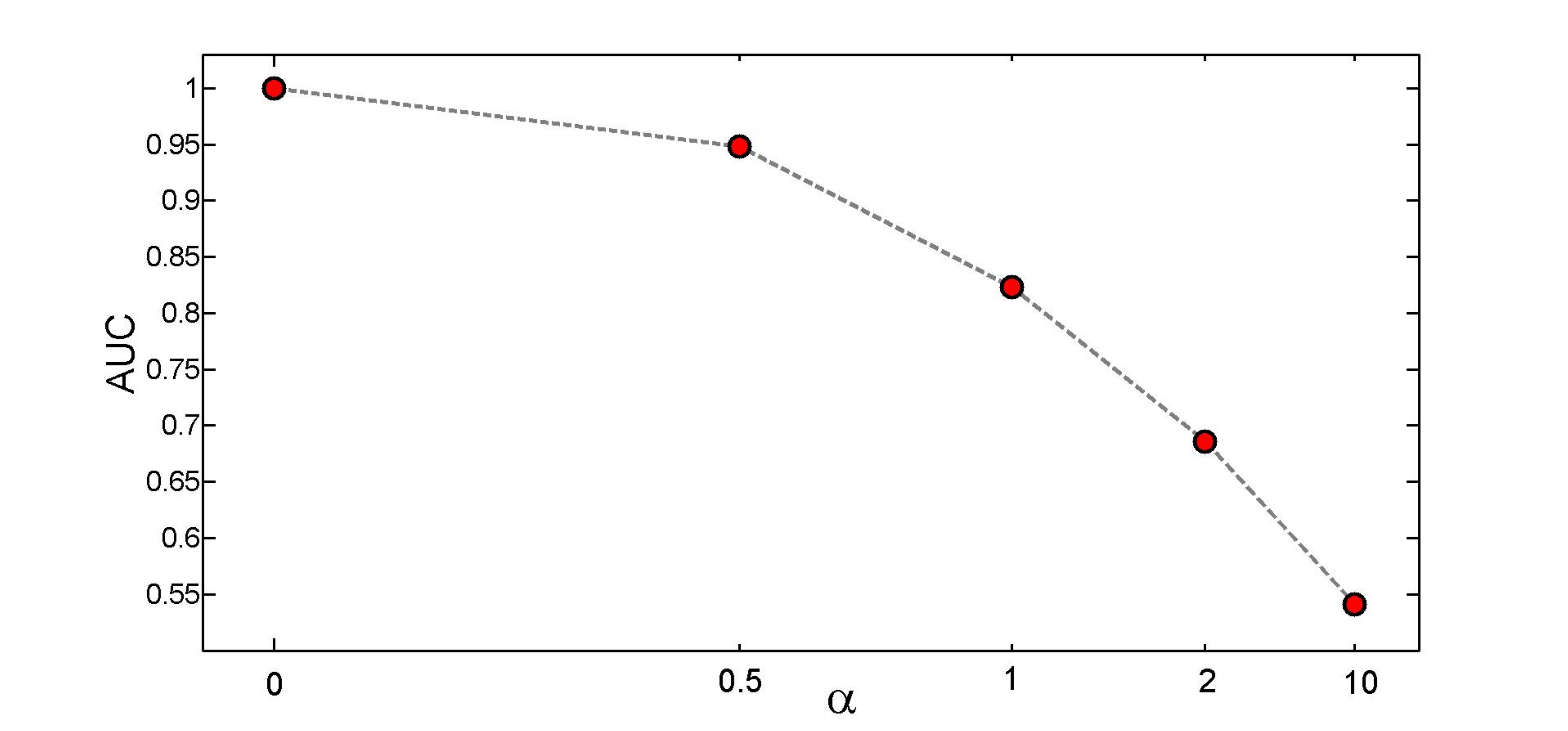}
\caption{Area Under the ROC Curve (AUC) vs. the randomness parameter $\alpha$. The different values of AUC  associated with different values of $\alpha$ are computed from Figure \ref{fig-roc-alpha}. The area under the curve from a perfect classification ($\alpha=0$) changes from AUC = 1 to a completely random input data ($\alpha \rightarrow \infty$) with AUC = 0.5.}
\label{fig-auc-alpha}
\end{figure}

We obtain all AUC values associated with the different $\alpha$ values and plot these in Figure \ref{fig-auc-alpha}. As can be seen, the area under the curve varies from a perfect classification (at $\alpha=0$) with a value of 1 to an almost random result of 0.55 (at $\alpha$=10). Thus, the AUC statistic strongly correlates with the true `signal' in the data, in this case $\Delta$SFR. Higher randomness or noise leads to lower AUC values. 

Our analysis techniques are now ready for exploitation on real data. In order to determine which galaxy properties modulate the quenching of star formation, we consider each variable from Table 1 in turn (and combinations thereof) as input to the ANN, and quantify how well they discriminate the passive and star forming populations. As described above, successful discrimination is characterized by a large AUC for that variable (or set of variables). The AUC results can then be ordered to give a quantitative ranking of the parameters' relative importance in determining whether or not a galaxy is star forming. In the following section we describe our results for central galaxies.

\section{Results}

In this section we describe our results for central galaxies, following the method outlined in Section 3. ANN pattern recognition training is performed on 50,000 SF and 50,000 PA galaxies for each configuration of science variables considered. Once the ANN model is constructed, it is tested on a new verification set of 50,000 SF and 50,000 PA galaxies. The output probability distributions, ROC curves and AUC parameters are determined for each case. From this, a ranking of how important different galaxy properties, and sets of two and three properties, are for determining the passive state of galaxies is constructed.

\subsection{Single Parameters}

 \begin{figure*}
\centering
\includegraphics[width=10cm,height=6.2cm,angle=0]{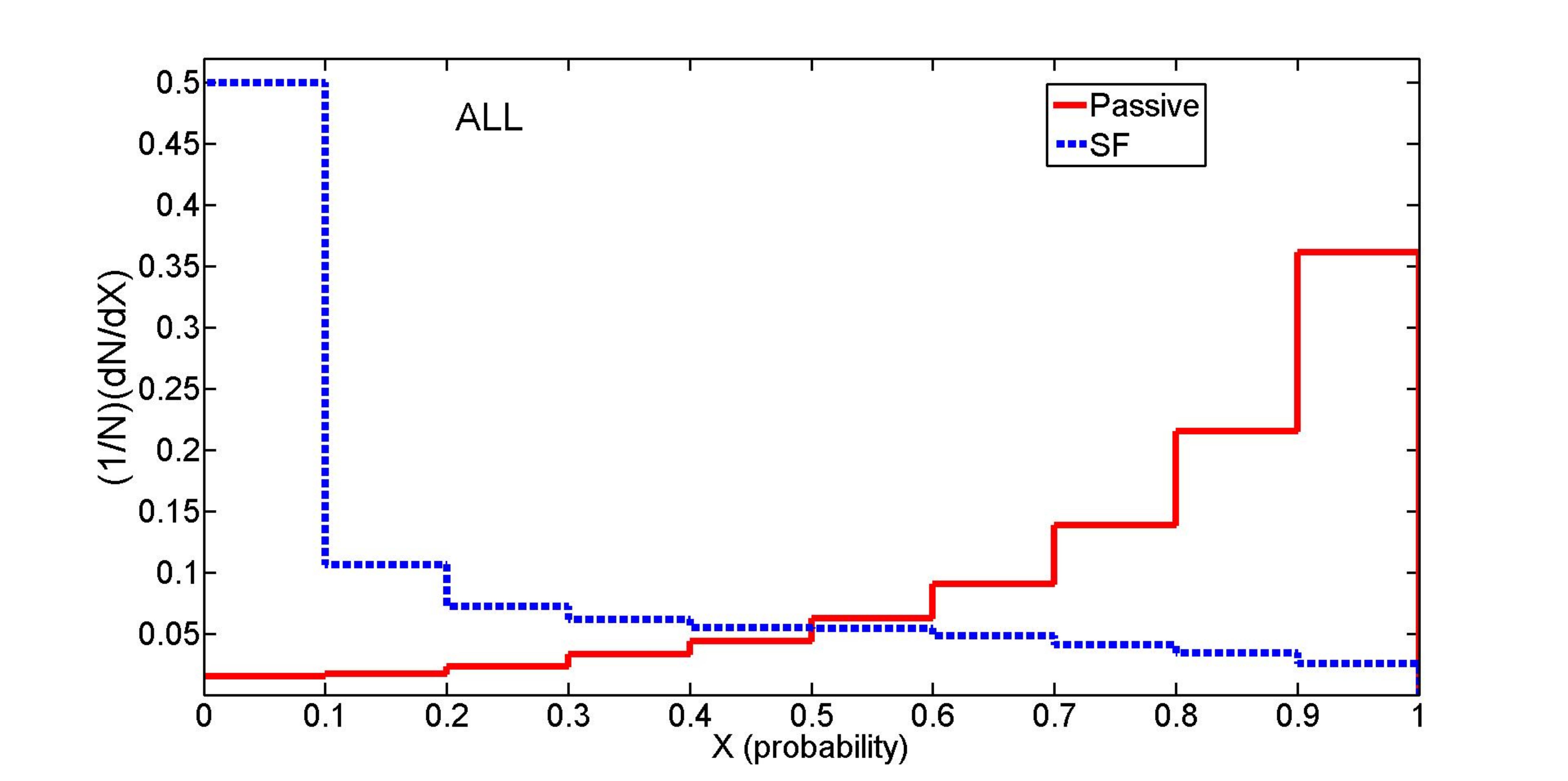}
\includegraphics[width=8cm,height=4cm,angle=0]{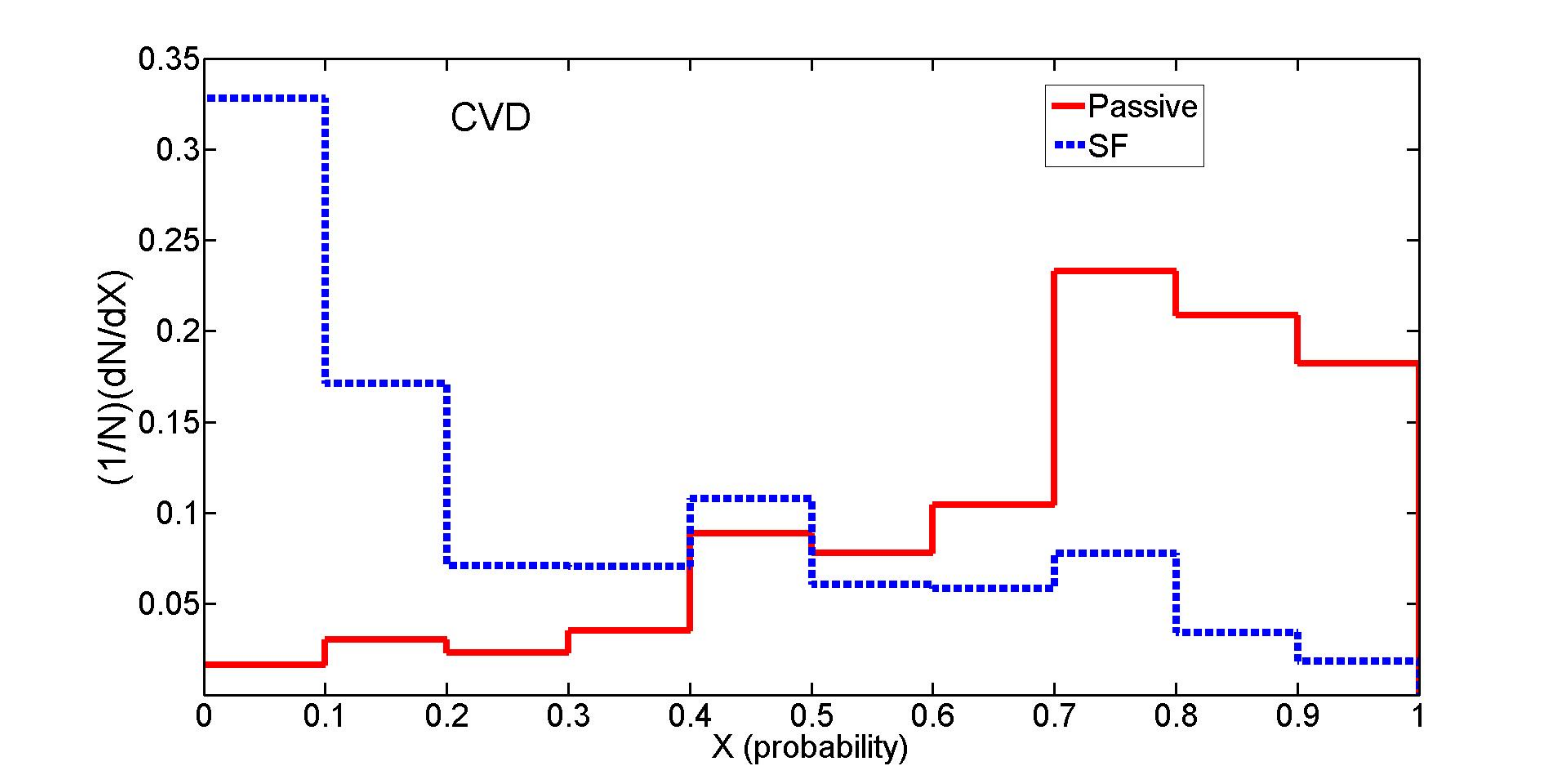}
\includegraphics[width=8cm,height=4cm,angle=0]{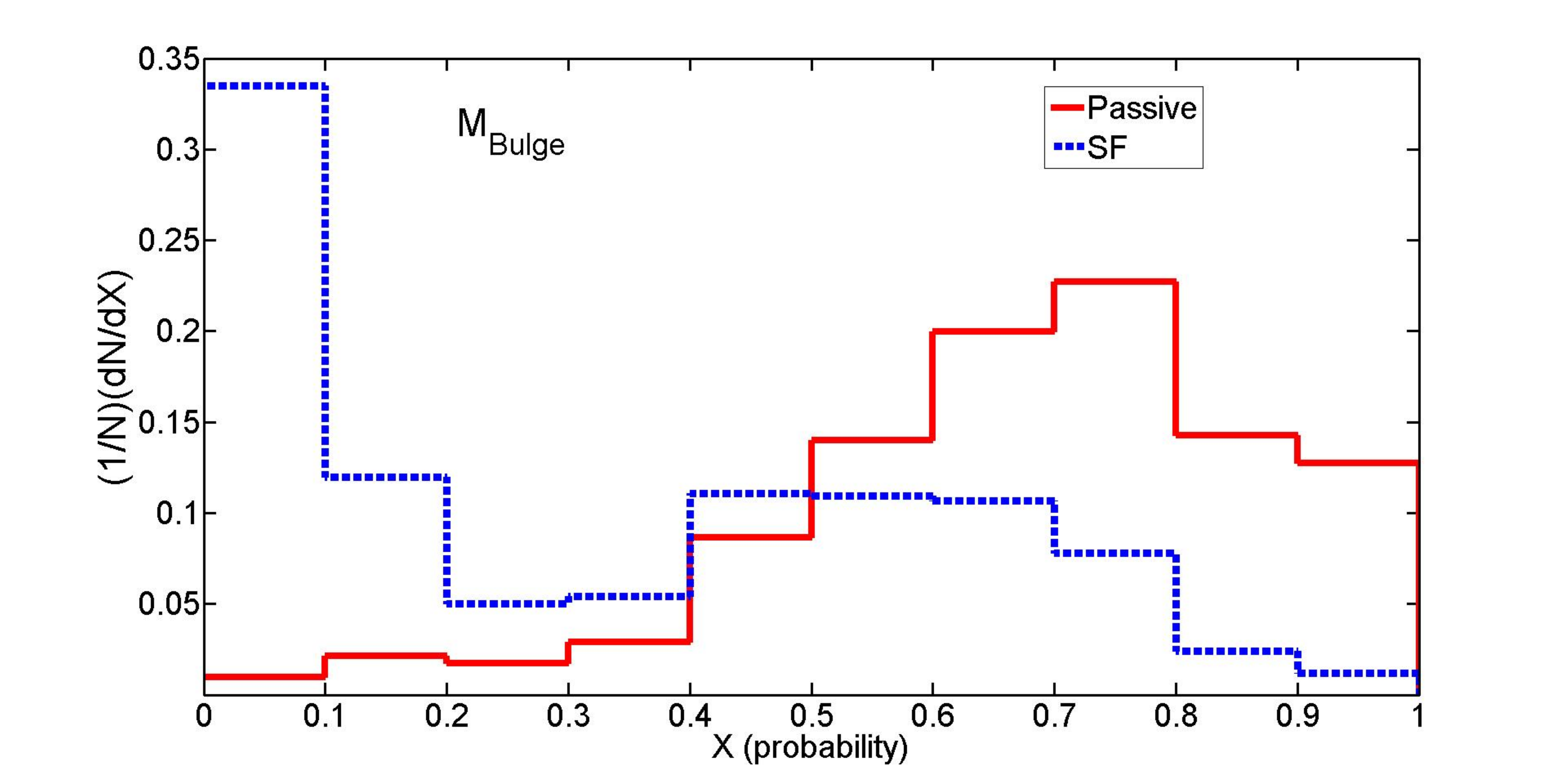}
\includegraphics[width=8cm,height=4cm,angle=0]{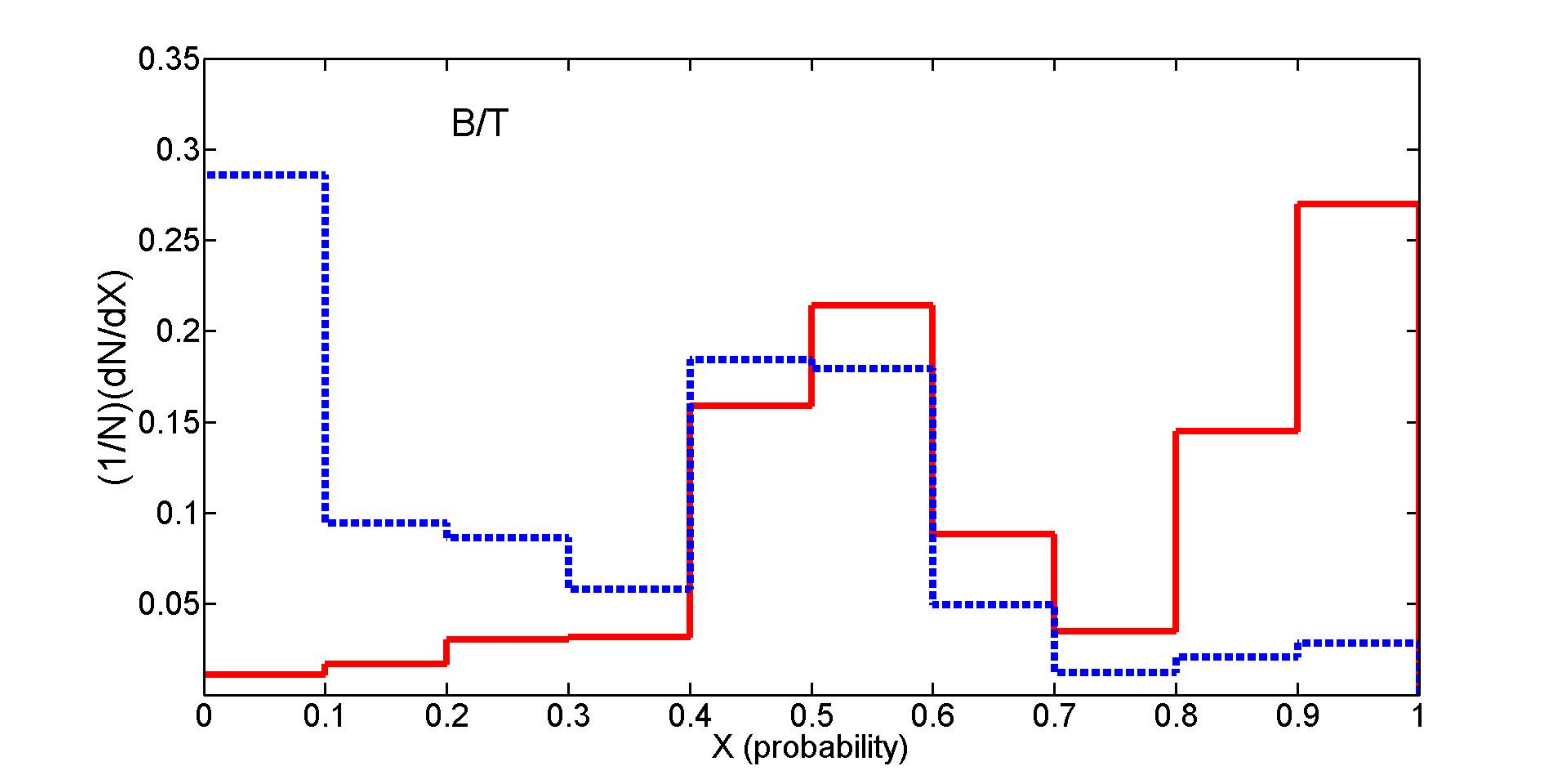}
\includegraphics[width=8cm,height=4cm,angle=0]{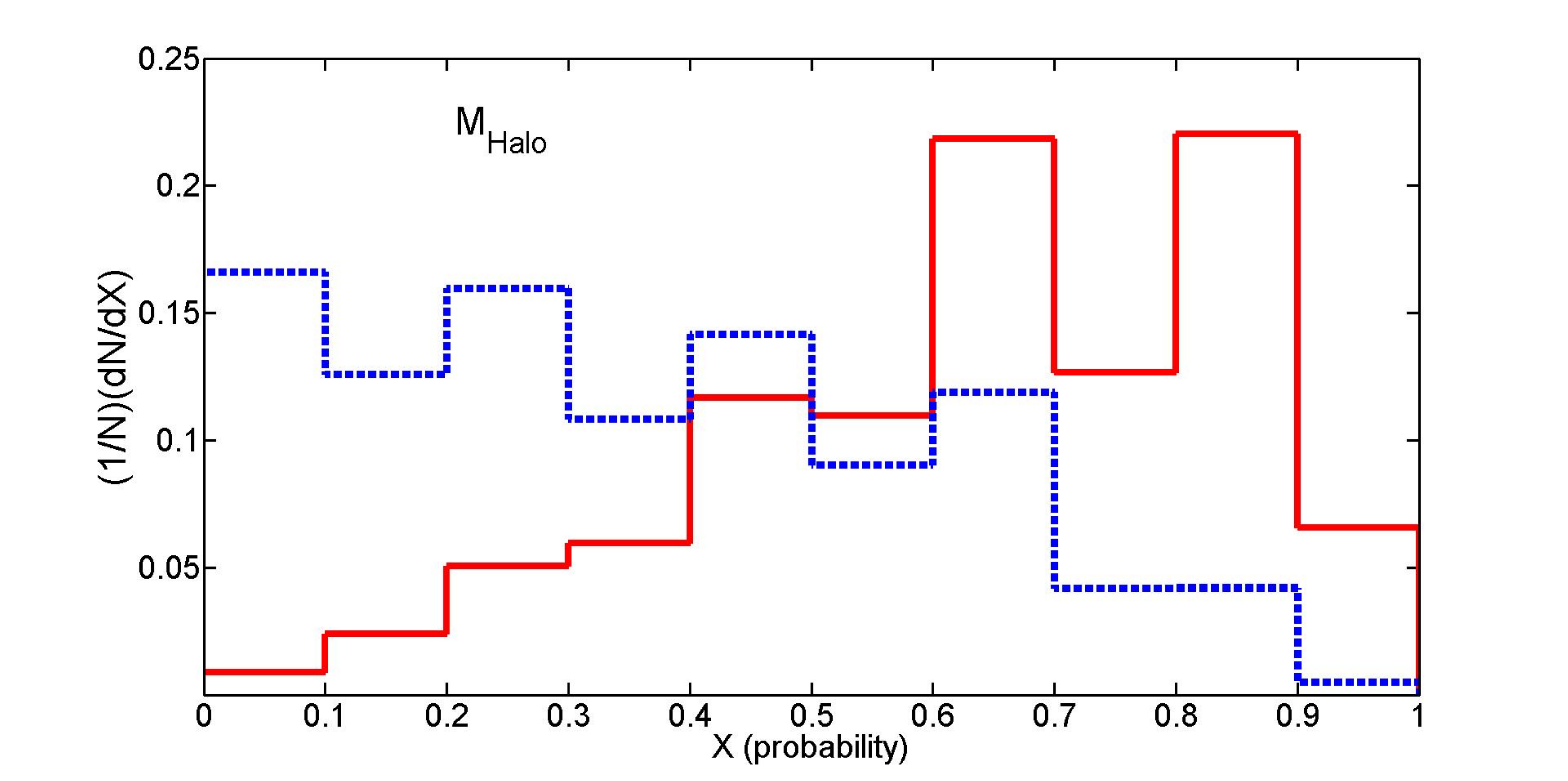}
\includegraphics[width=8cm,height=4cm,angle=0]{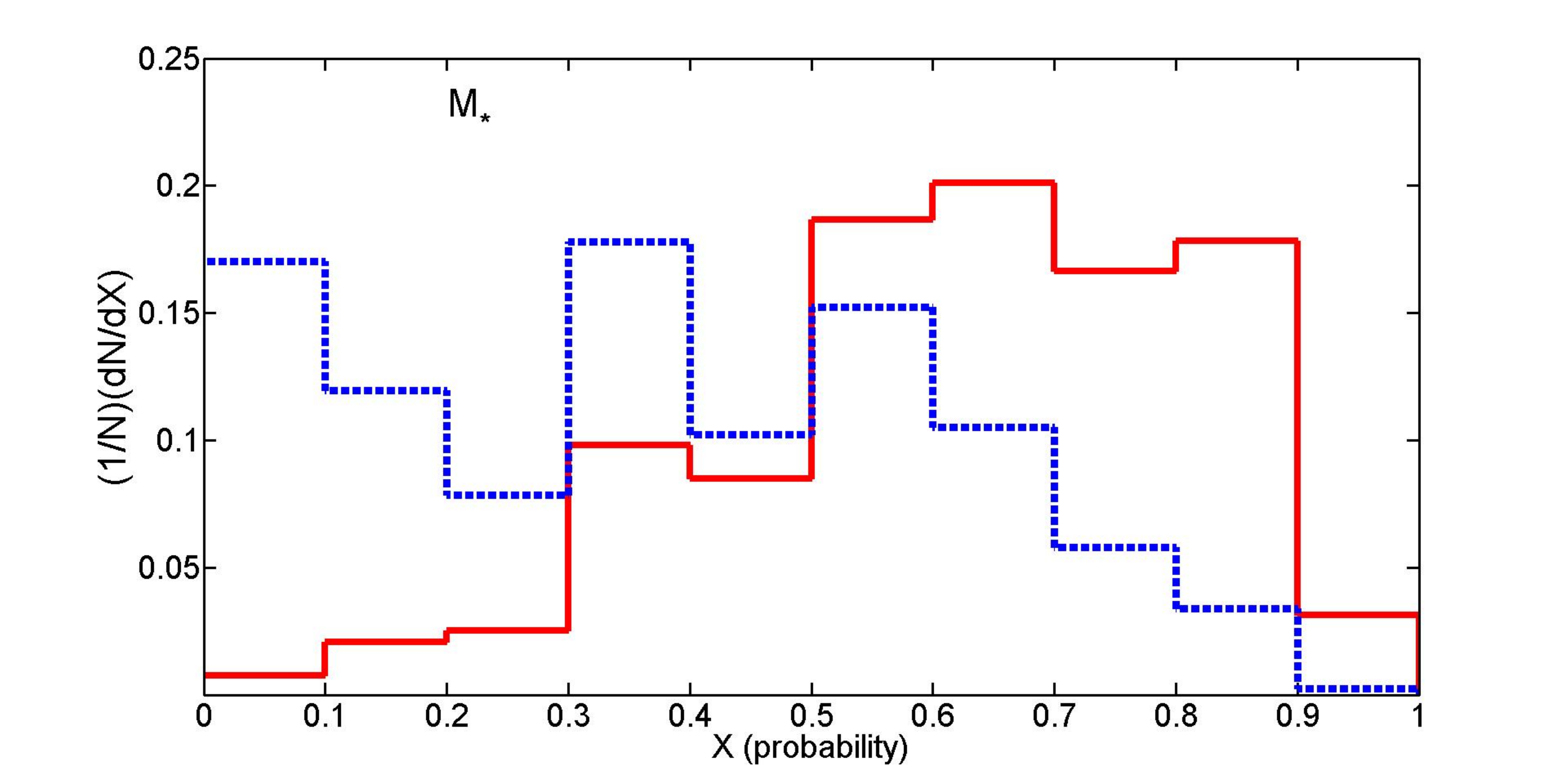}
\includegraphics[width=8cm,height=4cm,angle=0]{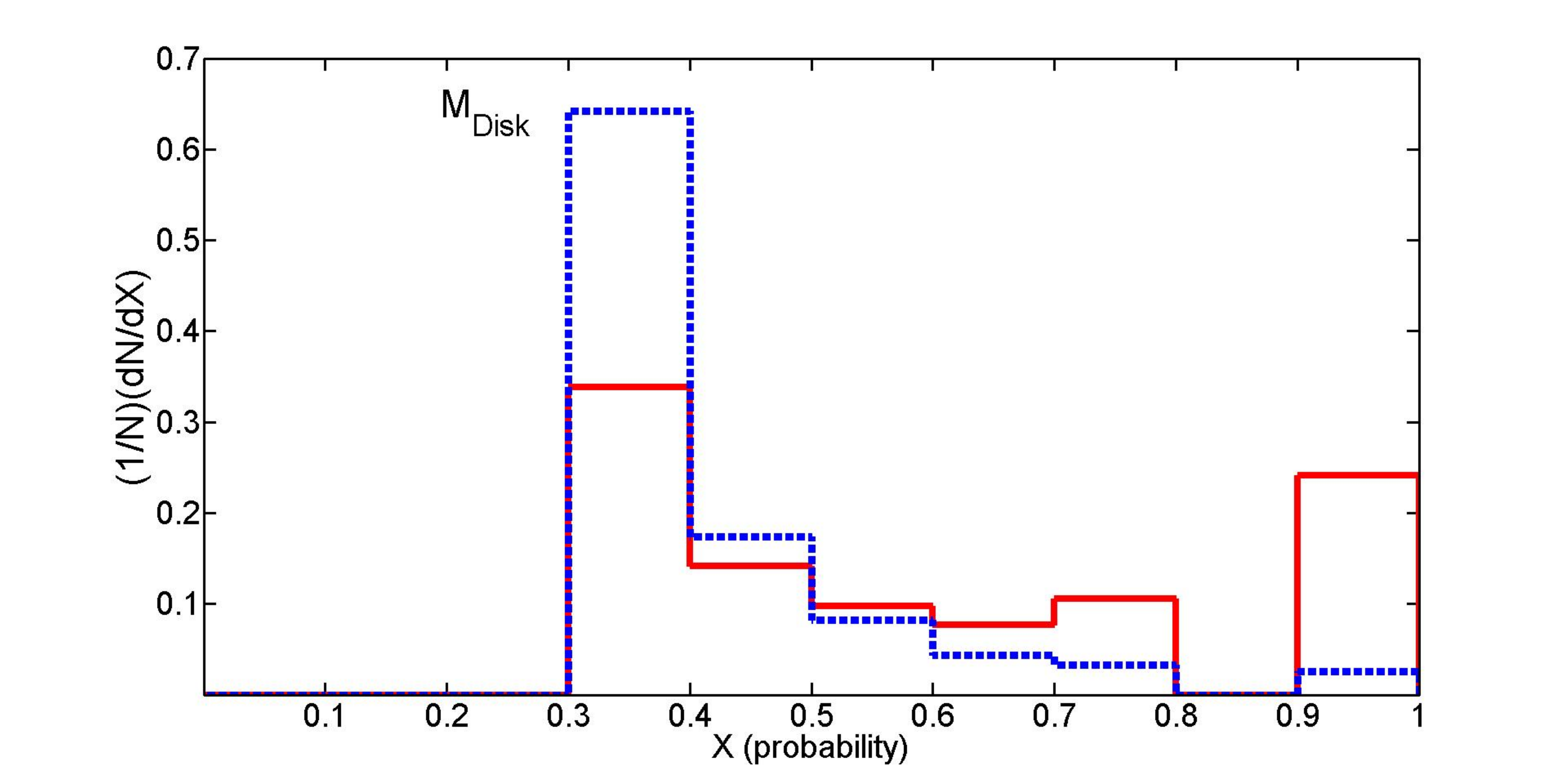}
\includegraphics[width=8cm,height=4cm,angle=0]{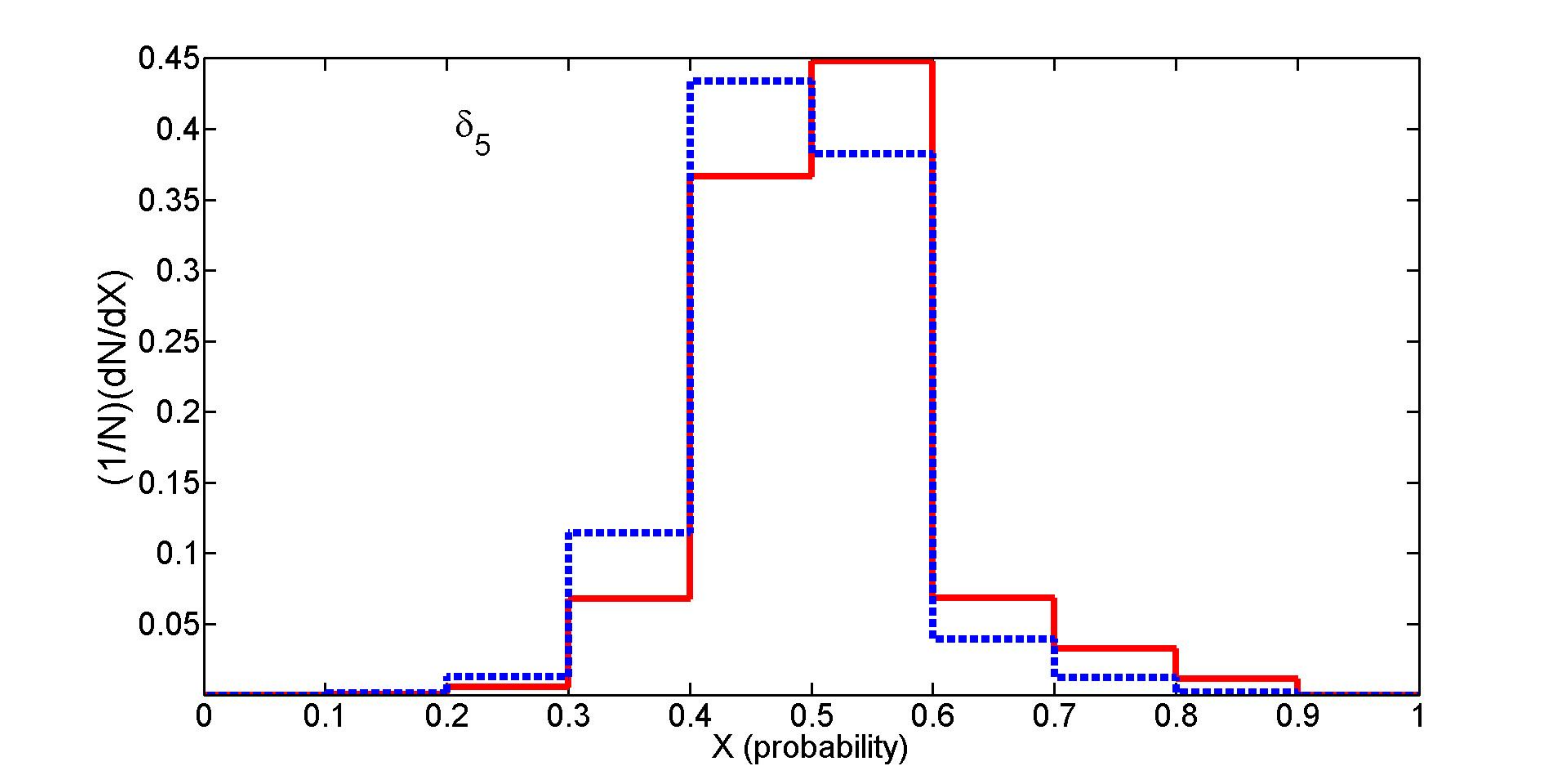}
\includegraphics[width=8cm,height=4cm,angle=0]{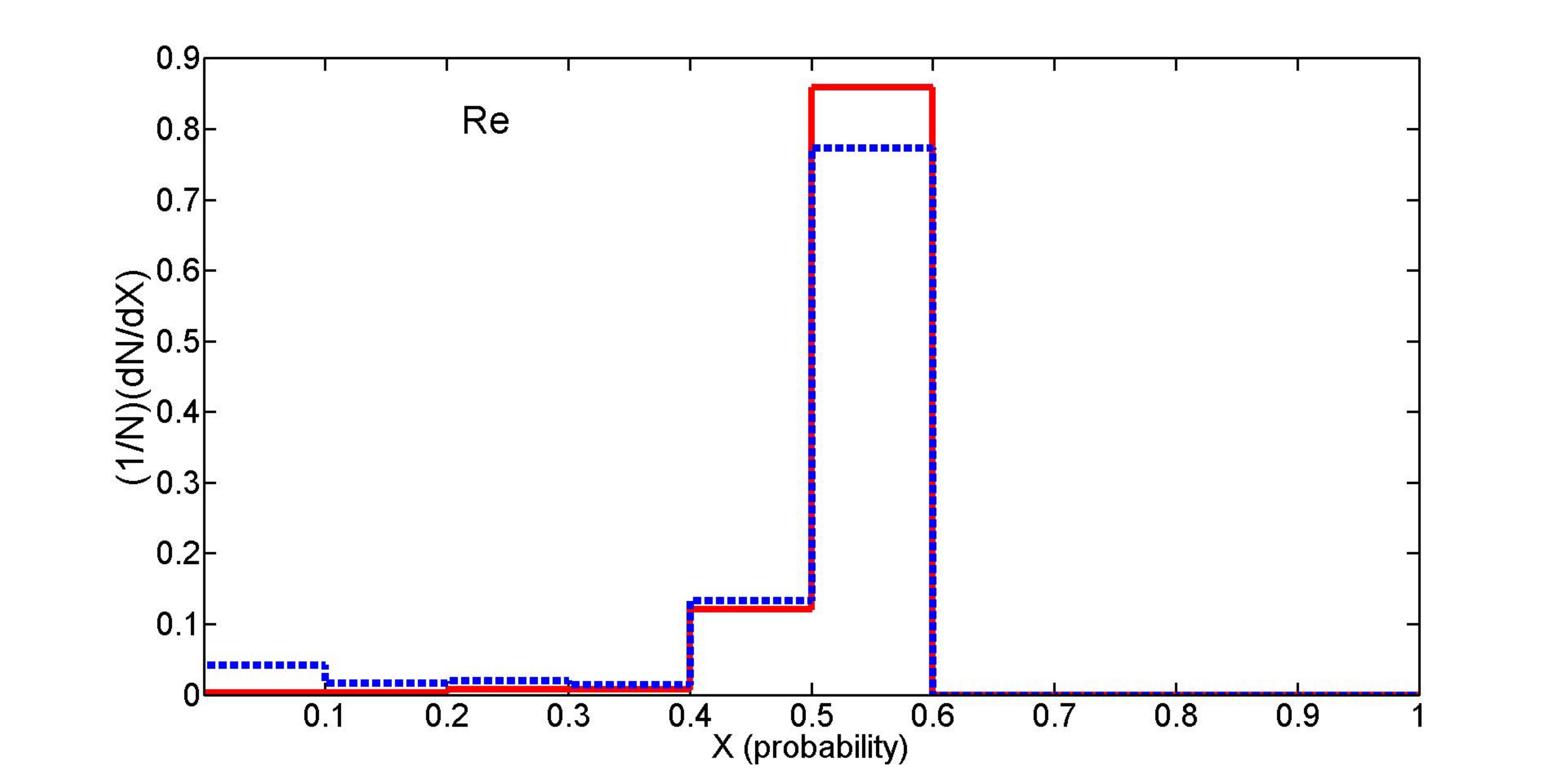}
\caption{Distributions of output probabilities (0 = SF, 1 = PA) from the ANN minimisation procedure for galaxies which are originally classified as star forming (blue lines) and passive (red lines). The top plot shows the distributions related to the ANN run where all of the parameters in Table 1 are used simultaneously as input data. The distributions of the eight single runs (single input data) are shown below. The parameters are organised from most predictive (top left) to least predictive (bottom right). These distributions can be compared to the trial case (for varying randomness, $\alpha$), shown in Figure \ref{fig-dist-area-2}.}
\label{fig-area}
\end{figure*}

Here we use the single parameters drawn from Table \ref{tab-data} as input data to the ANN pattern recognition algorithm. Initially, to show the maximum potential of our data and the ANN classifier, we perform a run in which all of the parameters are used {\it simultaneously} as input data. The distribution of the output probabilities for the two original classes for this case is shown at the top of Figure \ref{fig-area}. Two simple monotonic distributions are seen, one peaked at zero (for star formers, shown in blue) and one peaked near unity (for passive galaxies, shown in red). For example, if we choose a probability threshold at X = 0.5 we see that there are some misclassifications. Since we do not have a perfect classification, any single (or multiple) run should be compared to this run, which we hereafter label as `ALL'. 
However, the success rate of the ALL run is formally `outstanding' (see Table \ref{tab-auc}, and Hosmer \& Lameshow 2000), classifying $>$ 90\% of cases in the validation set correctly.

The rest of the panels in Figure \ref{fig-area} show the distributions of the ANN probabilities for each galaxy being passive for originally classified passive (red) and star forming (blue) galaxies, for each of the parameters in Table 1 treated singly. In general, the histograms in Fig. 8 for single runs can be compared to the test-data histograms in Fig. 3 to build some intuition for how the physical parameters perform compared to different levels of known degradation of information on the passive state. Central velocity dispersion and bulge mass perform qualitatively well, with simple monotonic distributions for each class, as with the ALL variables run. This behaviour is not seen for B/T, however, where there are many uncertain cases around probability X=0.5, although strong `correct' peaks at the extremes of the distribution are also present (we consider whether this could be a result of ambiguous `green valley' galaxies in Section A3). 

Particularly poorly separated distributions are seen for disk mass, bulge effective radius and $\delta_5$. For the former two parameters, the poor separation of star-forming and passive distributions is due to having a discrete value of zero for disk mass related to pure bulge (elliptical) galaxies. It is useful for the ANN code to know that there is no disk (this usually indicates a passive galaxy); however, knowing that it contains a disk does not determine the passive state with any kind of accuracy. Similarly, a very small bulge radii almost always indicates a star forming galaxy, but higher bulge radii can lead to a variety of masses, due to the underlying structure of the bulge. We test what impact spurious bulges or disks may have on our rankings in Appendix A7.   When we use the density parameter, $\delta_5$, as the input data the output is very similar to the case where $\alpha=10$ in the previous section. The two distributions are not distinguishable indicating that this parameter acts like a random number and has no connection to passivity.

\begin{figure}
\centering
\includegraphics[width=9cm,height=6.5cm,angle=0]{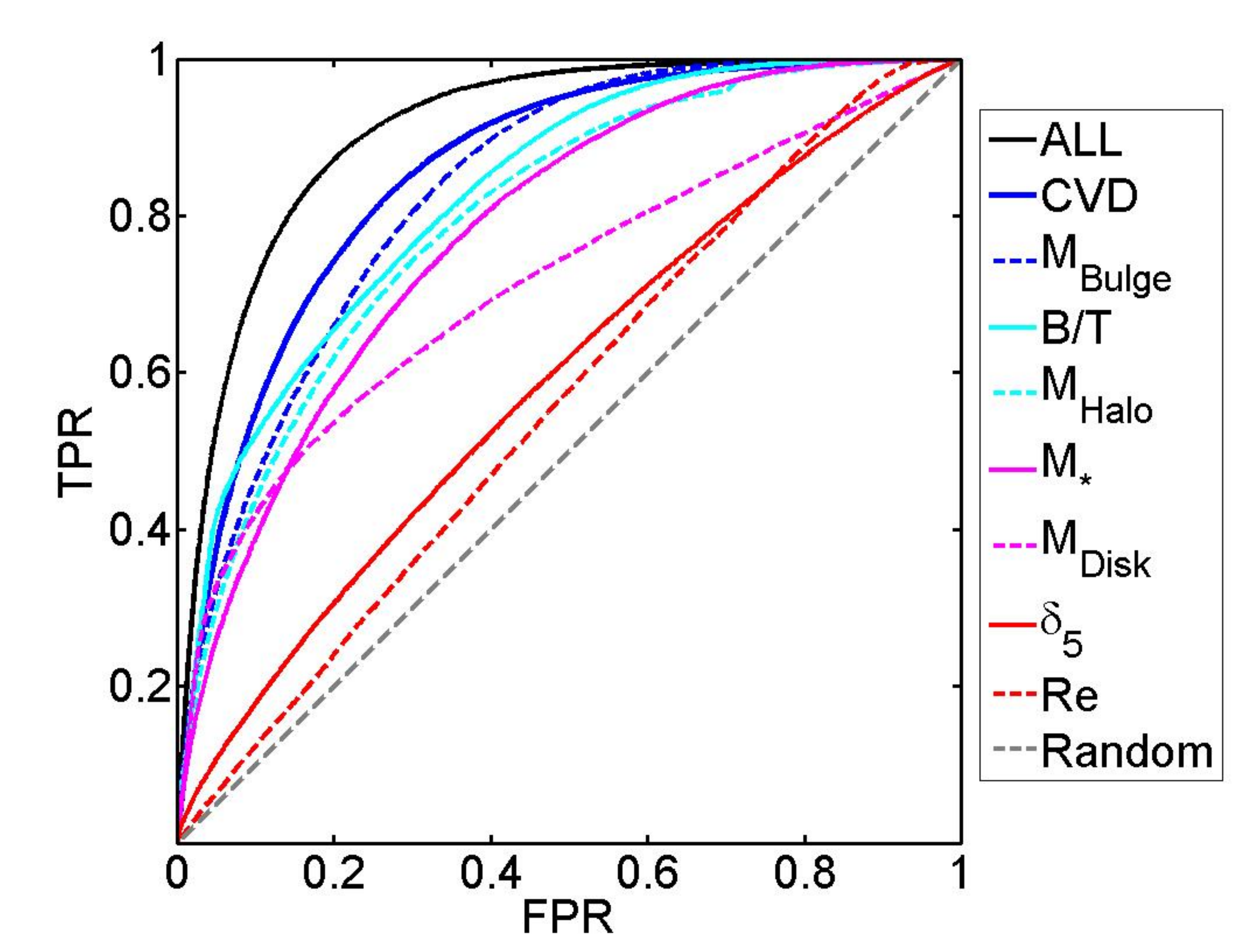}
\caption{Receiver Operating Characteristic (ROC) curves for each of the distributions shown in Figure \ref{fig-area}, for the galaxy parameters in Table 1, plotting True Passive Rate (TPR) vs. False Passive Rate (FPR), see Section 3.3 for details. The best performance (largest area under the curve, AUC) is achieved for all variables used together, `ALL', shown as a black solid line. The next best (and best single variable) is CVD followed by $M_{\rm bulge}$, i.e. it is parameters related to the inner-most regions of galaxies which perform best. An example random result is shown as the dashed black line, which performs only slightly worse than the local density ($\delta_{5}$) parameter or bulge effective radius.}
\label{fig-roc-comp}
\end{figure}

\begin{figure}
\centering
\includegraphics[width=9cm,height=6.cm,angle=0]{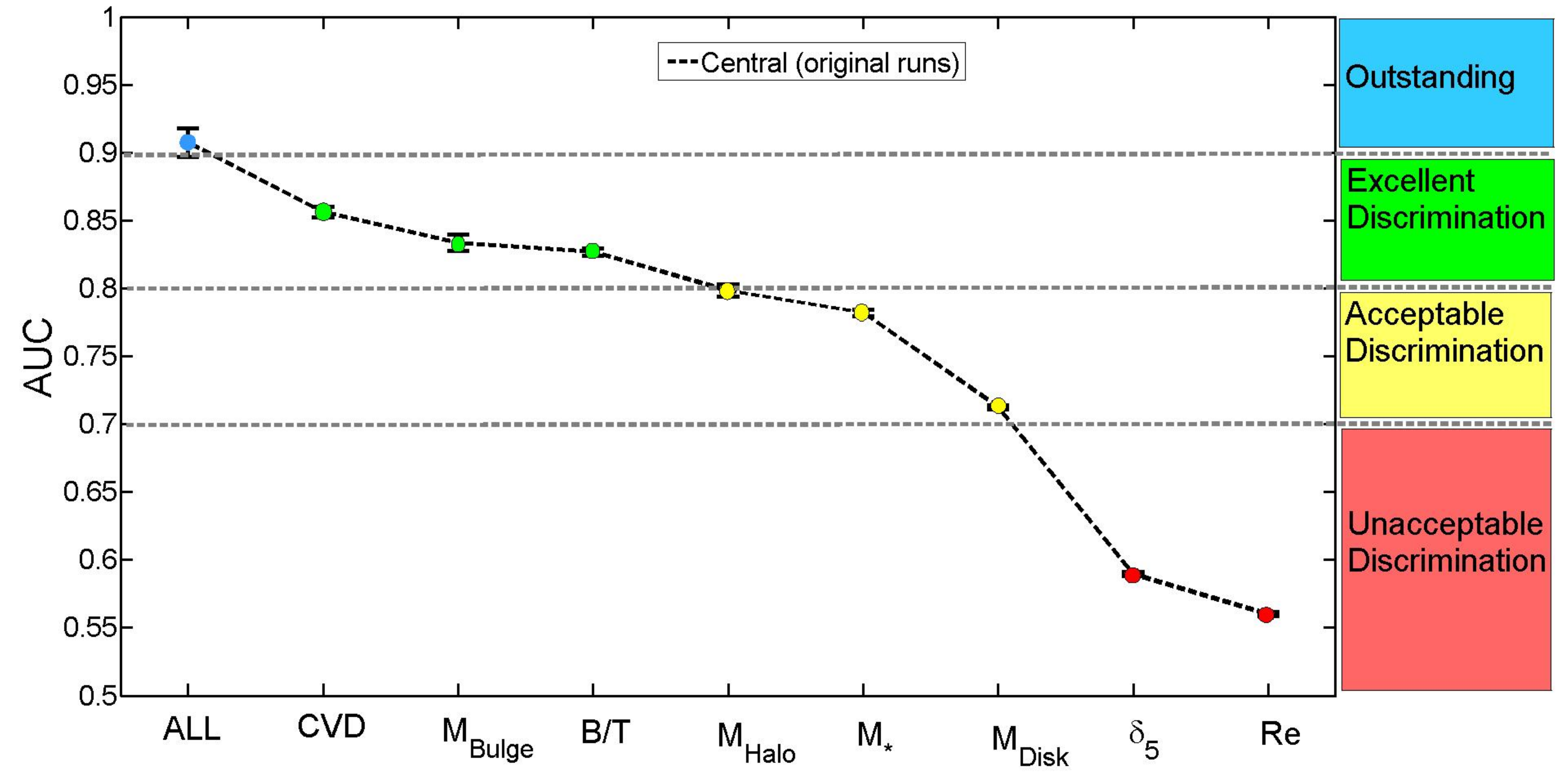}
\caption{Area Under the Curve (AUC) - single parameter plot. This plot illustrates the area under each ROC curve (see Figure \ref{fig-roc-comp}) with respect to the single galaxy parameters input data, given in Table 1. The parameters on the X-axis are sorted in terms of their AUC values, from highest (most predictive of the passive state of galaxies) to the lowest (least predictive of the passive state of galaxies). The errors are given as the standard deviation across the best ten runs, with the data points taken as the mean of the set. The points are colour coded by their success labels, as indicated on the plot (see Table 2). Clearly, parameters related to the inner regions of galaxies perform systematically better than parameters related to the whole galaxy, the outer regions, or environments of galaxies.}
\label{fig-single}
\end{figure}

We show the ROC plot (defined in Section 3.3) associated with each of the single parameters, as well as the ALL parameter run, in Figure \ref{fig-roc-comp}.  The black solid line is related to ALL, which has the best performance and the largest AUC.  We estimate the associated AUC for each of the single variables and show them in Figure \ref{fig-single}. To obtain the errors we perform many ANN runs for each single parameter and obtain the mean and standard deviation from the best well trained networks (top 10 results out of 15 total runs, each selecting a random 50,000 PA and 50,000 SF galaxies for training and a different random 100,000 galaxies for validation), ensuring an optimal solution has been found. The parameters on the X-axis of Figure \ref{fig-single} are ordered by their AUC values, i.e. showing most to least constraining variable. See Table 3 for the rankings and AUC values for centrals.

The physical galaxy properties are ordered in Table 3 by their AUC values, and hence by how predictive they are of whether a galaxy will be forming stars or not. The ordering is largely similar to Table 1, which is sorted by the scale at which each property is measured. Thus, there is a broad (but not perfect) trend from inner to outer regions in terms of quenching predictivity.  CVD, M$_{\rm bulge}$ and B/T are all ranked as "excellent" by our performance metric, with CVD being the single best performing property.  This result is in agreement with previous papers (e.g. Cheung et al. 2012, Fang et al. 2013, Bluck et al. 2014, Lang et al. 2014, Omand et al. 2014, Woo et al. 2015) that properties associated with central mass, or mass density, are the most important for determining the passive fraction.  Parameters associated with the galaxy's outer region or environmental metrics perform significantly less well. Such parameters include total stellar mass and halo mass which have frequently been used in the literature to parameterise the quenching of centrals (e.g. Peng et al. 2010, 2012; Woo et al. 2013, 2015). Interestingly, the \textit{size} of the bulge is the worst performing parameter, possibly suggesting that it is the mass and/or density of the inner region not its scale that affects star formation quenching. It is also interesting to note that bulge size is a particularly poor correlator to dynamical measurements of central black hole mass (e.g. Hopkins et al. 2007), whereas central velocity dispersion and bulge mass are tightly correlated to black hole mass (e.g. Ferrarese \& Merritt 2000, McConnell \& Ma 2014).
Taken together, Fig. 10 and Table 3 provide compelling evidence for the process that quenches central galaxies originating in the inner regions of galaxies.

\begin{table}
\begin{center}
\caption{ANN AUC ranking of single parameters for central galaxies.}
\begin{tabular}{c|l|c|l}
\hline\hline
Rank & Property & AUC & Success Label$^{*}$ \\
\hline
   & ALL                & 0.9074 $\pm$ 0.0106 & Outstanding    \\
1  & CVD  		&	0.8559 $\pm$ 0.0039  & Excellent	\\
2  & M$_{\rm bulge}$   	&	0.8335$\pm$ 	0.0060  & Excellent 	\\
3  &  B/T	& 0.8267 $\pm$ 0.0028  & Excellent	\\
4  &  	M$_{\rm halo}$ 	&0.7983 $\pm$ 0.0045  & Acceptable\\
5  & M$_{*}$ 	& 0.7819 $\pm$ 0.0025  & Acceptable	\\
6  & M$_{\rm Disk}$	 	& 0.7124 $\pm$0.0016  & Acceptable	\\
7  & $\delta_{5}$	& 0.5894 $\pm$ 0.0015  & Unacceptable 	\\
8  &  Re	& 0.5599$\pm$ 0.0013 & Unacceptable 	\\
\hline
\end{tabular}
\label{tab-sat}
\end{center}
$^{*}$ see Table 2 and associated text for definition. The errors are quoted as the standard deviation across the best 10 (out of 15) ANN runs, ensuring convergence.
\end{table}

\subsubsection{Implications of the Single Runs}

It is interesting that halo mass performs significantly better at predicting the passive state of central galaxies than local density, even though they are both ostensibly environmental parameters. Ellison, Patton \& Hickox (2015) find that halo mass is strongly correlated with the presence of radio loud AGN, whereas local density is not, which may offer us an explanation through the AGN driven quenching paradigm. Additionally, there are well known strong correlations between internal galaxy properties (e.g. stellar mass, B/T ratio, $M_{BH}$) and halo mass which are much weaker for local density (e.g. Moster et al. 2010). Further, Woo et al. (2013) argue that local density is a less useful parameter for measuring environment than halo mass or cluster-centric distance because it can exist in two distinct modes: inter-halo and trans-halo, and thus its relevance to a galaxy's star formation is unclear. 
In any case, halo mass is certainly not the most constraining single variable, performing significantly worse than properties related to the central regions of galaxies. Thus, it is possible that its relative success over local density (and stellar mass) is a result of `reflected glory' in that it is not a direct link to quenching but rather a result of its close correlation with inner galaxy properties.  

Our ANN rankings are broadly in agreement with the internal rankings of parameters made in the literature to date. However, this is the first attempt to rank the importance of all of these variables in a fully quantitative and objective manner. Specifically, we find that stellar mass has a much higher AUC than local density for centrals, in qualitative agreement with Peng et al. (2012). We also find that halo mass (derived indirectly from abundance matching) has a higher AUC than stellar mass, in agreement with Woo et al. (2013, 2015). Furthermore, we find that bulge mass is superior to all of the above in determining the passive fraction, as argued for in Bluck et al. (2014) and Lang et al. (2014). 
Bulge mass is also, slightly, superior to B/T structure in constraining the passive state of galaxies, as first pointed out in Bluck et al. (2014). However, bulge mass is not the best single variable found here in the ANN minimisation procedure: centralised velocity dispersion yields significantly higher AUC values (and hence tighter correlations to the star forming state of galaxies) than bulge mass. This was also argued for previously through an analysis of the passive fraction - (estimated) black hole mass relation in Bluck et al. (2015), and is consistent with the importance of central density or velocity dispersion found in several other works (e.g. Cheung et al. 2012, Wake et al. 2012, Fang et al. 2013, Woo et al. 2015).

Figure \ref{fig-single} may require a reformulation of the classic `mass-quenching' of Peng et al. (2010, 2012) and even the proposed updates to `bulge-mass-quenching' of Bluck et al. (2014) or `halo-mass-quenching' of Woo et al. (2013). We suggest that `inner-region-quenching', or most probably `black-hole-quenching' (i.e. AGN feedback) might be more appropriate given our results; we discuss this further in Section \ref{discussion_sec}. Clearly environmental properties, including those from the halo, are not the most constraining single variables for regulating quenching of central galaxies, nor is stellar mass or galaxy morphology ($B/T$), all of which have been previously claimed to be the dominant correlators to the passive fraction (e.g. Baldry et al. 2006, Cameron et al. 2009, Cameron \& Driver 2009, Peng et al. 2010, 2012, Woo et al. 2013). However, our result does agree with a complementary analysis, based on the area of the passive fraction relationships, presented in Bluck et al. (2015).  Furthermore, the finding by Bell et al. (2008, 2012) that essentially all truly passive systems have a high S\'{e}rsic index bulge (see also Wuyts et al. 2011) is in qualitative agreement with our finding that a high central velocity dispersion and hence central density is the best predictor that a galaxy will be quenched (out of our chosen list of physical galaxy parameters).  We have not included S\'{e}rsic index in our main parameter set, since it is not strictly a physical quantity, but we investigate it separately in Section \ref{discussion_sec}.  

Finally in this sub-section, it remains interesting that the mass of the galactic disk is so un-correlated with the passive state of the system (ranked 7/8), given that in most galaxies undergoing `normal' star formation it is the disk which is the site of gas being converted into stars. Thus, it seems that, even though disks are the sites of star formation, they are certainly not the regions from which quenching takes effect. This fact must present a serious challenge to models of galaxy quenching utilising feedback from stellar winds or supernovae in central galaxies.  Out of the list of physically motivated and plausible quenching scenarios considered in this work (see Section 1), AGN feedback suggests itself as a particularly attractive explanation since it is expected to originate in the central-most regions of galaxies, and hence it is a natural (and obvious) fit to our observed ranking of single galaxy parameters. In most models which apply AGN feedback, the energy available to quench central galaxies is directly proportional to the black hole mass (e.g. Croton et al. 2006, Henriques et al. 2014, Vogelsberger et al. 2014b, Schaye et al. 2015) and this is known empirically to be tightly correlated with central velocity dispersion and bulge mass (e.g. Ferrarese \& Merritt 2000, Haring and Rix 2004, Hopkins et al. 2007, McConnell et al. 2011, McConnel \& Ma 2014). However, other explanations may still exist (e.g. Carollo et al. 2013) and we examine some possibilities for these in the discussion (Section 5), alongside the, perhaps more obvious, contender of AGN-feedback.

We consider whether systematics from our initial sample selection can lead to a significant change in the ordering of these variables in the Appendices (see Sections A1 - A8). Generally, we find that the exact AUC values can change for the single parameters, up or down, as a result of sample selection (e.g. removing AGN, excluding green valley galaxies, restricting the sample to lower redshifts or higher velocity dispersions) but that our rankings are almost entirely unaffected and hence are highly stable to sample variation. In the next sub-section we consider multiple parameters acting in concert as predictors of the star forming state of central galaxies.

\subsection{Multiple Parameters}

\begin{figure}
\centering
\includegraphics[width=9cm,height=6.cm,angle=0]{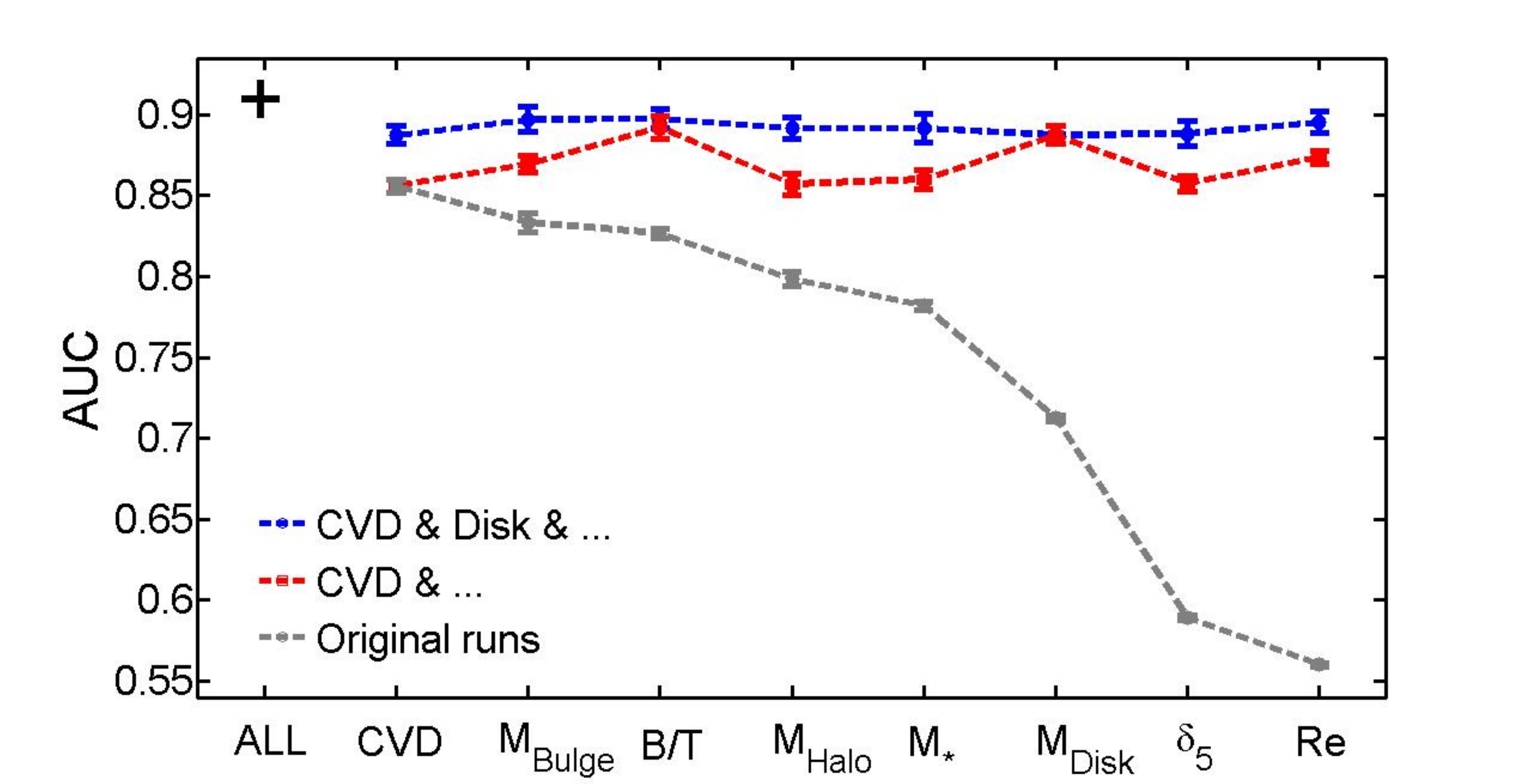}
\caption{f AUC - parameter plot for multiple runs. The grey line is the same as in Figure \ref{fig-single} which shows the results for single variables. The red line shows the result for CVD + each of the rest of the variables in turn, and the blue line shows CVD + $M_{\rm disk}$ + each of the other variables in turn. Note that CVD is the single best variable and $M_{\rm disk}$ is the best secondary variable in conjunction with CVD. No tertiary variable gives significant improvement over CVD and $M_{\rm disk}$, although $Re$ does perform formally the best. The black cross represents the AUC performance for all variables used simultaneously, shown for comparison. Note that the lines intersect where there are duplications of variables (i.e. for CVD and $M_{\rm disk}$), as they should.}
\label{fig-cvd-2-3}
\end{figure}

\begin{figure*}
\centering
\includegraphics[width=20cm,height=11.5cm,angle=0]{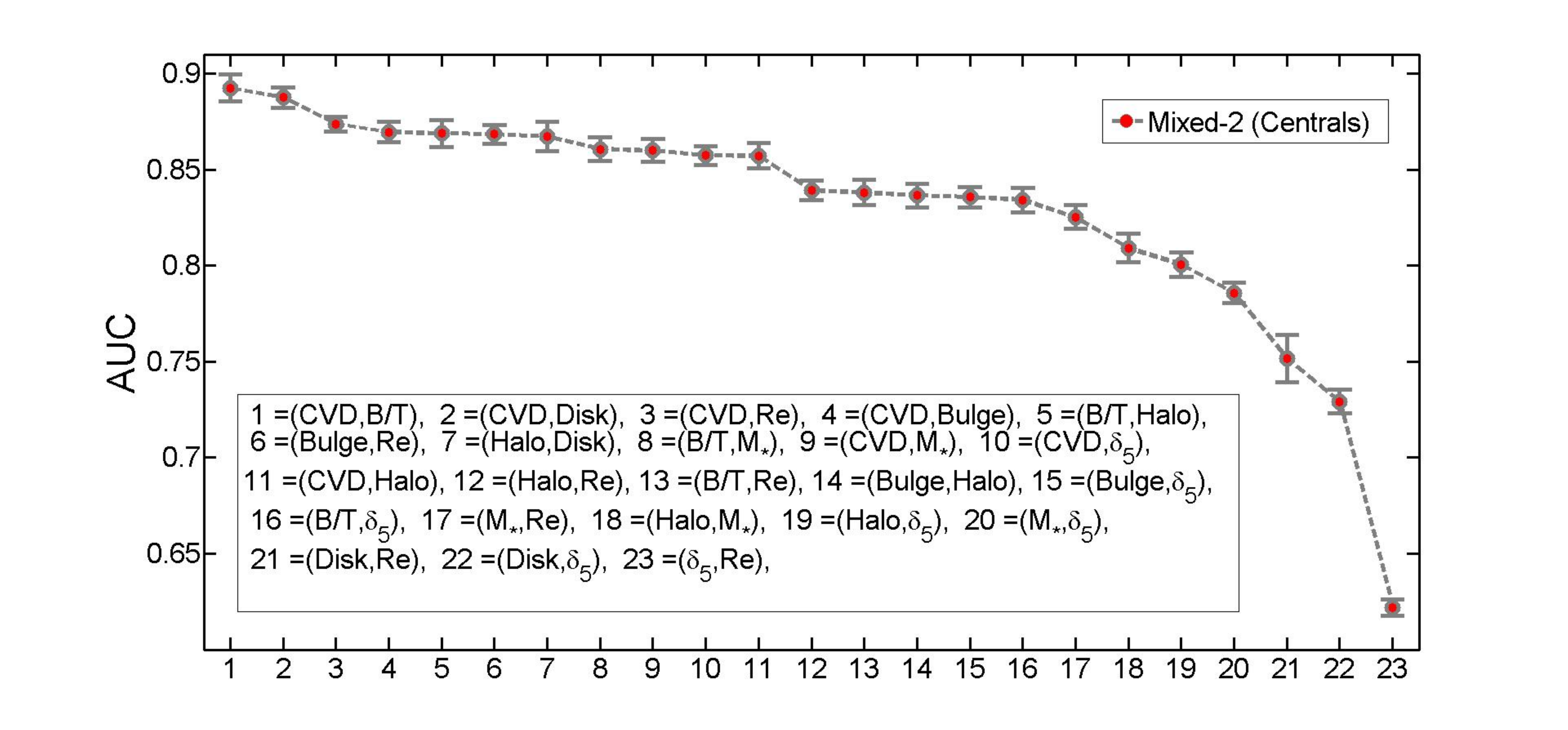}
\includegraphics[width=20cm,height=10cm,angle=0]{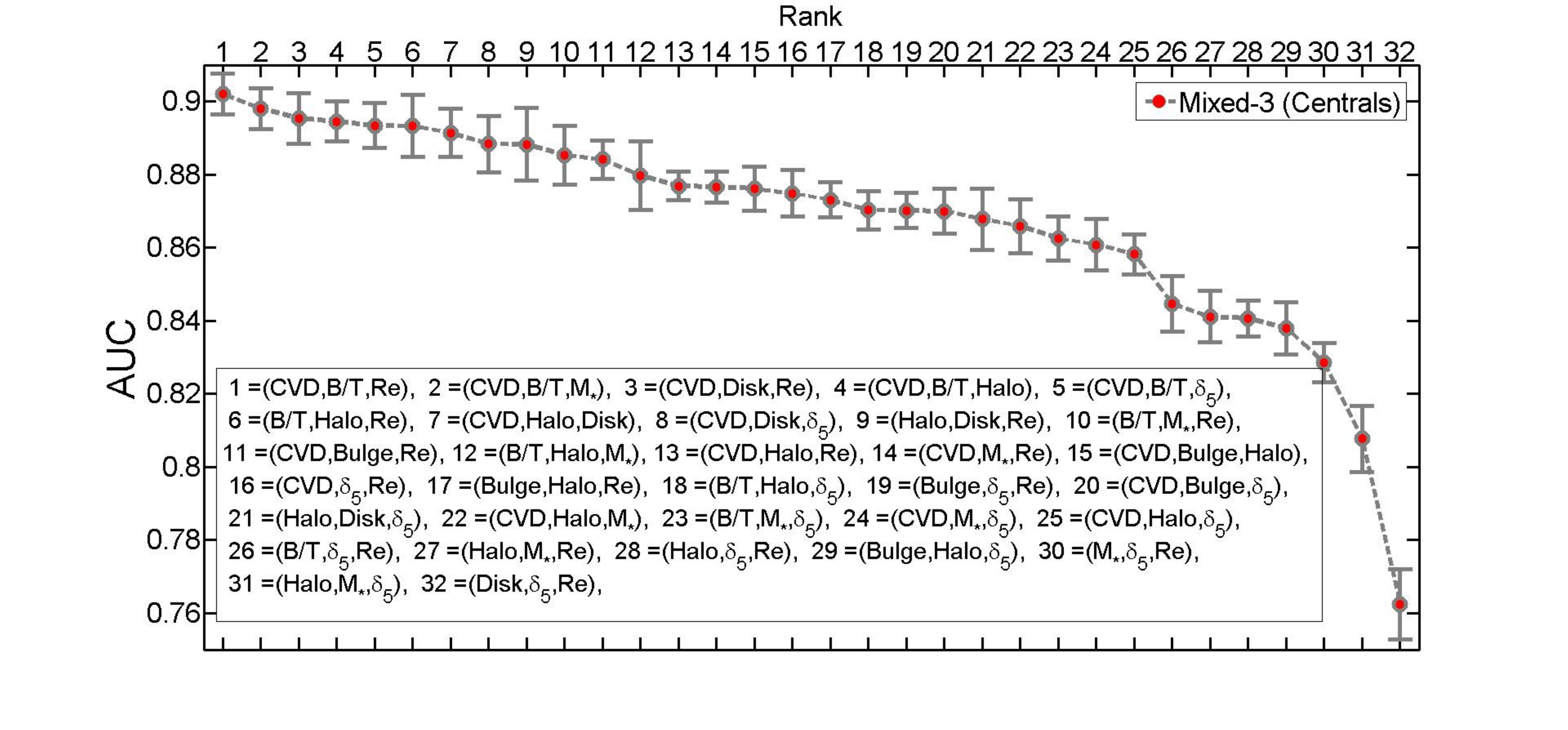}
\caption{ AUC - parameter plot for all unique set of multiple runs for central galaxies. The top panel shows all possible unique combinations of two parameters as input data, and the bottom plot shows all possible unique combinations of three parameters as input data. They are both ordered from most to least predictive at determining the passive state of galaxies. See Fig. A10 for all variables, regardless of uniqueness (i.e. containing various duplicates).}
\label{fig-mix}
\end{figure*}

Galaxy formation and evolution is a highly complex and non-linear problem, hence there is a limited amount of information and ultimately insight that can be gleaned from assessing how a single variable affects another single variable (e.g. the predictivity of the parameters of Table 1 in determining $\Delta$SFR). To improve on this picture one must seek to understand how galaxy properties interact together to constrain other variables, or sets of variables. Much pioneering work has already been attempted in the direction of multi-variable analysis of galaxy quenching. For example, Baldry et al. (2006) and Peng et al. (2010, 2012) find that the passive fraction of galaxies is a function of two variables, $M_{*}$ and $\delta_{N}$, and that these are in principle separable. 
Further work has found that galaxy morphology (e.g. B/T) has a strong influence at fixed $M_{*}$ and $\delta_{N}$ (Bluck et al. 2014, Lang et al. 2014) and that halo mass and central density can both affect the passive fraction of galaxies at fixed values of the other parameter (Woo et al. 2015). However, to date, no systematic ranking of two variable approaches for parameterizing the passive fraction exists, and certainly no higher (e.g. three variable) analyses exist. ANN techniques are ideal for problems of this type. 

 In this sub-section we perform a systematic analysis of the predictive power of all unique sets of two and three variables drawn from Table 1. But before we discuss our results for these 55 ANN runs, we start by considering a simplified case. Our goal here is to ascertain what the second, and the third, most important variable from Table 1 is for predicting the quenching of central galaxies. We start by always giving the ANN code our estimate for the central velocity dispersion (which is found to be the best single case, see Section 4.1). We then run an ANN minimisation for each pair of variables \{CVD, $X$\}, where $X$ represents each of the remaining variables in Table 1. 
We show the results on an AUC plot as a red line in Figure \ref{fig-cvd-2-3}. We find that adding any of the other variables leads to some incremental improvement in the predictive power over CVD alone, but this improvement is much smaller than for the other way around, i.e. adding CVD to any of the other parameters (compare the difference between the red and grey lines in Figure \ref{fig-cvd-2-3}). Disk mass and $B/T$ are the most successful secondary parameters, with similarly high AUC results, which amount to more or less the same thing given the strong relationship between CVD and $M_{\rm bulge}$.

 This is a surprising result because disk mass was found to be one of the worst parameters for a single variable and yet in conjunction with central velocity dispersion it performs better than any other 2-variable set containing CVD (this is a similar result to what is found with bulge mass in Bluck et al. 2014). There is no contradiction here, however, it just reflects that having complementary information about both the central and outer regions of galaxies is useful. Nonetheless, if one must choose only one region, the inner region is much more important for constraining central galaxy quenching than the outer. To explore this further, we consider the directionality of this trend: increasing disk mass at fixed central velocity dispersion actually {\it decreases} the probability of a galaxy being passive. Thus, it is likely that the inner region (i.e. CVD) gives us information about the quenching power (most probably the AGN, given the strong correlations between CVD and $M_{BH}$) and the outer region gives us information on what remains to be quenched (e.g. gas mass or gas fraction, both of which correlate with disk mass).

 We continue by giving the ANN codes \{CVD, $M_{\rm disk}$\} in conjunction with each of the other remaining variables (i.e. 1st + 2nd best + each of the rest). This is shown as a blue line in Figure \ref{fig-cvd-2-3}. Here most of the added variables offer some small improvement again, but with no clear sign of any single variable giving the highest improvement. It is particularly interesting to note that halo mass and local density (environmental parameters) lead to no significant improvement over \{CVD, $M_{\rm disk}$\}. Thus, even as a tertiary parameter environment is not significantly constraining of the passive state of central galaxies. This fact suggests that the quenching of galaxies is not strongly related to their dark matter haloes, once the inter-correlations with, e.g., black hole mass, central density and B/T are accounted for. 
However, it is not necessarily true that the best combination of two variables will contain the best single variable, nor is it necessary that the best combination of three variables will contain the best single or secondary variables. It is to the full list of unique possibilities we turn to next.

In Figure \ref{fig-mix} we show the AUC results for all unique combinations of 2-variable (top) and 3-variable (bottom) sets of parameters drawn from Table 1, i.e. we remove sets of variables which are equivalent (for example, \{$B/T$, $M_{*}$\} is identical to \{$M_{\rm bulge}$, $M_{\rm disk}$\}). The interested reader is referred to Fig. A10 in the appendix for the full rankings of all possible combinations of variables, which we warn contains repetitious content. The top four pairs of variables (top panel, Figure \ref{fig-mix}) all contain central velocity dispersion, with parameters related to the disk or galaxy morphology being the best additional combinations. This result is qualitatively similar to what we found in Figure \ref{fig-cvd-2-3}. The worst pairs frequently contain the local density parameter ($\delta_{5}$), often in conjunction with an outer region or whole galaxy parameter (e.g. $M_{\rm disk}$). These tend to perform significantly worse than variables which include information on the inner region of galaxies. Halo mass does perform quite well in combination with galaxy morphology, although it is significantly less predictive than some sets containing CVD.

We note that the pair of variables \{$M_{*}$, $\delta_{5}$\} ranks very poorly as 20/23 couplings of variables from Table 1, even though this has previously been considered the main dual-input for parameterizing galaxy quenching (Baldry et al. 2006, Peng et al. 2010, 2012). That said, it is important to emphasize that the use of galaxy density to constrain quenching is mostly applied to satellites in these prior works and here we focus solely on central galaxies. Also the set \{$M_{*}$, $B/T$\}, which was considered as a possible optimal ranking in Bluck et al. (2014), performs only near the middle of the possible sets determined here (8/23). We do not have a central density parameter in our set of variables, however, it is likely closely coupled to CVD (as indeed is suggested in Woo et al. 2015). If this is so the combined variables of the halo and the CVD can be compared to the result of Woo et al. (2015) for halo plus central density. 
This combination does not perform particularly well, with a rank of 11/23. Our brief comparison to the literature should serve as a caution to anyone planning to model the quenching of galaxies via conventional techniques, these are clearly not optimal. If a two parameter fitting technique is required, the best choice, out of the variables we consider, is \{CVD, B/T\}.

The lower panel of Figure \ref{fig-mix} shows our results from 32 ANN runs for all unique combinations of three variable sets of the parameters in Table 1. To our knowledge, this is the first attempt to construct a systematic ranking of three-variable parameterizations of galaxy quenching. All of the top five sets contain central velocity dispersion. Thus, parameters related to the centre of galaxies are essential for predicting quenching even in sets of two and three parameters. Environmental metrics  ($\delta_{5}$, $M_{\rm halo}$) are rare in the top ten, whereas amongst the lowest ranked sets these are much more common; the very worst sets often contain two environmental metrics, further highlighting their lack of predictivity for central galaxy quenching. 
The best three-variable parameterization from our data is \{CVD, $B/T$, $Re$\}, although it performs comparably well with all of the top five or so combinations, again containing no evironmental metric. 

The results from these mixed runs point in a similar direction to the single variable run: whatever quenches central galaxies is mostly connected with the inner-most regions of galaxies probed in our dataset. These are the parameters which are most tightly correlated with supermassive black hole mass, and hence AGN feedback energy (see Section 5). Parameters related to the halo mass (or local galaxy density) are significantly less predictive in constraining the passive state of central galaxies, which must present a serious challenge to models of central galaxy quenching arising from the halo, or the environment generally (e.g. Dekel \& Birnboim 2006, Dekel et al. 2009, Woo et al. 2013, Dekel et al. 2014).

\section{Discussion -- What drives central galaxy quenching?}\label{discussion_sec}

From a theoretical perspective, there are numerous physical processes associated with galaxy formation and evolution that can lead to a gradual or more sudden impact on star formation, in some cases leading to total cessation or quenching. Broadly speaking, all of these scenarios can be described as varying types and degrees of `baryonic feedback'. There are two essential questions here: 1) why do galaxies stop forming stars (especially given that there is plenty of gas remaining in the Universe for them to convert)? and 2) why do so few baryons end up residing in galaxies, i.e. at the local gravitational minima (current estimates of $\sim$ 10\%, Shull et al. 2012)? These two questions are highly likely to be related, with a common (set of) explanation(s). 
Our aim in this paper has been to identify the key parameter(s) associated with different quenching scenarios and assess how effective they are at predicting whether galaxies will be passive or star forming. From this we can give evidence for or against different models.

In Section 4.1 we find that properties related to the central regions of central galaxies are most predictive of the passive state of the system, with properties related to the entirety of the galaxy or the outer regions and environments being significantly less constraining (see Figure \ref{fig-single} and Table 3). This immediately suggests that the source of the energy needed for quenching central galaxies might originate (or be closely coupled with) the centre of these galaxies. This is exactly as expected for the AGN-feedback driven quenching scenarios (e.g. Croton et al. 2006, Bower et al. 2008, Hopkins et al. 2008, 2010, Vogelsberger et al. 2014a,b, Schaye et al. 2015). 
On the other hand, with virial shock heating driven quenching we would anticipate halo mass to be the most significant parameter (e.g. Dekel \& Birnboim 2006, Dekel et al. 2009, Woo et al. 2013, 2015); with supernova feedback driven quenching we would expect stellar mass to be key (e.g. Dalla \& Schaye 2009, Guo et al. 2011); and with environmental quenching we would expect a more significant dependence on both local density and halo mass (e.g. van den Bosch et al. 2008, Tasca et al. 2009, Wetzel et al. 2013, Hirschmann et al. 2013).

Our conclusion that quenching originates in the centre of galaxies is somewhat different to that reached by several papers in the field (e.g. Baldry et al. 2006, Cameron et al. 2009, Peng et al. 2010, 2012, Woo et al. 2013) although we do find accord with the conclusions of several other more recent papers (e.g.  Wake et al. 2012, Bell et al. 2012, Bluck et al. 2014, Lang et al. 2014, Omand et al. 2014, Bluck et al. 2015, Tacchella et al. 2015).  In earlier work, Bell et al. (2008) presented some of the first evidence for the central bulge component being the most significant indicator of quenching by noting that a high S\'{e}rsic index bulge is ubiquitous in passive systems. The reason for most of the tension between our results and some of the literature is that we consider a more complete list of parameters than these earlier works. The internal rankings seen in the literature are recovered precisely by our analyses, we just extend this prior work by including more parameters. We are also the first to make a systematic ranking of the predictivity of pairs and triplets of variables (see Section 4.2). 
Here we find that parameters related to the central regions of central galaxies are still crucial to include in the most successful sets, indicating that the importance of the central region is not an artifact of multiple (other) processes acting in concert. 

Although AGN feedback driven quenching of central galaxies is a natural explanation of our results, it is not necessarily the only good explanation. In the remainder of this discussion we will focus on plausible alternative explanations (our conclusions are only as good as our input assumptions). 

A key assumption we have made in this investigation is that the quenching of galaxies is binary in nature, i.e. galaxies are either star forming or they are quenched. We do consider the possibility of intermediate (green valley) cases in the appendices (see Section A3), although even here the implicit assumption is that these are rare or non-representative cases, most probably transitory in nature. Broadly speaking this same assumption is inherent in any approach which uses passive fractions (as with much of the literature on the subject, e.g. Baldry et al. 2008, Peng et al. 2010, 2012, Woo et al. 2013, Bluck et al. 2014, 2015). However, there is mounting evidence that the specific star formation rates (sSFR) of galaxies might change as a function of halo mass, without significantly affecting the passive fraction (Woo et al. 2015) and that different galaxies can migrate through the green valley at different rates depending on their morphologies (Schawinski et al. 2014). These types of subtle effects would not be noticeable in our current ANN analysis, although it would be possible and interesting to additionally train a network for predicting sSFR values (and green valley transition times) in addition to the binary quenched : star forming designation. This notwithstanding, we expect these non-binary extensions to be only minor perturbations on our general trends since galaxies do separate out convincingly into two clearly separable sub-sets in terms of their star formation rates and colours, suggesting that successful binary classification is the most important step in understanding quenching.

One possibility for further consideration is that the success of a given variable (or combination of variables) at predicting whether a central galaxy will be star forming or passive is primarily a function of how accurately measured that variable is. Thus, in this scenario, well measured parameters would perform better. This is certainly true if all physical galaxy parameters are fundamentally equally predictive of quenching. However, we note that this is unlikely to be the main driver of our trends here. To illustrate this, consider bulge and disk mass. These two sub-components of galaxies are measured with more or less equal precision in the bulge disk decompositions of Mendel et al. (2014) and Simard et al. (2011). 
However, bulge mass is significantly more predictive of quenching than disk mass (see Fig. 10 and Table 3).  One exception to this is perhaps halo mass which is inferred indirectly. It is certainly possible that improved measurements of the masses of central galaxy haloes in Sloan might improve the overall ranking of halo mass. That said, our conclusion that AGN feedback is the most probable explanation of our trends rests on the tight relationship between CVD and $M_{BH}$. If we were to estimate $M_{BH}$ from CVD it is unlikely we would measure this with any greater precision than $M_{\rm halo}$. If this is true then it is still most likely that black hole quenching dominates over halo mass quenching for low redshift central galaxies. Nonetheless, it would certainly be interesting to revisit these analyses with dynamically measured halo and black hole masses, when sufficient numbers of each become available.

Another interesting potential explanation for the apparent dominance of central velocity dispersion to quenching is that it is not the current set of galaxy properties which matter for quenching but the set of parameters at the time (or before) quenching takes effect (e.g. Carollo et al. 2013). This is unarguably true; however, estimating the parameters a galaxy had at an earlier epoch is fraught with difficulty (e.g. Torrey et al. 2015). In this first work on applying ANN techniques to the problem of galaxy quenching we choose to focus on directly measurable physical galaxy parameters. That said, by following a few lines of empirical reasoning we may conclude that a galaxy of a given stellar mass which quenched earlier than another similar mass galaxy would be denser (and hence have a higher central velocity dispersion). This follows directly from the assertion that we are looking at a same mass galaxy and a simple application of the size Ñ- mass relation as a function of redshift (Carollo et al. 2013). Arguments of this type provide an important equivocation to our interpretation: {\it correlation does not imply (nor necessitate) causation}. Thus, there are any number of possible explanations for the observed trends found in this work, of which the current example is just one possibility. It is therefore necessary to ask the follow-up question: given the observed rankings of galaxy parameters in quenching, what is the most likely physical explanation for this? To aid in answering this question detailed comparisons to semi-analytic models (e.g. Henriques et al. 2014, Somerville et al. 2015) and cosmological hydrodynamical simulations (e.g. Vogelsberger et al. 2014a, Schaye et al. 2015) must be made. Bluck et al. (in prep.) will begin this process for central galaxies. 

It is, of course, also conceivable that some new physical parameter or set of parameters will do much better than central velocity dispersion, and may ultimately reveal a link between quenching and some other physical process than considered here. At this point it is important to reiterate that in this work we have focused exclusively on {\it physical} galaxy parameters, e.g. masses, velocities, densities and sizes. In this manner we have disregarded other parameters of potential interest, such as S\'{e}rsic indices. Wuyts et al. (2011) showed that the two peaks of the star formation rate Ñ- stellar mass plot are divided cleanly by S\'{e}rsic index, with quenched galaxies having higher values of $n$ than star forming galaxies (a result previously considered in Bell et al. 2008). In fact, running our ANN method for $n$ we find that it performs slightly better even than CVD (with AUC = 0.891 $\pm$ 0.003), confirming these prior results. However, we exclude $n$ from our main analysis in this paper for two reasons: 1) $n$ is a parameter in a fitting model, not a physical galaxy parameter, and hence it does not fit within the remit of this paper to examine which physical parameters are most tightly correlated with central galaxy quenching; and 2) $n$ is measured in a single optical wave-band and thus can be significantly affected by ongoing star formation (or absence of star formation) in its measurement. The second point is very important to highlight, since the excellent performance of $n$ in predicting quenching could be no more significant than attesting that star formation typically happens in disk structures. Bright blue new stars in a disk lower $n$ and the absence of these stars in a galaxy in general yields a higher value of $n$. Thus, for these reasons we find the S\'{e}rsic parameter, $n$, to be less interesting to focus on than the other (physical) parameters in our study. Nevertheless, we mention here its excellent performance in AUC, should this be of use or interest to further research.

Finally, it is interesting that the set of physical galaxy properties listed in Table 1 is not sufficient, even acting together as inputs for a sophisticated pattern recognition algorithm, to correctly determine the star forming state of all central galaxies (with $\sim$ 8\% misclassified). There are a number of possible explanations for this effect, including, of course, inaccuracies in the measured parameters and observational errors. However, it seems likely that this set of variables is simply not an exhaustive list of all galaxy properties relevant to central galaxy quenching. A similar conclusion is made for a slightly different set of data in Knobel et al. (2015), where they conclude that galactic conformity (the tendency for passive satellites to orbit passive centrals) is evidence for `hidden variables' in galaxy formation. 
Whilst this may well be true, it is also possible that there is an irreducibly probabilistic nature to whether a given galaxy will be passive or not, based in part on the chaotic evolutionary history of individual galaxies. In any case, this motivates the need to explore more variables in future statistical studies of the relationship between galactic star formation, quenching, and galaxy properties. However, global parameters may never be sufficient to be perfectly predictive of quenching, thus it may be necessary to consider more complex sets of sub-galactic variables.

\section{Conclusions}

In this paper we present a novel technique for assessing which galaxy properties impact the quenching of central galaxies. We train an artificial neural network (ANN) non-linear model to recognise star forming and passive galaxies (for a training and verification set each containing 100,000 galaxies). The network is provided with each of the physical galaxy parameters shown in Table 1 as input data, singly and in groups of two and three. A higher success rate of predicting whether galaxies will be star forming or passive from a given variable, or set of variables, is taken to imply a greater causal link between that parameter (or set) and the quenching mechanism(s). We quantify the performance of the network for each parameter and group of parameters by computing the area under the ROC curve (see Section 3.3), with higher AUC values signalling greater predictive power. We summarise our main contributions here:\\

\noindent $\bullet$ For single variables, we find the highest AUC values, and hence predictive power, for central velocity dispersion, followed by bulge mass and B/T.  All of these parameters formally rank as `excellent' predictors of passivity in galaxies.

\noindent $\bullet$ Parameters related to larger scale galaxy properties (e.g.  $M_{*}$, $M_{\rm disk}$) or environment ($M_{\rm halo}$,  $\delta_{5}$) perform significantly less well.

\noindent $\bullet$ The general trend in predictivity from central internal parameters to outer or external parameters provides evidence for the quenching of central galaxies originating in the mass concentration of inner regions, and being largely unrelated to their extended structures or environments (see Figure \ref{fig-single} and Table 3).  

\noindent $\bullet$  We suggest that the predictive success of inner-region galaxy parameters reflects the source of the quenching energy, most probably originating from black hole accretion and AGN feedback. However, we do consider other possibilities to this explanation in the discussion (Section 5).

\noindent $\bullet$ Bulge effective radius is the worst performing parameter amongst those tested.  This is not inconsistent with AGN-driven quenching, since bulge size is not strongly correlated with black hole mass, whereas bulge mass and central velocity dispersion are.

\noindent $\bullet$ For dual and triple variable sets, inner-galaxy properties are very common amongst the best configurations, with environmental properties being rarely seen. This indicates that the importance of the inner-region parameters over outer region or environmental parameters does not diminish with the more inclusive multi-variate analysis.  

\noindent $\bullet$ Although we exclude the S\'{e}rsic index parameter, $n$, from our main analysis since it is not a physical galaxy property per se, we note that it performs particularly well at predicting whether galaxies will be star forming or not. This could, however, just be an artefact of this parameter tracing the light from star formation directly.

We perform many tests and investigations of the effects of sample variation and potential biases and systematics on our results in the Appendix (Sections A1 - A8). We find that our rankings are very stable to issues of this type (including exclusion of green valley galaxies or AGN, volume weighting or restricting to a volume limited sample, and additional axis ratio, mass, redshift or data quality cuts).\\

\section*{Acknowledgments}

We thank J. Trevor Mendel, David R. Patton, Jillian Scudder and Luc Simard for helpful discussions on this work. We are particularly grateful to Luc Simard and Trevor Mendel for much assistance with using the structural, morphological and mass parameters in the Simard et al. (2011) and Mendel et al. (2014) catalogues. We gratefully acknowledge funding from the National Science and Engineering Research Council (NSERC) of Canada, particularly for a Discovery Grant awarded to SLE. 

Funding for the SDSS and SDSS-II has been provided by
the Alfred P. Sloan Foundation, the Participating Institutions, the National Science Foundation, the U.S. Department of Energy, the National Aeronautics and Space Administration, the
Japanese Monbukagakusho, the Max Planck Society, and the
Higher Education Funding Council for England. The SDSS
Web Site is http://www.sdss.org/

The SDSS is managed by the Astrophysical Research Consortium for the Participating Institutions. The Participating
Institutions are the American Museum of Natural History, Astrophysical Institute Potsdam, University of Basel, University of Cambridge, Case Western Reserve University, University of Chicago, Drexel University, Fermilab, the Institute for Advanced Study, the Japan Participation Group, Johns
Hopkins University, the Joint Institute for Nuclear Astrophysics, the Kavli Institute for Particle Astrophysics and
Cosmology, the Korean Scientist Group, the Chinese Academy
of Sciences (LAMOST), Los Alamos National Laboratory,
the Max-Planck-Institute for Astronomy (MPIA), the Max-Planck-Institute for Astrophysics (MPA), New Mexico State
University, Ohio State University, University of Pittsburgh, University of Portsmouth, Princeton University, the United States Naval Observatory, and the University of Washington.\\\\

\noindent \textbf{Author contributions:} HT wrote the ANN code, performed the ANN analysis, led the technical direction of the project, produced the figures, and contributed to the text. AFLB generated the observational samples, led the intellectual direction of the project, and wrote the majority of the paper. SLE conceived the project and contributed to the text and intellectual direction.

\appendix

\section{Sample Variation and Possible Systematics}
To demonstrate the robustness and stability of our rankings of the single galaxy parameters to sample variation, we perform different ANN runs for different carefully selected sub-samples, similar to what is shown in Figure \ref{fig-single}.

\subsection{Lower Redshift Cut}

\begin{figure}
\centering
\includegraphics[width=9cm,height=6.cm,angle=0]{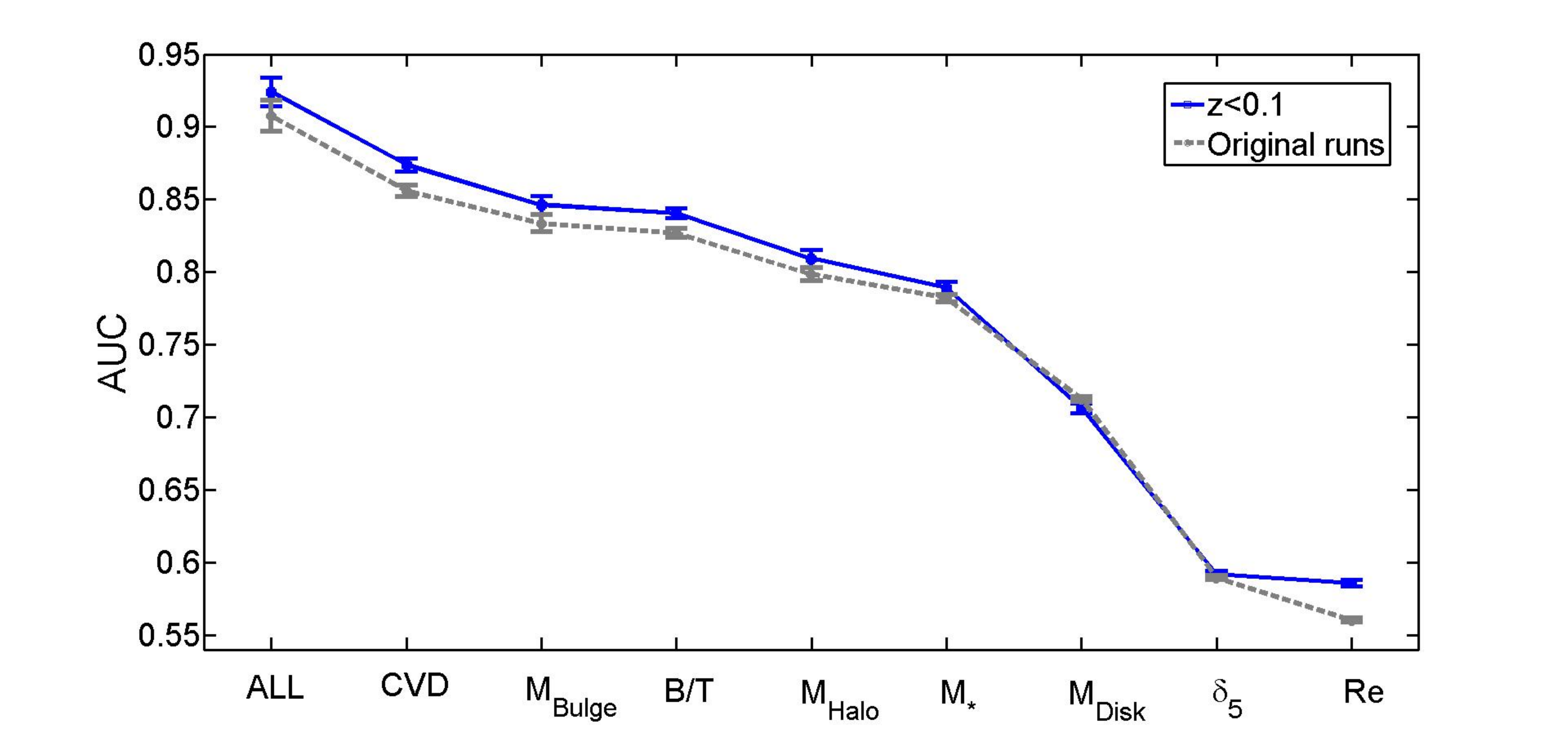}
\caption{AUC - single parameter plot. The grey dashed line is the same as Figure \ref{fig-single} which comprises all redshifts up to our original limit of $z_{\rm spec} <$ 0.2. The blue solid line is for a restrictive sub-sample of galaxies with $z_{\rm spec} <$ 0.1. }
\label{fig-zl01}
\end{figure}

For the main sample we use a redshift cut of $z_{\rm spec} <$ 0.2 (as in Bluck et al. 2014, 2015). Here we consider restricting the sample to $z_{\rm spec} <$ 0.1 where we will have more reliable bulge + disk decompositions (due to higher surface brightness features at a given mass) and a higher S/N of emission lines (used for SFR and AGN determination) and the spectral continuum aiding absorption line measurements (used in calculating velocity dispersions and estimating $M_{BH}$). This sub-sample also has a higher mass/ colour completeness than the higher redshift sample (but see Section A7 for a more thorough treatment of completeness). We re-run the ANN codes for ALL and each of the single runs, and go through the methodology exactly as in Section 4.1.

We show the AUC performance indicator for this more restrictive sample in Figure \ref{fig-zl01} (shown as a blue line), and overplot the previous result from Figure \ref{fig-single} (shown as a grey line). In general, the performance of the ANN is improved by the lower redshift cut, indicated by higher AUC values for ALL and most single cases. This is as we might expect from increasing the S/N of our average data. However, importantly, the ranking of single variables (i.e. their ordering in terms of AUC and thus how effective they are at constraining the passive state of galaxies) is left completely unchanged. This implies that our results are robust to changes in the surface brightness of galaxy components and to the S/N of emission and absorption lines, lending more confidence to our rankings.

\subsection{Excluding AGN}

\begin{figure}
\centering
\includegraphics[width=9cm,height=6.cm,angle=0]{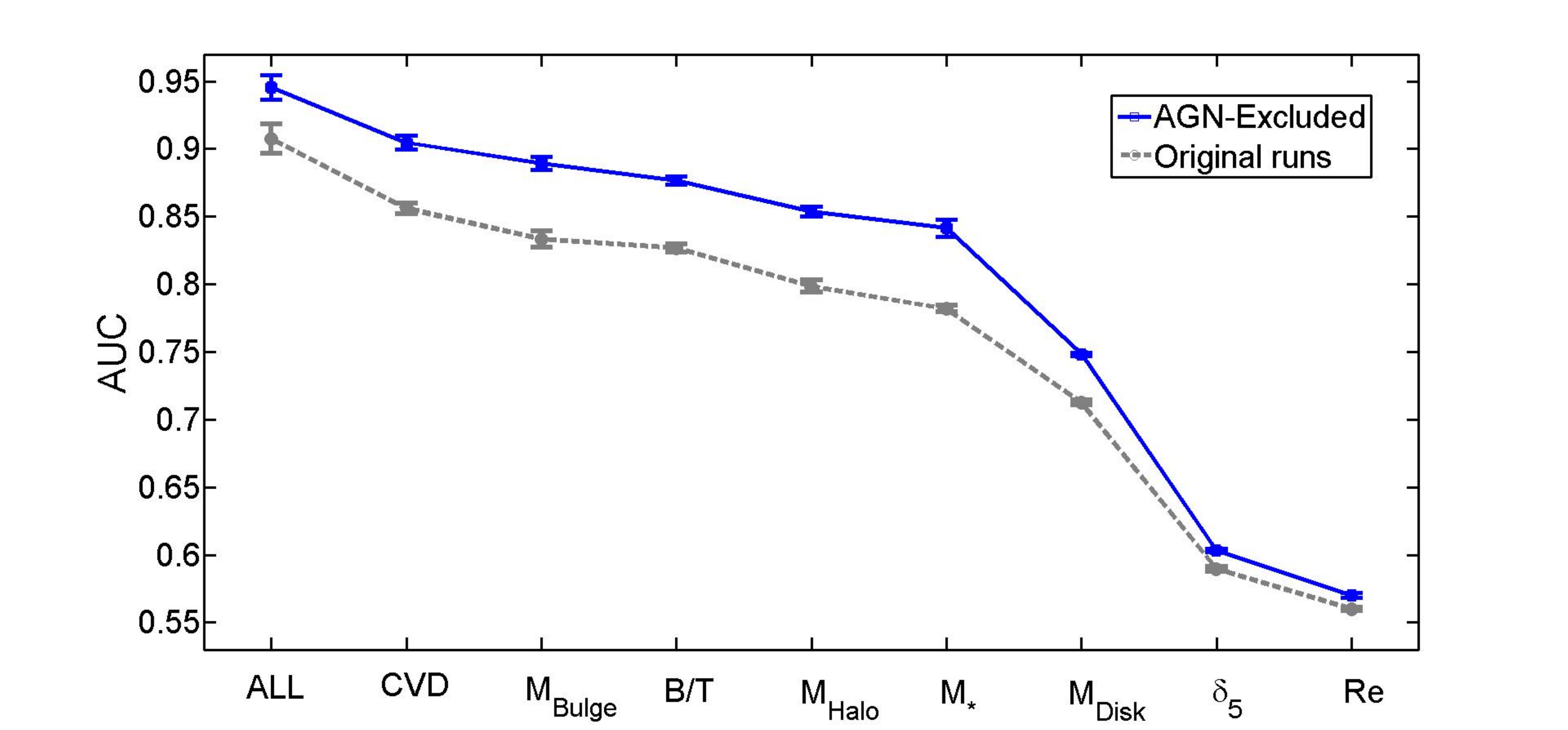}
\caption{AUC - single parameter plot. The grey dashed line is the same as Figure \ref{fig-single} which comprises all galaxies including AGNs. The blue solid line is for the sample in which AGN galaxies are excluded. We define AGN for this analysis in \S A2.}
\label{fig-agn}
\end{figure}

We use indirect means for determining the SFRs for AGN based on the empirical relationship between the strength of the 4000 \AA \hspace{0.1cm} break and the sSFR of the galaxy (see Section 2). This is necessary because AGN contribute flux to the emission lines used to determine SFRs. However, the errors in the SFRs of AGN can be significant (Rosario et al. 2015), potentially leading to misclassifications of star forming or passive systems in our training sample. Here we consider the effect of removing all AGN from our sample. We define AGN to be any galaxy which lies above the Kauffmann et al. (2003) line on the BPT diagram, at a S/N $>$ 1. We then redo our ANN analysis for single variables and ALL. 

We plot the AUC result for the non-AGN sample in Figure \ref{fig-agn} (blue line) and overplot the result for the original sample (grey line). A significant improvement in performance is seen (AUC values are generally higher). However, we find no difference in the ordering by AUC of these variables. So, whilst removing AGN from our sample (hence restricting to more reliable SFRs) improves the ANN performance, it does not affect the results of Section 4.1 in any way.

\subsection{Excluding the Green Valley}

\begin{figure}
\centering
\includegraphics[width=9cm,height=6.cm,angle=0]{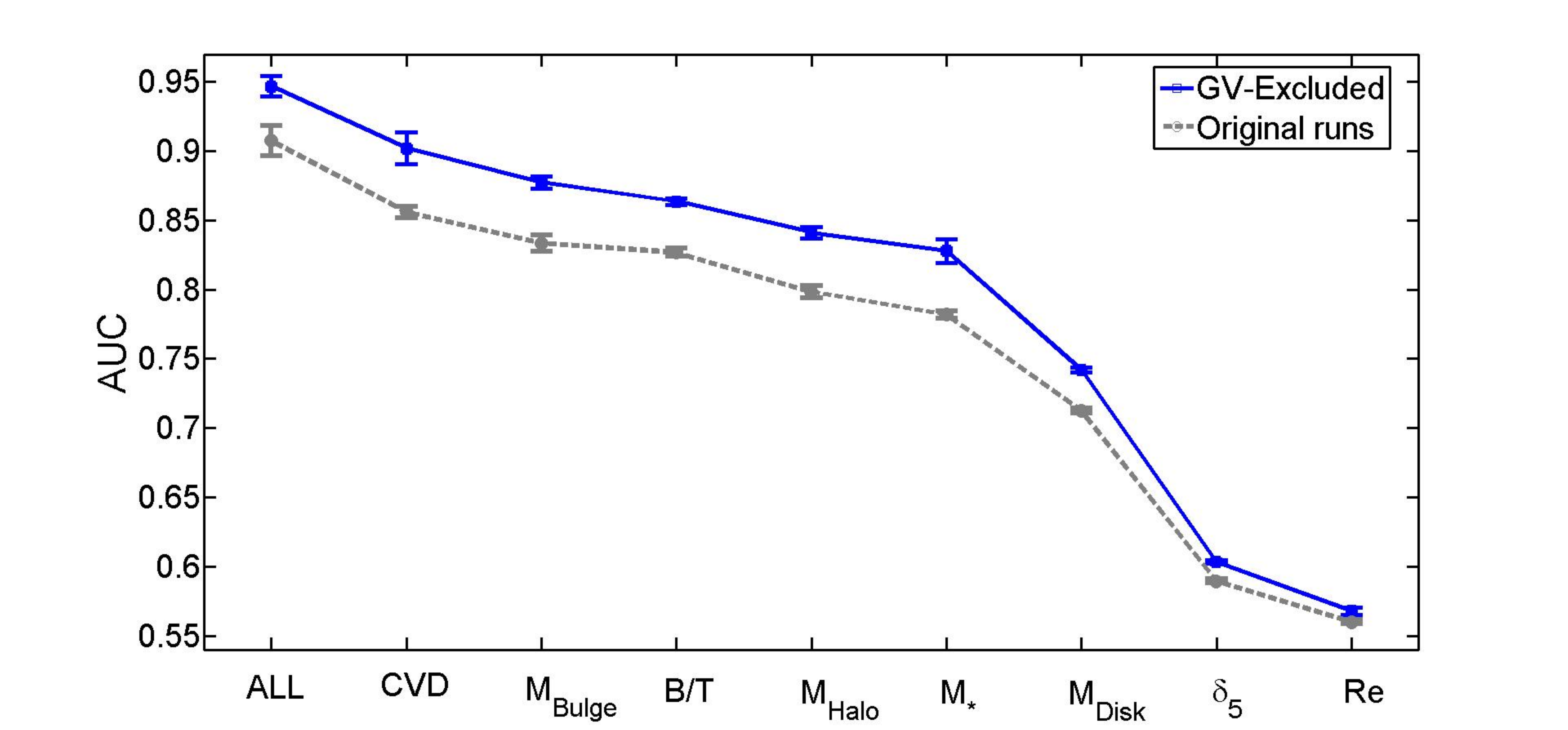}
\caption{AUC - single parameter plot. The grey dashed line is the same as Figure \ref{fig-single} which comprises all galaxies including green valley galaxies. The blue solid line is for the sample in which green valley galaxies are excluded, from both training and verification.}
\label{fig-gvx}
\end{figure}

One possible source of serious systematic error in our ANN analysis can come from our initial assumption that galaxies can be decomposed cleanly into just two (binary) states in terms of their star formation, i.e. passive or star forming. This ignores the possibility that some galaxies belong in neither of these categories. In particular, galaxies lying in the `valley' between the two peaks of $\Delta$SFR in Figure 1 are hard to place in either of these two categories. Here we follow many authors (e.g. Strateva et al. 2001, Driver et al. 2006, Schawinski et al. 2014) in considering a third case, that of the `Green Valley'. The definition for this class in terms of $\Delta$SFR is given in Section 2.3.

Figure \ref{fig-gvx} shows the result in terms of AUC for the sample with these green valley galaxies excluded (blue line), for comparison we overplot the original result for all galaxies (grey line). As with restricting the redshift range and excluding AGN, a significant improvement in the ANN performance is seen. This is as we might expect, since we are deliberately `cleaning' the sample of ambiguities. However, this restriction does not lead to any difference in the ordering by AUC of the single variables, and hence does not have any impact on the ranking of how important these variables are for quenching.

\subsection{Restricting the Velocity Dispersions}

\begin{figure}
\centering
\includegraphics[width=9cm,height=6.cm,angle=0]{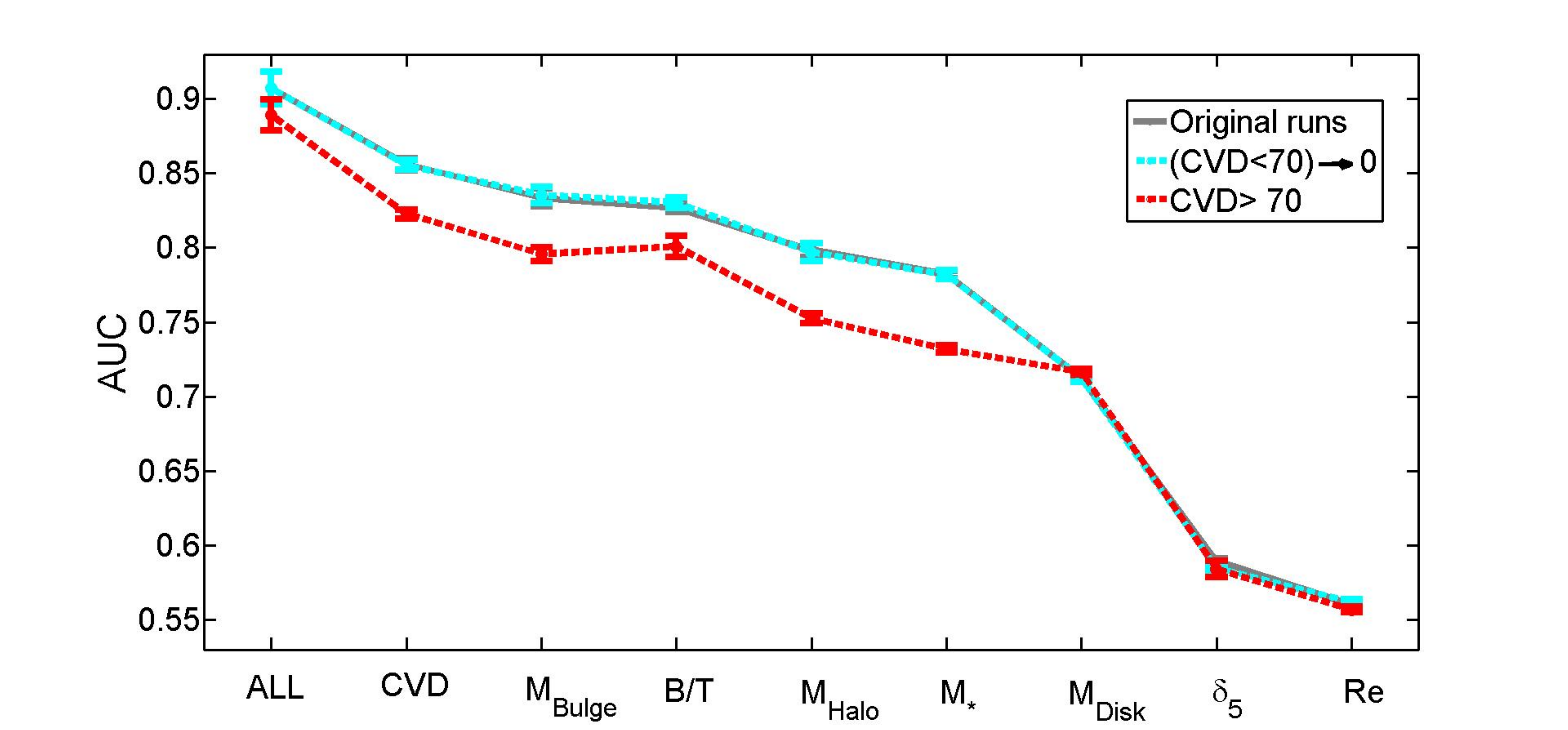}
\caption{AUC - single parameter plot. The grey dashed line is the same as Figure \ref{fig-single} which includes all velocity dispersions. The red line shows the results for a sample where velocity dispersions with $\sigma <$ 70 km/s are excluded. The light cyan line shows the result for galaxies with $\sigma <$ 70 km/s having their input values changed to 0 km/s.}
\label{fig-cvd}
\end{figure}

Velocity dispersions with values less than 70 km/s are intrinsically less reliable than those with higher values, due to the resolution of the SDSS spectra. However, placing a cut in velocity dispersion (in addition to the cut in stellar mass) would lead to a highly biased sample, where only bulge dominated galaxies (presumably more likely to be passive) are detectable at low stellar masses. So, for the initial sample we included all velocity dispersions, provided they pass our basic data quality checks (presented in Section 2). This could potentially leave us with a bias, and we investigate this possibility here.

First we restrict our sample to $\sigma >$ 70 km/s, and redo our ANN analysis for all single variables. The result of this procedure in terms of AUC is shown in Figure \ref{fig-cvd} as a red line, overplotted in grey is the original result for all $\sigma$. It is interesting to note that restricting the sample by velocity dispersion actually {\it lowers} the performance of the ANN, even for velocity dispersion itself! This is because a powerful piece of information is lost in this case. It seems that the presence of a compact (pressure supported) bulge is essential for a central galaxy to be quenched, and thus the opposite (where there is no bulge and hence low velocity dispersion) leads to a near certain classification of star forming. Removing the low  $\sigma$ cases removes the ability of the ANN code to correctly assign these cases. 
Therefore, we suggest that it is better to leave them in even though this could lead to a higher uncertainty of the ranking of $\sigma$. Nonetheless, the only change to the ranking caused by excluding the low velocity dispersions is the ordering of B/T and $M_{\rm bulge}$ (both of which are independent of the spectral resolution of the SDSS since they are determined from the photometry alone), everything else remains unchanged.

Since we are not sure of the exact values of $\sigma <$ 70 km/s due to the instrumental resolution of the SDSS, and that we posit that it is just that these values are {\it low} that is useful for the ANN code, we try setting all low velocity dispersions to zero, i.e.:

\begin{equation}
\sigma_{c} \longmapsto \hspace{0.1cm} 0 \hspace{0.1cm} ({\rm if \hspace{0.1cm} \sigma < 70km/s}) \hspace{0.1cm} || \hspace{0.1cm}  \sigma_{c} \hspace{0.1cm} ({\rm if \hspace{0.1cm} \sigma \geq 70km/s})
\end{equation}

\noindent We show the result for this sample in Figure \ref{fig-cvd} as a light cyan line. Note that it is almost identically coincident with the original sample (shown in grey). This demonstrates that no information is being derived by the ANN codes from the actual values of $\sigma <$ 70 km/s velocity dispersions, only that they are low. Thus, we conclude that including these low values in the sample is not biasing our results, and moreover is actually essential to get the most optimal (and reliable) performance (given that the grey and cyan lines lie above the red line).

\subsection{Restricting LTGs to Face-On}

\begin{figure}
\centering
\includegraphics[width=9cm,height=6.cm,angle=0]{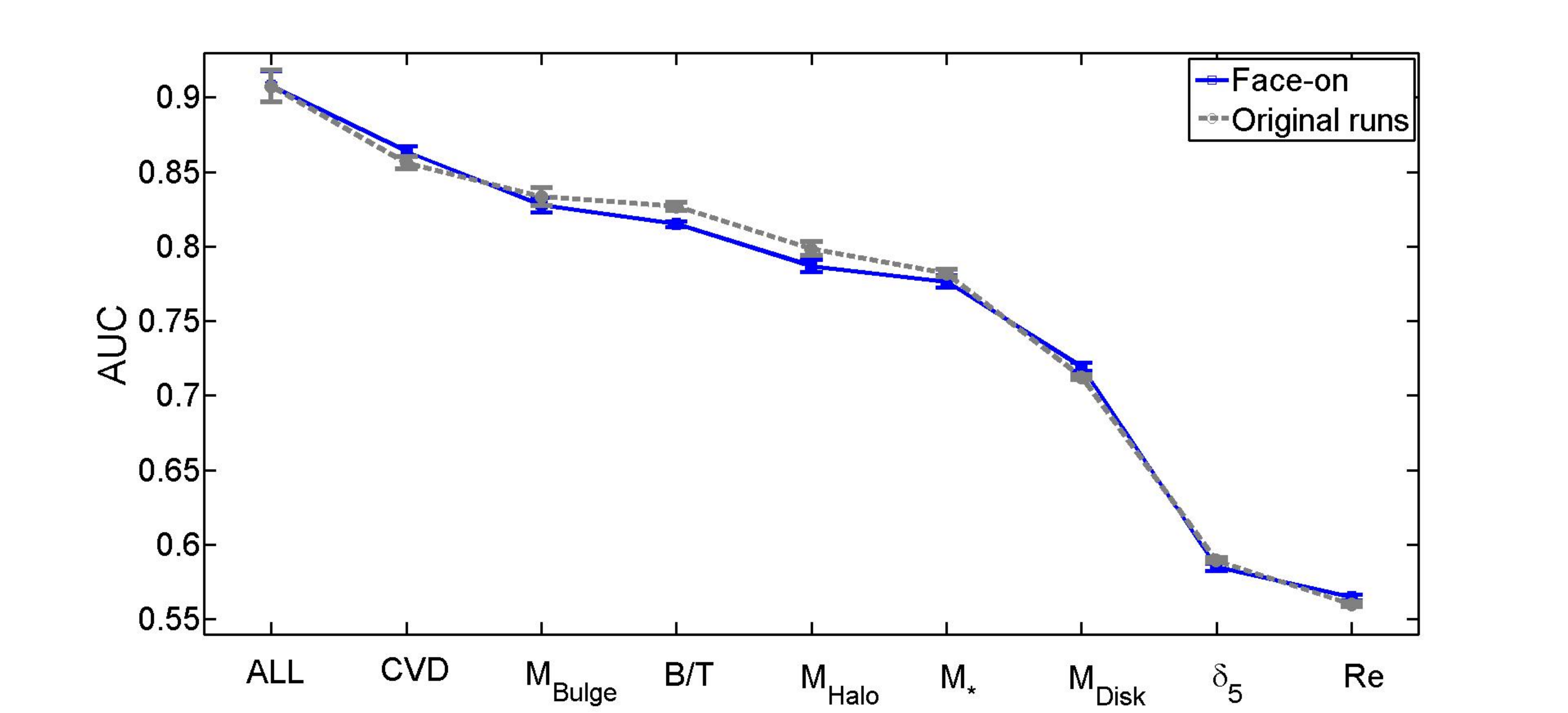}
\caption{AUC - single parameter plot. The grey dashed line is the same as Figure \ref{fig-single} which comprises all galaxies regardless of disk axis ratio. The blue solid line is for the sample in which late-type galaxies (with B/T $<$ 0.5) are restricted to being face-on (b/a $>$ 0.9).}
\label{fig-face-on}
\end{figure}

For velocity dispersions there is an ambiguity as to the source of the kinetic energy when measured via aperture spectroscopy, i.e. contributions to $\sigma$ can be made by a pressure supported bulge and/ or from disk rotation into the plane of the sky. Given that the SDSS fibre is generally centred on the middle of the galaxy light profile, for cases where the bulge dominates (and/or for very low redshifts) this effect will be small. However, where the bulge is not the dominant component of the stellar mass budget of the galaxy, significant kinematic contamination from the rotating disk can affect the measurement of $\sigma$, if the disk is inclined relative to Earth. Thus, the success of $\sigma$ and $M_{BH}$ (which is based in part on $\sigma$) in determining the passive state of galaxies in Section 4.1 could potentially be partially attributed to measuring the disk rotation in galaxies, i.e. not actually (solely) associated with the central region. We consider this possibility in this sub-section.

Here we construct a new sub-sample requiring all late-type galaxies (LTGs, defined as B/T $<$ 0.5) to be `face-on' with b/a $>$ 0.9 (inc$_{\rm disk} <$ 25$^{o}$). We take these values from photometric bulge-disk decompositions performed in Simard et al. (2011). This removes $\sim$ 90\% of LTGs but leaves the bulge dominated early-type galaxies (ETGs) unchanged. Since our purpose in this paper is to probe galaxy quenching we must carefully correct for this new bias before continuing. For this sub-sample we weight each galaxy in the ANN code by the inverse of the probability of its inclusion (which is a function of its structure, B/T), specifically we calculate (as in Bluck et al. 2015):

\begin{equation}
w_{i} = \frac{1}{1 - f_{\rm rem}(B/T)}
\end{equation}

\noindent where $f_{\rm rem}(B/T)$ is the fraction of galaxies removed from our sample due to the b/a cut of LTGs, which varies as a function of galaxy morphology for LTGs and is of value unity for ETGs (because they are not removed). This corrects for any bias in the passive : star forming ratio of the sample, but leaves us with only face-on disks, for which $\sigma$ is solely a probe of the bulge kinematics.

We show the AUC plot for this sub-sample in Figure \ref{fig-face-on} as the blue line, with the original result being shown in grey for comparison. The two lines are close to being identical, and the ordering of the variables is largely the same as well. The performance of CVD does slightly better, however, indicating that it is indeed the bulge kinematics (not disk contribution) which yields the predictive power of this variable in assigning the passive state of galaxies. Some of the other single variables perform slightly less well for this sub-sample, which is most probably explained by this dataset being statistically less rich due to the removed LTGs.

\subsection{Higher Stellar Mass Cut}

\begin{figure}
\centering
\includegraphics[width=9cm,height=6.cm,angle=0]{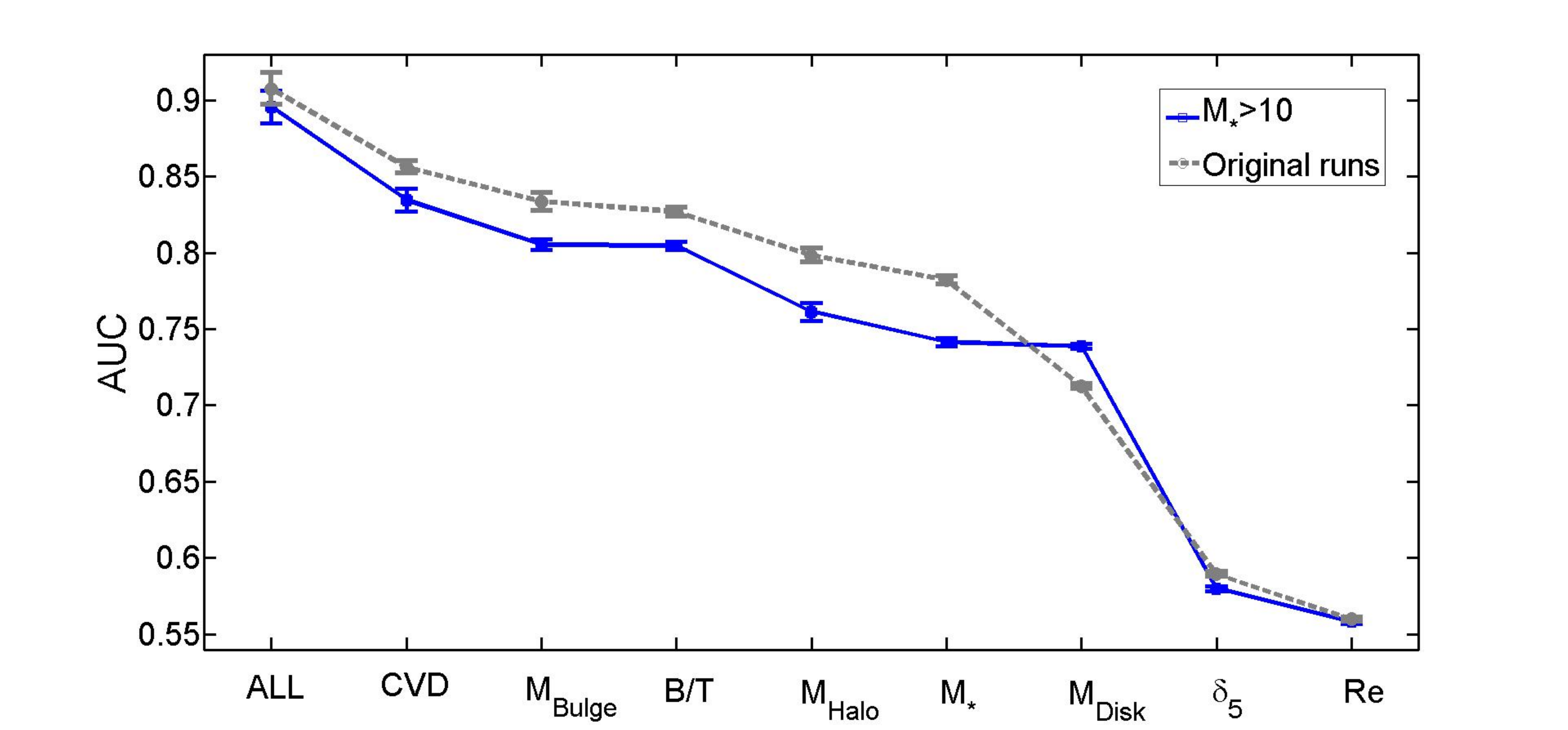}
\caption{AUC - single parameter plot. The grey dashed line is the same as Figure \ref{fig-single} which comprises galaxies with $M_{*} > 10^{9} M_{\odot}$. The blue solid line is for a restrictive sub-sample with $M_{*} > 10^{10} M_{\odot}$.}
\label{fig-m10}
\end{figure}

Our primary data sample is restricted in stellar mass to $M_{*} > 10^{9} M_{\odot}$, which is due to the relative scarcity of galaxies with lower stellar masses in the SDSS volume. For centrals, this cut in mass is probably low enough to include almost all passive galaxies (see the $1/V_{\rm max}$ weighted passive fraction - stellar mass relation presented in Fig. 8 of Bluck et al. 2014). However, it is possible that the performance of the galaxy parameters presented in Table 1 vary as a function of the stellar mass range considered. Obviously, at the extremes this will be uninteresting because all galaxies will be either passive or star forming, but at intermediate masses there may be some additional insights to be found.

In this sub-section we consider the effect on the ANN ranking of single parameters of a higher mass cut of $M_{*} > 10^{10} M_{\odot}$. Our new result is shown as a blue line in Figure \ref{fig-m10}, with the original result (for $M_{*} > 10^{9} M_{\odot}$) shown in grey for comparison. There are a few subtle differences between the mass cuts, such as disk mass performing better than stellar mass and B/T performing better than bulge mass in this sample. However, the general trend is the same, with galaxy parameters related to the inner regions of galaxies performing the best, and parameters related to the outer regions of galaxies or the local environment performing significantly worse. 
These changes do not in any way affect our conclusions, but it is interesting to note that the results from an ANN analysis of this type can in principle be affected by the range in masses of the input parameters.

\subsection{Pure Disks and Spheroids}

\begin{figure}
\centering
\includegraphics[width=9cm,height=6.cm,angle=0]{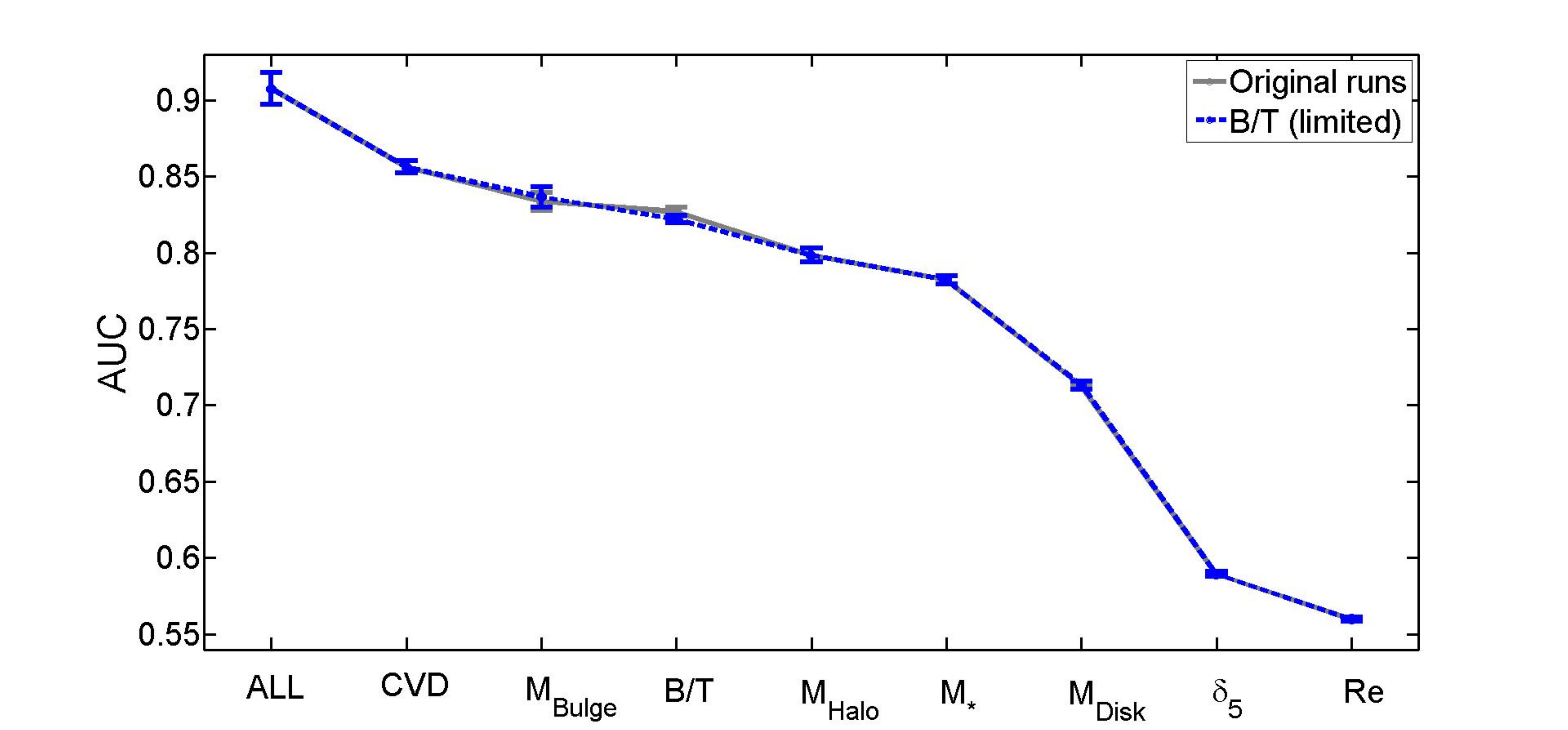}
\caption{AUC - single parameter plot. The grey dashed line is the same as our fiducial result, shown in Figure \ref{fig-single}, which takes all bulge - disk parameters at face value. The blue solid line is for a sample with low B/T galaxies re-categorised to pure disks and high B/T galaxies re-categorised to pure spheroids. The results for these two samples are identical within the errors, and thus the rankings remain unchanged.}
\label{fig-bot}
\end{figure}

 Our morphological and structural parameters come from bulge Ñ- disk decompositions performed in Simard et al. (2011) and Mendel et al. (2014). We define the structure of a galaxy to be the continuous variable B/T, which is the bulge-to-total stellar mass ratio, which is equal to one minus the disk-to-total stellar mass ratio (i.e. B/T = 1 - D/T). In this subsection we consider whether the ranking  by AUC of galaxy properties is affected by the possibility that some pure disk or pure spheroid galaxies are misclassified as composite systems. The average error on an individual B/T value is $\sim \pm$ 0.1 (see appendices in Bluck et al. 2014 for their determination from fitting of model galaxies). Thus, we allow all galaxies with B/T $<$ 0.1 to be set to pure disks and all galaxies with B/T $>$ 0.9 to be set to pure spheroids, which is permitted within their errors. Specifically, we define the following two mappings:

\begin{equation}
{\rm If} (B/T \leq 0.1) \longmapsto (B/T = 0) \& (M_{\rm bulge} = 0) \& (M_{\rm disk} = M_{*})
\end{equation}

\noindent and

\begin{equation}
{\rm If} (B/T \geq 0.9) \longmapsto (B/T = 1) \& (M_{\rm bulge} = M_{*}) \& (M_{\rm disk} = 0)
\end{equation}

\noindent We compare the AUC results for this new sample to the original runs in Figure \ref{fig-bot}. All of the parameters, including $M_{\rm bulge}$, $M_{\rm disk}$ and B/T perform identically within the errors to the original run, and hence there is no change to the ranking by AUC from possible misclassifications of pure disks or spheroids.

\subsection{Volume Limits and Weighting}

\begin{figure}
\centering
\includegraphics[width=9cm,height=6.cm,angle=0]{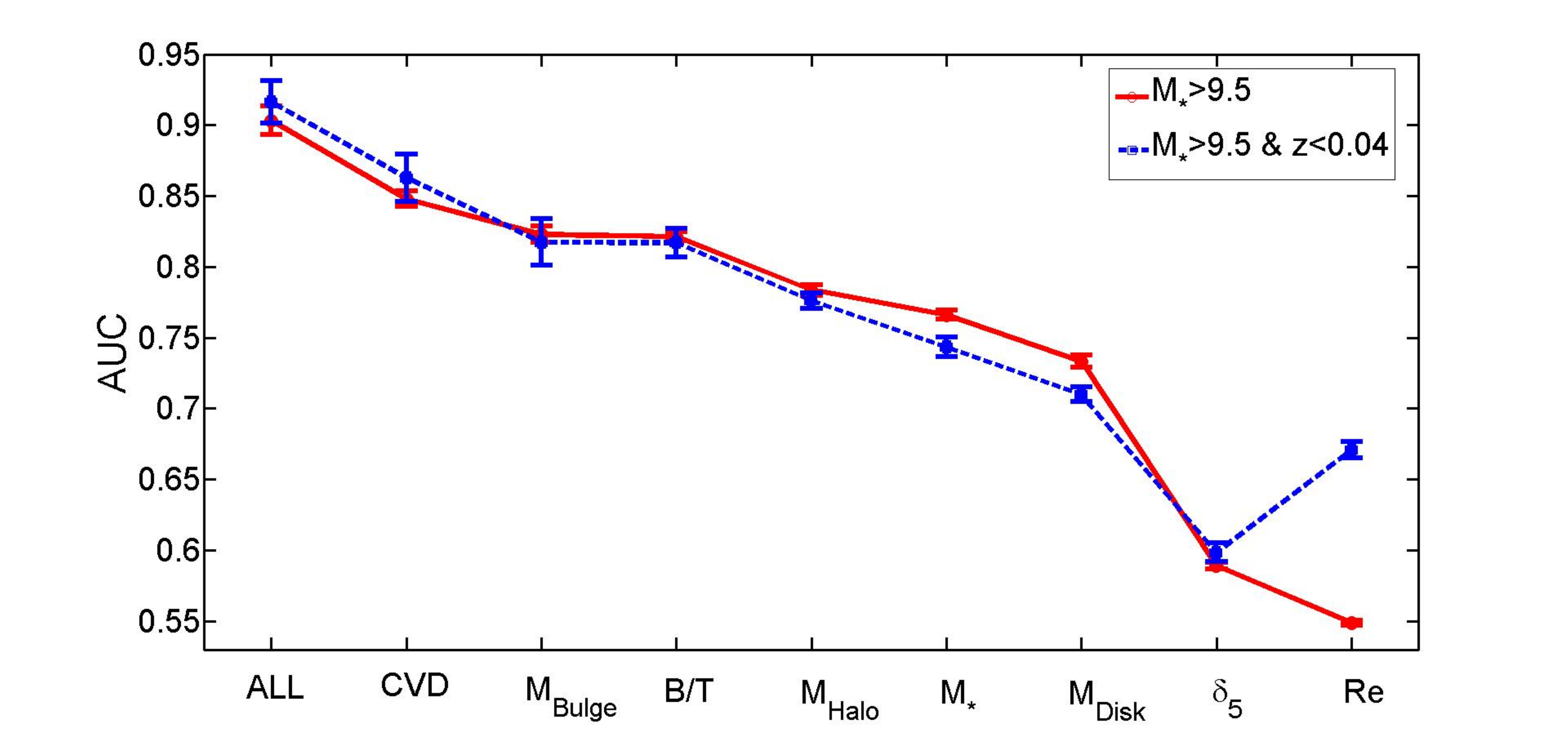}
\caption{AUC - single parameter plot. The blue line is for a mass cut of $M_{*} > 10^{9.5} M_{\odot}$. This sample is not restricted in redshift and extends out to $z_{\rm spec} <$ 0.2. The blue dashed line is for volume limited sub-sample ($z_{\rm spec} <$ 0.04).}
\label{fig-mz}
\end{figure}

\begin{figure}
\centering
\includegraphics[width=9cm,height=6.cm,angle=0]{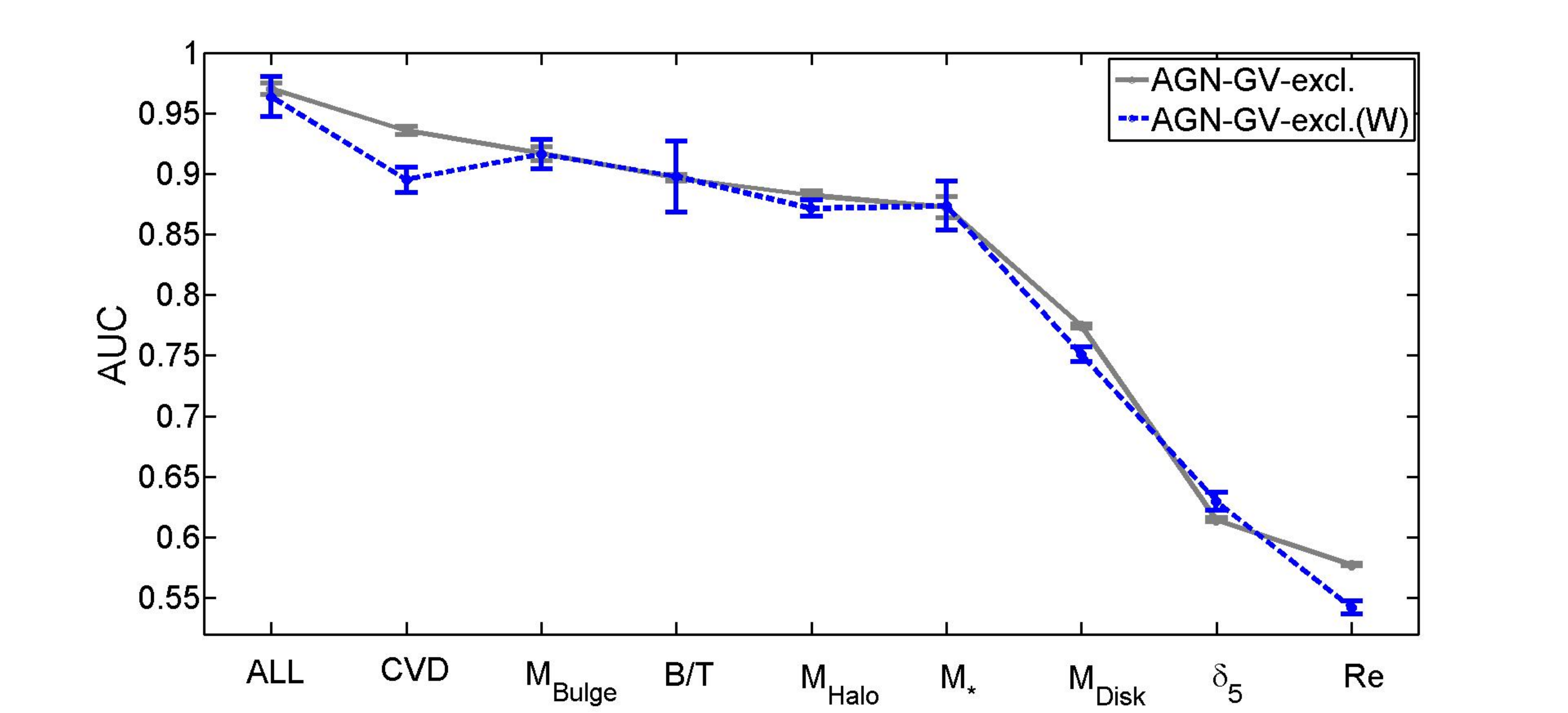}
\caption{AUC - single parameter plot. The grey line shows the original (un-weighted) sample with green valley and AGN galaxies removed (see Sections A2 and A3). The blue line shows the same sample (no GV or AGN) but now weighted by $1/V_{\rm max}$ in both the training and verification sets (see Section A7). The two lines are almost identical, with only very minor differences.}
\label{fig-agngvx-w}
\end{figure}

Due to the flux limit of the SDSS, galaxies of different masses and colours are visible in the survey to different maximum redshifts, which can lead to a bias on the ANN input sample. The usual way to deal with these effects is via volume weighting of statistics such as the passive fraction (as in, e.g., Peng et al. 2010, 2012, Woo et al. 2013, Bluck et al. 2014, 2015). The dependence of the maximum redshift, $z_{\rm max}$, each galaxy can be detected at in the SDSS on both stellar mass and (g-r) colour is presented in Figure 9 of Mendel et al. (2014). From this a maximum detection volume, $V_{\rm max}$, can be computed for each galaxy. Weighting any given statistic by $1/V_{\rm max}$ corrects for the flux limit bias. 
The alternative to weighting is, the more familiar approach of, constructing a volume limited sample, i.e. restricting the survey to a volume at which completeness is achieved for a given stellar mass (and technically colour) limit. In this sub-section we consider both of these approaches to test whether our flux limited input sample leads to any bias in the rankings of variables.  

In Figure \ref{fig-mz} we consider a slightly higher mass cut to the fiducial sample considered throughout the rest of this paper, of $M_{*} > 10^{9.5} M_{\odot}$, shown in red. This sample is not restricted in redshift and extends out to $z_{\rm spec} <$ 0.2. The blue line in Figure \ref{fig-mz} shows the AUC results for single variables for a volume limited sample where we are complete at the stellar mass limit, and at the average colour (for that mass) of the red sequence. The redshift cut for this sample is $z_{\rm spec} <$ 0.04. Generally, we find that restricting to a volume limited sample does not change our results significantly. The general trend of inner galaxy properties being more predictive of quenching than outer galaxy or environmental parameters still holds true for all samples. 
There are, however, a few small changes. The most prominent of these is that bulge effective radius performs significantly better than local density in the volume limited case but significantly worse in the flux limited case. With this one exception, the ordering of all of the rest of the parameters is identical between the flux limited and volume limited sample, thus our ranking of galaxy parameters in quenching is highly stable to issues of completion in the input sample.

Our restriction to a volume limited sample is imperfect, however, because 1) we have to assume a colour limit (here taken as the mean of the red sequence at the lower mass cut) and 2) this process necessarily reduces our sample size significantly, which impairs the power of the ANN technique. Volume weighting is a viable alternative, although there are also some issues with this approach to consider. Since the ANN procedure concentrates on finding patterns in the data, and is carefully tuned to avoid over-fitting, introducing a weight (often very large $\sim$ 100 - 1000 in some cases) can result in amplifying outliers to the status of significant sub-patterns. Thus, before weighting we must be careful to use the `cleanest' data set available, with the fewest `bad' data points. Given the results of Sections A2 and A3, where we find that excluding the green valley and AGN from our sample improves the ANN performance, we also remove these galaxies from our sample before volume weighting here.

In Figure \ref{fig-agngvx-w} we show the result of our ANN minimisation procedure for un-weighted galaxies with green valley and AGN cases removed (grey line), and the same sample weighted by $1/V_{\rm max}$ (blue line). Here weighting indicates the number of times each galaxy is included in the parent sample, and hence is closely related to the probability of inclusion in the ANN training and verification sets. The two samples agree almost identically, giving the same trend in AUC performance from inner-galaxy properties to environmental properties, seen throughout the appendix and Section 4.1. 
The biggest difference is a noticeably worse performance of CVD relative to bulge mass and ratio. This can be explained by the fact that in the volume weighted sample greater emphasis is placed on lower values of CVD, which are intrinsically less reliable (see Section 2 and A4). In the volume limited case (above) we still see CVD performing best, and this is likely because by restricting to lower redshifts we can accurately constrain CVD to lower values. However, the directionality of the trend from inner to outer regions is left unchanged by weighting, hence we conclude that our method is not significantly affected by the initial sample setup.

The primary invariance of our method to volume effects is likely a result of us selecting the same number of PA and SF galaxies for both training and verification. This reduces the effect of colour (or SFR) on our sample selection, and hence also reduces the impact of stellar mass detection thresholds, due to the strong correlation between $M_{*}$ and SFR or (g-r) colour. In any case, the impact of volume weighting, or restricting to a volume limited sample, is very minor on our rankings and results.

\begin{figure*}
\centering
\includegraphics[width=20cm,height=11.5cm,angle=0]{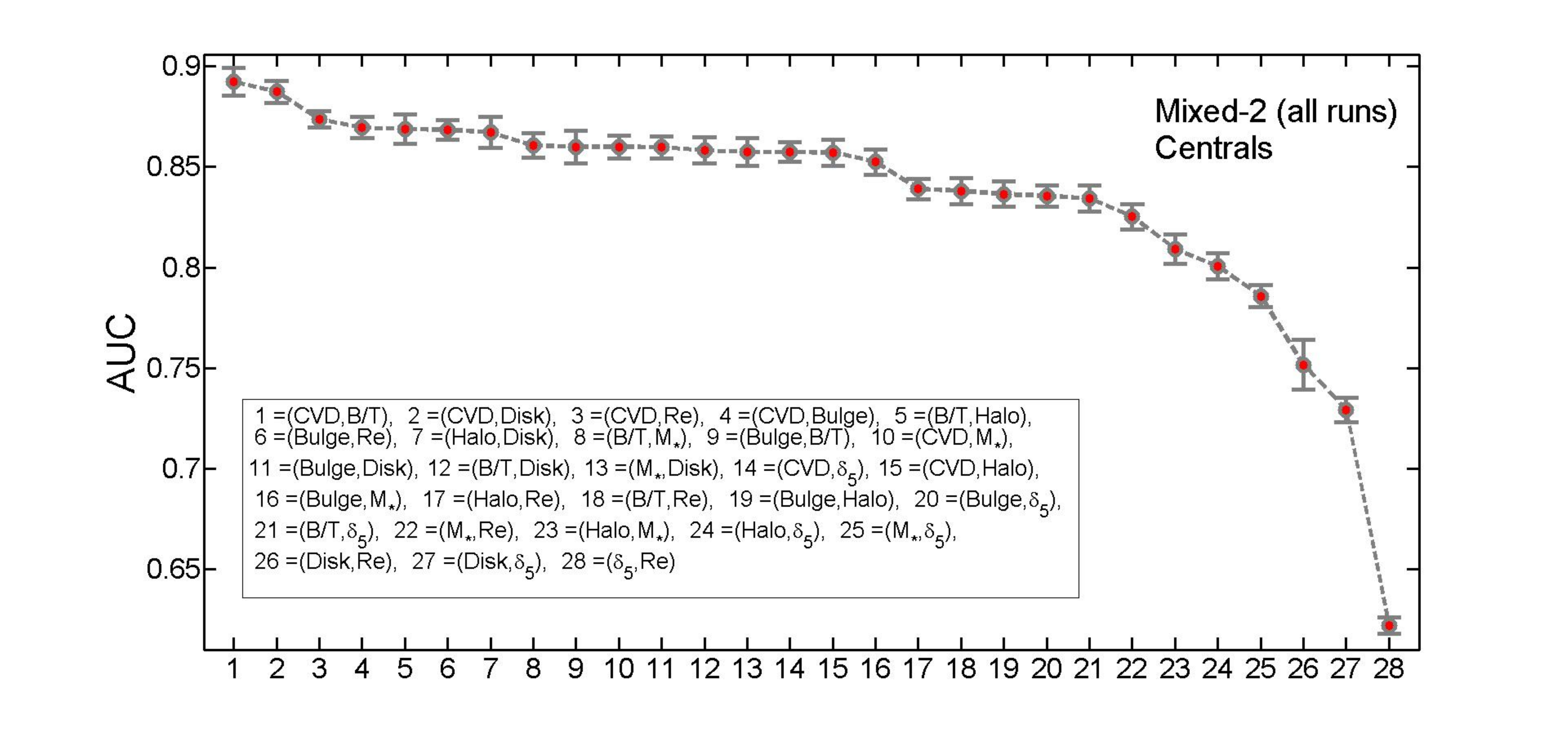}
\includegraphics[width=20cm,height=10cm,angle=0]{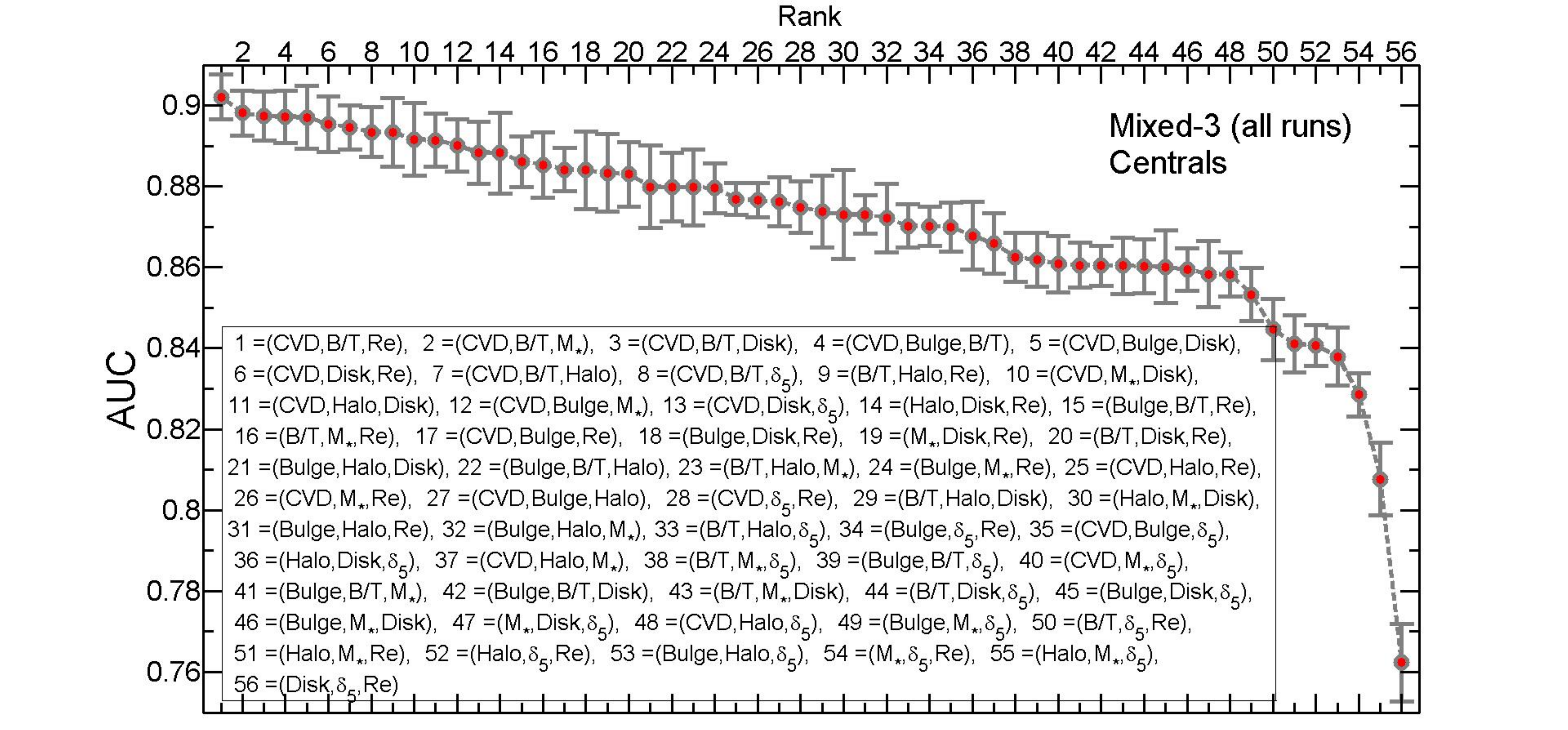}
\caption{AUC - parameter plot for all multiple runs for centrals. The top plot shows all possible combinations of two parameters as input data, and the bottom plot shows all possible combinations of three parameters as input data. They are both ordered from most to least predictive at determining the passive state of galaxies.}
\label{fig-mix-all}
\end{figure*}


\begin{thebibliography}{}
\bibitem[ et al. ()]{}
Abazajian K. et al., 2003, AJ, 126, 2081
\bibitem[ et al. ()]{} 
Abazajian K. et al., 2009, ApJS, 182, 543
\bibitem[ et al. ()]{}
Abramson L. E. et al., 2014, ApJ, 785L, 36
\bibitem[ et al. ()]{}
Andreon S. et al., 2000, MNRAS, 319, 700
\bibitem[ et al. ()]{}
Baldry I. K., Glazebrook K., Brinkmann J., Ivezic Z., Lupton R. H., Nichol R. C., Szalay A. S., 2004, ApJ, 600, 681
\bibitem[ et al. ()]{}
Baldry I. K., Balogh M. L., Bower R. G., Glazebrook K., Nichol R. C., Bamford S. P., Budavari T., 2006, MNRAS, 373, 469
\bibitem[ et al. ()]{}
Baldwin J. A., Phillips M. M., Terlevich R., 1981, PASP, 93, 5
\bibitem[ et al. ()]{}
Ball N. M. et al., 2004, MNRAS, 348, 1038
\bibitem[ et al. ()]{}
Balogh M. L., Baldry I. K., Nichol R., Miller C., Bower R., Glazebrook K., 2004, ApJ, 615L, 101
\bibitem[ et al. ()]{}
Barnes J. E., Hernquist L., 1992, ARA\&A, 30, 705
\bibitem[ et al. ()]{}
Bell E. F., 2008, ApJ, 682, 355
\bibitem[ et al. ()]{}
Bell E. F. et al., 2012, ApJ, 753, 167
\bibitem[ et al. ()]{}
Bernardi M. et al., 2003, AJ, 125, 1817
\bibitem[ et al. ()]{}
Bernardi M., Hyde J. B., Sheth R. K., Miller C. J., Nichol R. C., 2007, AJ, 133, 1741
\bibitem[ et al. ()]{}
Bishop, C. M. 1995, Neural Networks for Pattern Recognition (Oxford: Oxford University Press)
\bibitem[ et al. ()]{}
Bluck A. F. L., Conselice C. J., Almaini O., Laird E., Nandra K. \& Gruetzbauch R., 2011, MNRAS, 410, 1174
\bibitem[ et al. ()]{}
Bluck A. F. L., Mendel J. T., Ellison S. L., Moreno J., Simard L., Patton D. R., Starkenburg E., 2014, MNRAS, 441, 599
\bibitem[ et al. ()]{}
Bluck A. F. L. et al., 2015, MNRAS submitted (arXiv:1412.3862B)
\bibitem[ et al. ()]{}
Bower R. G. et al., 2006, MNRAS, 370, 645
\bibitem[ et al. ()]{}
Bower R. G., McCarthy I. G., Benson, A. J., 2008, MNRAS, 390, 1399
\bibitem[ et al. ()]{}
Bradley A.P., 1997, `The use of the area under the ROC curve in the evaluation of machine learning algorithms', Pattern Recog., 30 (7), 1145-1159
\bibitem[ et al. ()]{}
Brinchmann J., Charlot S., White S. D. M., Tremonti C., Kauffmann G., Heckman T., Brinkmann J., 2004, MNRAS, 351, 1151 
\bibitem[ et al. ()]{}
Burkert A., Naab T., 2004, Carnegie Observatories Astrophysics Series, Vol. 1: `Coevolution of Black Holes and Galaxies', ed. Ho L. C. (Pasadena:Carnegie Obs), (arXiv:astro-ph/0305076)
\bibitem[ et al. ()]{}
Butcher H., Oemler A., 1984, ApJ, 285, 426
\bibitem[ et al. ()]{}
Cameron E., Driver S. P., Graham A. W., Liske J., 2009, ApJ, 699, 105
\bibitem[ et al. ()]{}
Cameron E., Driver S. P., 2009, A\&A, 493, 489
\bibitem[ et al. ()]{}
Carollo C. M. et al., 2013, ApJ, 773, 112
\bibitem[ et al. ()]{}
Catinella B. et al., 2010, MNRAS, 403, 683
\bibitem[ et al. ()]{}
Cheung E. et al., 2012, ApJ, 760, 131
\bibitem[ et al. ()]{}
Cicone C. et al., 2013, A\&A in press, arXiv:1311.2595
\bibitem[ et al. ()]{}
Cole S., Lacey C. G., Baugh C. M., Frenk C. S., 2000, MNRAS, 319, 168
\bibitem[ et al. ()]{}
Cortese L., Gavazzi G., Boselli A., Franzetti P., Kennicutt R. C., O'Neil K., Sakai S., 2006, A\&A, 453, 847
\bibitem[ et al. ()]{}
Cortiglioni F. et al., 2001, ApJ, 556, 937
\bibitem[ et al. ()]{}
Croton D. J. et al., 2006, MNRAS, 365, 11
\bibitem[ et al. ()]{}
Darg D. W. et al., 2010a, MNRAS, 401, 1552
\bibitem[ et al. ()]{}
Dalla V. C., Schaye J., 2008, MNRAS, 387, 1431
\bibitem[ et al. ()]{}
De Lucia G., Springel V., White S. D. M., Croton D., Kauffmann G., 2006, MNRAS, 366, 499
\bibitem[ et al. ()]{}
De Lucia G., Blaizot J., 2007, MNRAS, 375, 2
\bibitem[ et al. ()]{}
Dekel A., Birnboim Y., 2006, MNRAS, 368, 2
\bibitem[ et al. ()]{}
Dekel A. et al., 2009, Natur., 457, 451
\bibitem[ et al. ()]{}
Dekel A., Burkert A., 2014, MNRAS, 438, 1870
\bibitem[ et al. ()]{}
Di Matteo T., Springel V., Hernquist L., 2005, Natur., 433, 604
\bibitem[ et al. ()]{}
Djorgovski S., Davis M., 1987, ApJ, 313, 59
\bibitem[ et al. ()]{}
Dressler A., 1980, ApJ, 236, 351
\bibitem[ et al. ()]{}
Dressler A., Lynden-Bell D., Burstein D., Davies R. L., Faber S. M., Terlevich R., Wegner G., 1987, ApJ, 313, 42
\bibitem[ et al. ()]{}
Drinkwater M. J. et al., 2010, MNRAS, 401, 1429
\bibitem[ et al. ()]{}
Driver S. P. et al., 2006, MNRAS, 368, 414
\bibitem[ et al. ()]{}
Driver S. P. et al., 2011, MNRAS, 413, 971
\bibitem[ et al. ()]{}
Dunn J. P. et al., 2010, ApJ, 709, 611
\bibitem[ et al. ()]{}
Ellison S. L., Patton D. R., Simard L., McConnachie A. W., 2008, AJ, 135, 1877
\bibitem[ et al. ()]{}
Ellison S. L., Mendel J. T., Patton D. R., Scudder J. M., 2013, MNRAS, 435, 3627
\bibitem[ et al. ()]{}
Ellison S. L., Patton D. R., Hickox R. C., 2015, MNRAS, 451L, 35
\bibitem[ et al. ()]{}
Ellison S. L., Fertig D., Rosenberg J. L., Nair P., Simard L., Torrey P., Patton D. R., 2015, MNRAS, 448, 221
\bibitem[ et al. ()]{}
Fabian A. C., 1999, 308L, 59
\bibitem[ et al. ()]{}
Fabian A. C., 2012, 2012, ARA\&A, 50, 455
\bibitem[ et al. ()]{}
Fang J. J., Faber S. M., Koo D. C., Dekel A., 2013, ApJ, 776, 63
\bibitem[ et al. ()]{}
Fawcett, T., 2006, `An introduction to ROC analysis'  Pattern Recognition Letters 27, 861-874
\bibitem[ et al. ()]{}
Ferrarese L., Merritt D., 2000, ApJ, 539L, 9
\bibitem[ et al. ()]{}
Gebhardt K. et al., 2000, ApJ, 539, 13 
\bibitem[ et al. ()]{}
Genzel R. et al., 2015, ApJ, 800, 20
\bibitem[ et al. ()]{}
Guo Q. et al., 2011, MNRAS, 413, 101
\bibitem[ et al. ()]{}
Haring N., Rix H., 2004, ApJ, 604L, 89
\bibitem[ et al. ()]{}
Henriques B. et al., 2014, MNRAS submitted, arXiv:1410.0365
\bibitem[ et al. ()]{}
Hirschmann M., De Lucia G., Iovino A., Cucciati O., 2013, MNRAS, 433, 1479
\bibitem[ et al. ()]{}
Hopkins P. F., Hernquist L., Cox T. J., Di Matteo T., Robertson B., Springel V., 2006a, ApJS, 163, 1
\bibitem[ et al. ()]{}
Hopkins P. F., Hernquist L., Cox T. J., Robertson B., Springel V., 2006b, ApJS, 163, 50
\bibitem[ et al. ()]{}
Hopkins, P. F., Hernquist L., Cox T. J., Robertson B., Krause E., 2007, ApJ, 669, 67
\bibitem[ et al. ()]{}
Hopkins P. F., Hernquist L., Cox T. J., Keres D., 2008, ApJS, 175, 356
\bibitem[ et al. ()]{}
Hopkins P. F. et al., 2010, ApJ, 715, 202
\bibitem[ et al. ()]{}
Hopkins P. F., Cox T. J., Hernquist L., Narayanan D., Hayward C. C., Murray N., 2013, MNRAS, 430, 1901
\bibitem[ et al. ()]{}
Hosmer D. W., Lameshow S., 2000, `Applied Logistic Regression' 2nd ed., Wiley \& Sons, Pp. 156 - 164 
\bibitem[ et al. ()]{}
Hou L., Wang Y., 2015, ARA\&A, 15, 651
\bibitem[ et al. ()]{}
Hung C. et al., 2013, ApJ, 778, 129
\bibitem[ et al. ()]{}
Ivezic Z. et al., 2008, LSST Document, arXiv:0805.2366
\bibitem[ et al. ()]{}
Jorgensen I., Franx M., Kjaergaard P., 1995, MNRAS, 276, 1341
\bibitem[ et al. ()]{}
Jorgensen I., Franx M., Kjaergaard P., 1996, MNRAS, 280, 167
\bibitem[ et al. ()]{}
Kauffmann G. et al., 2003, MNRAS, 346, 1055
\bibitem[ et al. ()]{}
Knobel C., Lilly S. J., Woo J., Kovac K., 2015, ApJ, 800, 24
\bibitem[ et al. ()]{}
Lang P. et al., 2014, ApJ, 788, 11
\bibitem[ et al. ()]{}
Magorrian J. et al., 1998, AJ, 115, 2285
\bibitem[ et al. ()]{}
Martig M., Bournaud F., Teyssier R., Dekel A., 2009, ApJ, 707, 250
\bibitem[ et al. ()]{}
McConnell N. J. et al. 2011, Nature, 480, 215
\bibitem[ et al. ()]{}
McConnell N. J., Ma C. P., 2013, ApJ, 764, 184
\bibitem[ et al. ()]{}
McNamara B. R. et al., 2000, ApJ, 534L, 135
\bibitem[ et al. ()]{}
McNamara B. R., Nulsen P. E. J., 2007, ARA\&A, 45, 117
\bibitem[ et al. ()]{}
Mendel J. T., Simard L., Palmer M., Ellison S. L., Patton D. R., 2014, ApJS, 210, 3
\bibitem[ et al. ()]{}
Moran S. M., Ellis R. S., Treu T., Smith G. P., Rich R. M., Smail I., 2007, ApJ, 671, 1503
\bibitem[ et al. ()]{}
Moreno J., Bluck A. F. L., Ellison S. L., Patton D. R., Torrey P., Moster B. P., 2013, MNRAS, 436, 1765
\bibitem[ et al. ()]{}
Moreno J., Torrey P., Ellison S. L., Patton D. R., Bluck A. F. L., Bansal G., Hernquist L., 2015, MNRAS, 448, 1107
\bibitem[ et al. ()]{}
Moster B. P., Somerville R. S., Maulbetsch C., van den Bosch F. C., Maccio, A. V., Naab T., Oser L., 2010, ApJ, 710, 903
\bibitem[ et al. ()]{}
Nulsen P. E. J., McNamara B. R., Wise M. W., David L. P., 2005, ApJ, 628, 629
\bibitem[ et al. ()]{}
Omand C. M. B., Balogh M. L., Poggianti B. M., 2014, MNRAS, 440, 843O
\bibitem[ et al. ()]{}
Patton D. R., Torrey P., Ellison S. L., Mendel J. T., Scudder J. M., 2013, MNRAS, 433L, 59
\bibitem[ et al. ()]{}
Peng Y. et al., 2010, ApJ, 721, 193
\bibitem[ et al. ()]{}
Peng Y., Lilly S. J., Renzini A., Carollo M., 2012, ApJ, 757, 4
\bibitem[ et al. ()]{}
Peng Y., Maiolino R., Cochrane R., 2015, Nature, 521, 192
\bibitem[ et al. ()]{}
Rosario D. J., Mendel J. T., Ellison S. L., Lutz D., Trump J. R., 2015, MNRAS, submitted
\bibitem[ et al. ()]{}
Saintonge A. et al., 2011, MNRAS, 415, 32
\bibitem[ et al. ()]{}
Salim S. et al., 2007, ApJS, 173, 267
\bibitem[ et al. ()]{}
Schawinski K. et al., 2014, MNRAS, 440, 889
\bibitem[ et al. ()]{}
Schaye J. et al., 2015, MNRAS, 446, 521
\bibitem[ et al. ()]{}
Scudder J. M., Ellison S. L., Torrey P., Patton D. R., Mendel J. T., 2012, MNRAS, 426, 549
\bibitem[ et al. ()]{}
Shull J. M., Smith B. D., Danforth C. W., 2012, ApJ, 759, 23
\bibitem[ et al. ()]{}
Silk J,. Rees M. J., 1998, A\&A, 331, 1
\bibitem[ et al. ()]{}
Simard L. et al., 2002, ApJS, 142, 1
\bibitem[ et al. ()]{}
Simard L., Mendel J. T., Patton D. R., Ellison S. L., McConnachie A. W., 2011, ApJ, 196, 11
\bibitem[ et al. ()]{}
Soltan A., 1982, MNRAS, 200, 115
\bibitem[ et al. ()]{}
Somerville R. S., Hopkins P. F., Cox T. J., Robertson B. E., Hernquist L., 2008, MNRAS, 391, 481
\bibitem[ et al. ()]{}
Somerville R. S., Dave R., 2015, ARA\&A, 53, 51
\bibitem[ et al. ()]{}
Springel V. et al., 2005, Nature, 435, 629
\bibitem[ et al. ()]{}
Springel V., Hernquist L., 2005, ApJ, 622L, 9
\bibitem[ et al. ()]{}
Strateva I. et al., 2001, AJ, 122, 1861
\bibitem[ et al. ()]{}
Tacchella S. et al., 2015, Science, 348, 314
\bibitem[ et al. ()]{}
Tasca L. A. M. et al., 2009, A\&A, 503, 379
\bibitem[ et al. ()]{}
Teimoorinia H., 2012, AJ, 144,172
\bibitem[ et al. ()]{}
Teimoorinia H., Ellison S., 2014, MNRAS, 439, 3526
\bibitem[ et al. ()]{}
Toomre A., Toomre J., 1972, ApJ, 178, 623
\bibitem[ et al. ()]{}
Torrey P. et al., 2015, MNRAS, 454, 2770
\bibitem[ et al. ()]{}
van den Bosch F. C. et al., 2007, MNRAS, 376, 841
\bibitem[ et al. ()]{}
van den Bosch F. C. et al., 2008, MNRAS, 387, 79
\bibitem[ et al. ()]{}
Vogelsberger M. et al., 2014a, Nature, 509, 177
\bibitem[ et al. ()]{}
Vogelsberger M. et al., 2014b, 2014, MNRAS, 444, 1518
\bibitem[ et al. ()]{}
Wake D. A., van Dokkum P. G., Franx M., 2012, ApJ, 751L, 44
\bibitem[ et al. ()]{}
Wetzel A. R., Tinker J. L., Conroy C., van den Bosch F. C., 2013, MNRAS, 432, 336
\bibitem[ et al. ()]{}
Wichchukit S., O'Mahony M., J. Food Sci., 2010, 75(9), R183-R193
\bibitem[ et al. ()]{}
Wuyts S. et al., 2011, ApJ, 742, 96
\bibitem[ et al. ()]{}
Woo J. et al., 2013, MNRAS, 428, 3306
\bibitem[ et al. ()]{}
Woo J., Dekel A., Faber S. M., Koo D. C., 2015, MNRAS, 448, 237
\bibitem[ et al. ()]{}
Yang X., Mo H. J., van den Bosch F. C., Pasquali A. L., Cheng B. M., 2007, ApJ, 671, 153
\bibitem[ et al. ()]{}
Yang X., Mo H. J., van den Bosch F. C., 2008, ApJ, 676, 248
\bibitem[ et al. ()]{}
Yang X., Mo H. J., van den Bosch F. C., 2009, ApJ, 695, 900


\end{thebibliography}
\end{document}